\RequirePackage{fix-cm}
\documentclass[twocolumn]{svjour3}          
\smartqed  
\usepackage{graphicx}
\usepackage[colorlinks=true, linkcolor=blue, urlcolor=blue, citecolor=blue]{hyperref}
\usepackage{booktabs}
\usepackage{comment}
\usepackage{color, colortbl}
\definecolor{Gray1}{gray}{0.7}
\definecolor{Gray2}{gray}{0.9}
\usepackage{arydshln}
\usepackage{subfigure}
\usepackage{amsmath}
\DeclareUnicodeCharacter{00A0}{ }

\journalname{Preprint: Submitted to ArXiv}
\begin{document}

\title{Computer-Aided Diagnosis of Thoracic Diseases in Chest X-rays using hybrid CNN-Transformer Architecture
}


\author{Sonit Singh
}

\institute{Sonit Singh \at
              School of Computer Science and Engineering \\
              UNSW Sydney, Australia \\
              \email{sonit.singh@unsw.edu.au}}


\maketitle

\begin{abstract}
    Medical imaging has been used for diagnosis of various conditions, making it one of the most powerful resources for effective patient care. Due to widespread availability, low cost, and low radiation, chest X-ray is one of the most sought after radiology examination for the diagnosis of various thoracic diseases. Due to advancements in medical imaging technologies and increasing patient load, current radiology workflow faces various challenges including increasing backlogs, working long hours, and increase in diagnostic errors. An automated computer-aided diagnosis system that can interpret chest X-rays to augment radiologists by providing actionable insights has potential to provide second opinion to radiologists, highlight relevant regions in the image, in turn expediting clinical workflow, reducing diagnostic errors, and improving patient care. In this study, we applied a novel architecture augmenting the DenseNet121 Convolutional Neural Network (CNN) with multi-head self-attention mechanism using transformer, namely \emph{SA-DenseNet121}, that can identify multiple thoracic diseases in chest X-rays. The work is based on attention augmented CNN~\cite{bello2019} which showed that combining convolutions and self-attention gives improved results on various computer vision tasks. We conducted experiments on four of the largest chest X-ray datasets, namely, ChestX-ray14, CheXpert, MIMIC-CXR-JPG, and IU-CXR. Experimental results in terms of area under the receiver operating characteristics (AUC-ROC) shows that augmenting CNN with self-attention has potential in diagnosing different thoracic diseases from chest X-rays. The proposed methodology has the potential to support the reading workflow, improve efficiency, and reduce diagnostic errors.

\end{abstract}

\section{Introduction}\label{intro}
With the rise of artificial intelligence (AI) technologies, health care sector is rapidly adopting to improve clinical workflow as well as improving the practices how health care professionals engage in their routine tasks~\cite{Raghupathi:2014:big_data_analytics_in_healthcare}. Major factors that contribute towards this include digitisation of health records, release of large-scale medical datasets, advancements in learning algorithms, and increasing compute power. Medical imaging accounts for about $90\%$ of total healthcare data. Among the various imaging modalities, chest X-ray imaging is currently one of the most widely used radiological examination for screening and diagnosis of various lung related diseases, including pneumothorax, cardiomegaly, and pneumonia. Chest X-ray is one of the most preferred chest imaging examination, owing to its affordable pricing, quick turnaround, less radiations, and wider availability across healthcare systems in the world. There are approximately 2 billion chest X-ray examinations per year for screening, diagnosis and management of various diseases~\cite{Raoof:Feigin:2012:interpretation_of}. According to an estimate by the World Health Organisation, two thirds of the global population lacks access to radiology diagnostics~\cite{Mollura:2010:white_paper,Rajpurkar:Irvin:2017:CheXNet}. In remote areas around the world, even though there are imaging equipment available, there is shortage of radiologists for interpreting those images~\cite{Kesselman:2016}.  This may lead to delayed diagnosis or even death of patient in rare cases. With increase in global population, the number of radiologists entering the workforce is quite low compared to increasing number of patient scans. Due to increasing workload, many radiologists have to read more than 100 X-ray studies daily~\cite{Gundel:2019:Learning_to_recognise_abnormalities}. Given interpreting chest X-rays is a challenging and time-consuming task, even experienced radiologists can make subjective assessment errors due to subtle differences in visual features of diseases. This makes it essential to develop computer-aided diagnosing and screening tools that can efficiently and accurate detect diseases on chest radiographs. These CAD tools help radiologists by providing actionable insights by providing them ``second opinion", support in expediting clinical workflow, enhancing the confidence of the radiologist by highlighting relevant region in images, and helping in prioritising the reading list where critical cases would be read first. 

Several researchers have studied thoracic disease identification and localisation. Most of these studies depend on convolutional neural networks and visual attention mechanism. Recently, the Natural Language Processing community has been shifting from recurrent models plus attention mechanism to attention-only based models such as transformers. The transformer network outperformed recurrent neural network (RNN) and temporal convolutional network (TCN) on various NLP tasks, including machine translation, speech recognition, and language modeling. The key module of the transformer network is \emph{multi-head attention} (MHA). The MHA block utilises multiple heads, with each employing an attention mechanism to capture both global and local features. In this work, we investigate multi-head self-attention convolutional network for the diagnosis of thoracic diseases. This is motivated by the following points: (1) multi-head attention network can capture both global and local features, which are important for the accurate identification of thoracic diseases; (2) self-attention based models proved to be faster for training and can leverage GPU parallelisation; (3) most NLP research has benefited from transformer based models.

\begin{figure*}
    \centering
    \includegraphics[width=\textwidth, height=8cm]{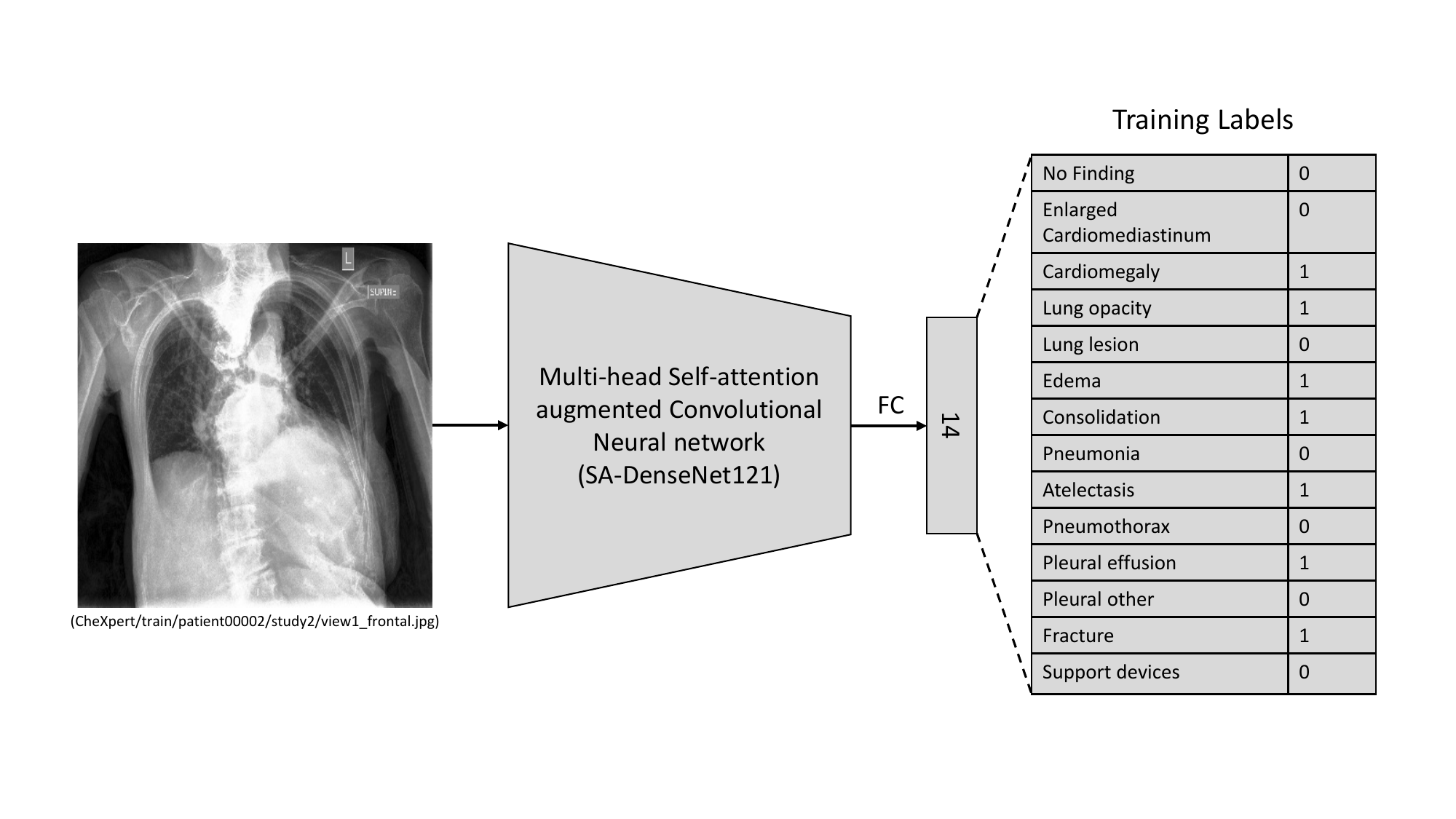}
    \caption{Block diagram of multi-head self-attention based convolutional network for thoracic diseases identification.}
    \label{fig:block_diagram}
\end{figure*}

\begin{figure*}[tb]
  \centering
 \subfigure[ChestX-ray14]{\includegraphics[width=3.25cm, height=3.5cm]{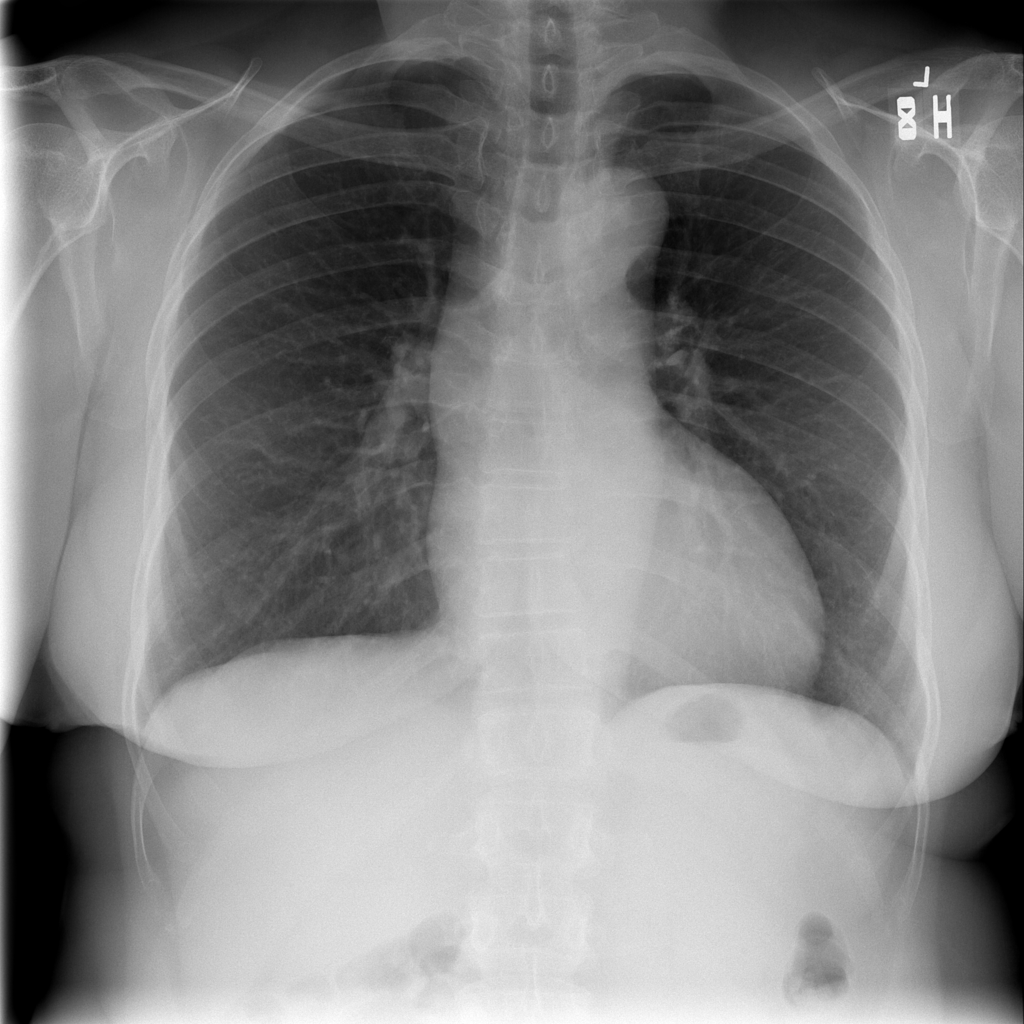}}\quad
  \subfigure[CheXpert]{\includegraphics[width=3.25cm, height=3.5cm]{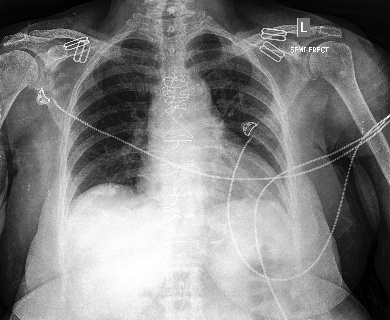}}\quad
  \subfigure[MIMIC-CXR-JPG]{\includegraphics[width=3.25cm, height=3.5cm]{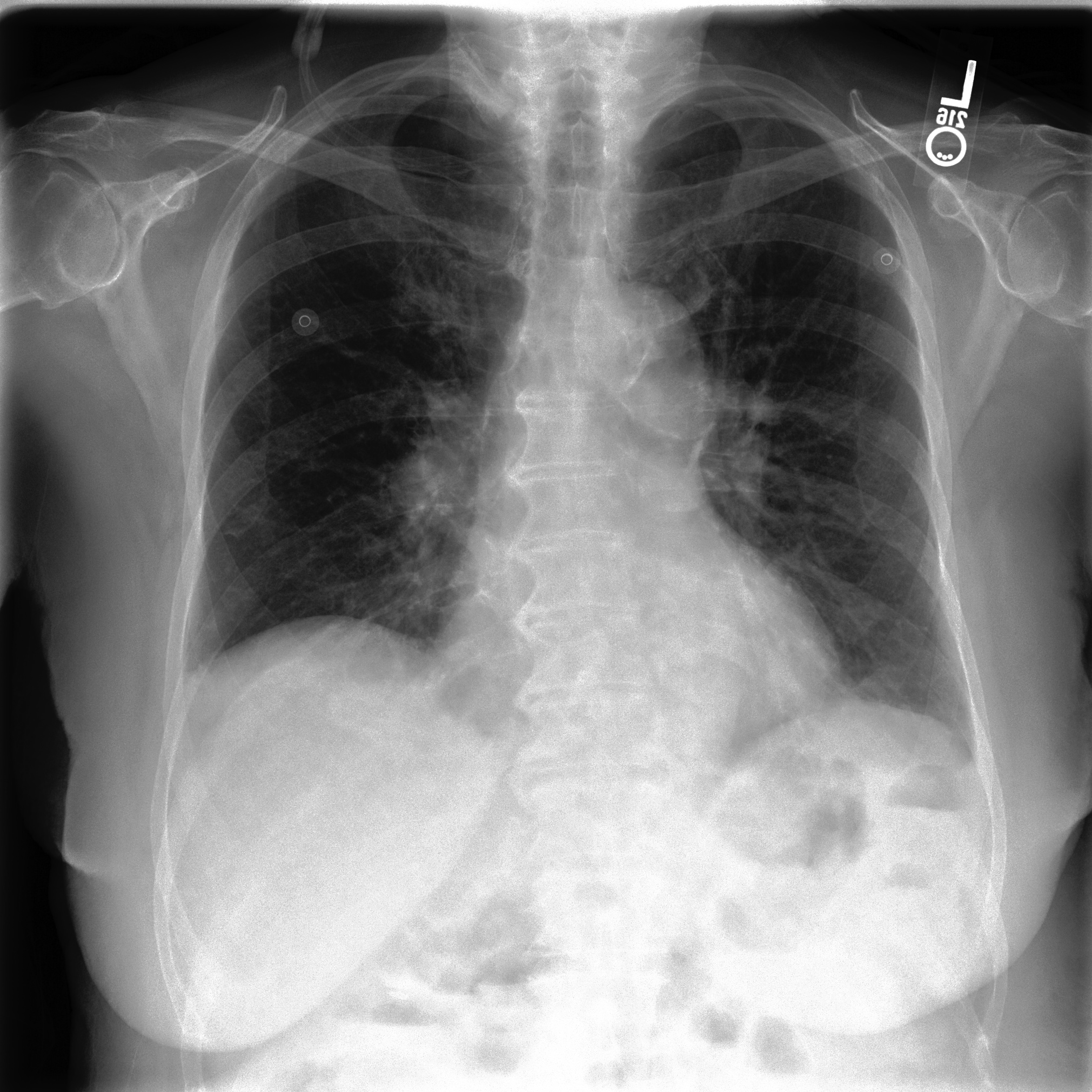}}\quad
  \subfigure[IU-CXR]{\includegraphics[width=3.25cm, height=3.5cm]{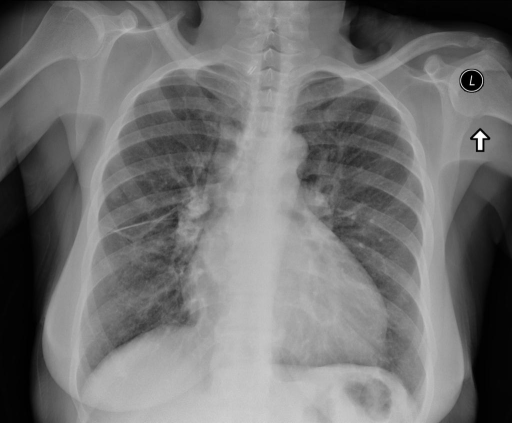}}\quad \\
  \caption{Randomly sampled chest X-ray having Cardiomegaly from each of the four datasets: ChestX-ray14, CheXpert, MIMIC-CXR-JPG, and IU-CXR.} \label{fig:sample-image-four-datasets}
\end{figure*}

In this work, we developed computer-aided diagnosis system that can automatically detect thoracic diseases present in Chest X-rays. Specifically, we developed a \emph{multi-head self-attention convolutional network} based \emph{multi-label classification} framework for predicting the presence of common thoracic diseases on four of the largest publicly available chest X-ray datasets, namely, ChestX-ray14, CheXpert, MIMIC-CXR-JPG, and IU-CXR. Fig.~\ref{fig:sample-image-four-datasets} shows sample chest X-ray from four datasets having positive cardiomegaly. The block diagram of our proposed model for the identification~\footnote{We will use the term identification and classification interchangeably.} of common thoracic diseases is shown in Fig.~\ref{fig:block_diagram}. 

\section{Related Work}
Thoracic diseases refers to collection of diseases in the chest area, comprising of structural abnormalities, infectious diseases, and their complications~\cite{Yang:Pan:2020:ChestWNet}. The Fleischner Society, which is an international multidisciplinary medical society for thoracic diseases of chest, provides list of common thoracic diseases. The four large-scale chest X-ray datasets, which we aim to work on in this study, follows Fleischner Society's glossary for getting structured labels from free-text radiology reports. Most of the common thoracic diseases include \emph{Atelectasis}, \emph{Cardiomegaly}, \emph{Effusion}, \emph{Infiltration}, \emph{Mass}, \emph{Nodule}, \emph{Pneumonia}, \emph{Pneumothorax}, \emph{Consolidation}, \emph{Edema}, \emph{Emphysema}, \emph{Fibrosis}, \emph{Pleural thickening}, \emph{Hernia}, \emph{Enlarged cardiomediastinum}, \emph{Lung lesion}, \emph{Lung opacity}, and \emph{Pleural other}. Apart from these abnormalities, there are labels for \emph{Fracture}, \emph{Support devices}, and \emph{No finding}.

\subsection{Chest X-ray datasets}
 To motivate research in the medical imaging community, various datasets have been released. Recently, large-scale chest X-ray datasets are released, which make it possible to apply deep learning to solve various tasks for disease classification.  The Indiana University Chest X-ray Collection (IU-CXR) dataset~\cite{Demner-Fushman:2015:Preparating_a_collection} consists of 7470 chest X-rays along with their accompanied 3955 radiology reports. The ChestX-ray14 dataset, released by \cite{Wang:2017:ChestX-ray14}, contains 112120 frontal-view chest X-rays from 30805 unique patients. Each radiography is labeled with one or more types of 14 common thorax diseases. The CheXpert dataset~\cite{Irvin:2019:CheXpert}, collected at Standford Hospital, consist of 224316 chest X-rays from 65240 patients. The MIMIC-CXR-JPG dataset~\cite{Johnson:2019:MIMIC-CXR-JPG} consists of 377110 chest X-rays and structured labels extracted from the associated radiology reports using natural language processing tool. With the availability of these large-scale datasets, research has begun towards applying deep learning models for the automatic detection of thoracic diseases from chest X-rays. 

\subsection{Thoracic diseases detection}
Early computer-aided detection systems were based on classical machine learning methods. \cite{Jun:2018:development_of_CAD} proposed SVM classifier trained on texture and shape features to diagnose lung diseases. Various specialised algorithms have been developed to detect specific thoracic diseases such as lung nodule~\cite{Xie:yang:2019:automated_pulmonary_nodule_detection}, pneumonia~\cite{Sirazitdinov:2019:dnn_ensemble_pneumonia_localization}, cardiomegaly~\cite{Sogancioglu:Murphy:2020:cardiomegaly_detection,Que:tang:2018:CardioXNet}, pneumothorax~\cite{Taylor:2018:automated_detection_pneumothorax}, and lung cancer~\cite{Chowdhury:2012:lung_cancer_detection}. However, accurately detecting the presence of multiple diseases from chest X-rays is still a challenging task. \cite{Wang:2017:ChestX-ray14} provides one of the first large-scale publicly available chest X-ray datasets with disease labels. \cite{Wang:2017:ChestX-ray14} formulated disease diagnosis problem as multi-label classification using class-specific image feature transformations. \cite{Mao:Yao:2018:deep_generative_classifiers} proposed a deep generative classifier to automatically diagnose thorax diseases from chest X-rays to make model robust to noise and reduce overfitting. \cite{Yan:Yao:2018:weakly_supervised_deep_learning} proposed a unified weakly supervised deep learning framework equipped with squeeze-and-excitation blocks to jointly perform thoracic disease classification and localisation on chest X-rays. To explore the correlation among the 14 diseases, \cite{yao:2018:weakly_supervised_medical_diagnosis} used a LSTM to repeatedly decode the feature vector from a DenseNet and produced one disease prediction at each step. \cite{Kumar:2018:Boosted_cascaded_convnets} explored loss functions to train a CNN from scratch and presented a boosted cascaded CNN for multi-label classification. \cite{Li:2018} used a pre-trained ResNet to extract features and divided them into patches which are passed through a fully connected network to obtain a disease probability map. 

Beyond hand-crafted features, \cite{Wang:2017:ChestX-ray14} applied four convolutional neural network in classifying thoracic diseases on images from the ChestX-ray14 dataset. \cite{Rajpurkar:Irvin:2017:CheXNet} proposed \emph{CheXNet} which takes a chest X-ray as an input and outputs the probability of pneumonia along with a heatmap localising the areas of the image most indicative of pneumonia. \cite{Rajpurkar:Irvin:2017:CheXNet} introduced CheXNet, a DenseNet based model that was trained on the ChestX-ray14 dataset, which achieved state-of-the-art performance on over 14 disease classes and exceeded radiologist performance on pneumonia using the F1 metric. \cite{Wang:Peng:2018:TieNet} proposed \emph{TieNet} model that extracts image and text representations. \cite{Allaouzi:Ben_Ahmed:2019:a_novel_approach} extracted features from chest X-rays using convolutional networks and then classifying the extracted features with multi-label classification algorithms. \cite{Sangeroki:Tjeng:2021:fast_and_accurate_thoracic} proposed a combination of Convolutional Block Attention Module (CBAM) and shuffleNetV2 for diagnosing thoracic diseases. \cite{Wang:Peng:2018:TieNet} developed a text-image embedding network, TieNet, based on an argument that not only the X-ray images but also the text reports are useful.

\subsection{Attention models in thoracic disease detection}
Given the subtle differences in the visual features of various thoracic diseases, the chest X-ray multi-label classification is really a challenging problem. The visual attention mechanism allows the model to adaptively focus on relevant regions of the image, providing an insight of model interpretability as well as improving effectiveness of the model. \cite{Zhou:Khosla:CAM} proposed the class activation mapping (CAM) method that enables deep convolutional neural network to perform object localisation. \cite{Selvaraju:2017:Grad-CAM} proposed the gradient-weighted class activation mapping (Grad-CAM) which combines feature maps with the gradient signal to produce visual explanations. \cite{Sangeroki:Tjeng:2021:fast_and_accurate_thoracic} proposed a modified version of Convolutional Block Attention Module (CBAM) due to its lightweight structure compared to the standard attention mechanism. Though the model is fast given light-weight CNN model is taken, but it is at the cost of model effectiveness. \cite{wang:jia:2020:thorax-net} proposed an attention regularised DNN model, \emph{Thorax-Net}, which consists of a classification branch and attention branch. The Grad-CAM method is embedded in stacked convolutional operations to perform attention. \cite{Rajpurkar:Irvin:2017:CheXNet} used the CAM method to produce heatmaps for the interpretation of CheXNet predictions. \cite{wang:2018:chestnet} also used the Grad-CAM method in the attention branch to exploit the correlation between class labels and relevant regions of pathological abnormalities via analysing learned feature maps. 

\section{Methodology}
This section describes our methodology for the diagnosis of thoracic diseases in Chest X-rays. It provides the rationale behind the choice of DenseNet as our backbone network and the need to augment self-attention with the DenseNet model. We also provides various techniques, including transfer learning, data augmentation, NLP data labeler, and handing uncertain labels generated by the NLP labeler. 

\subsection{Convolutional Neural Networks}
\emph{Deep learning}~\cite{Goodfellow:2016:DL_book}, a sub-field of machine learning, has seen a remarkable success in recent years. Various factors that made it possible, include improvement in learning algorithms, release of large-scale datasets, and increasing compute power. Deep learning has become a standard tool in various fields including computer vision, natural language processing, speech recognition, medical imaging, and health informatics. In the field of computer vision, analysing images and videos, convolutional neural networks (CNN) have proven to be powerful in providing state-of-the-art results, and even surpassing human performance on various image-based classification and recognition problems.  CNNs take raw data (\textit{e.g.}, images) as input and perform a series of convolutional and non-linear operations to hierarchically learn rich information about the image, in order to bridge the gap between high-level representation and low-level features. During the training phase, the CNNs adjust their filter values (weights) by optimising certain loss functions through forward passes and backpropagation procedures, so that the inputs are correctly mapped to the ground-truth labels. Remarkably, CNNs have recently been shown to match or exceed human performance in visual tasks such as natural image classification, skin cancer classification, diabetic retinopathy detection~\cite{tang:tang:2020:automated_abnormality_classification}. The superlative ability to automatically extract useful features from the inherent characteristics of data makes CNN the first choice for complex medical problem solving. To date, CAD systems embedded with deep-learning algorithms have worked efficiently for medical disease detection by effectively generating a range of high-quality diagnostic solutions while spotlighting suspicious features~\cite{Guo:Passi:2020:tuberculosis_diagnosis}.

\begin{figure*}
    \centering
    \includegraphics[scale=0.4]{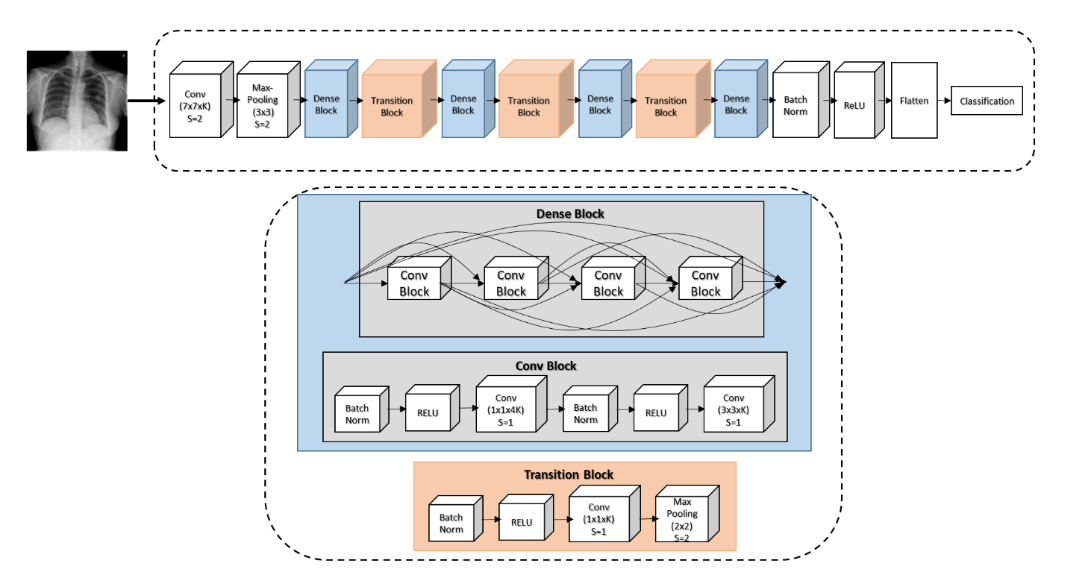}
    \caption{DenseNet-121 model architecture~\cite{Ho:gwak:2019:multi-feature_integration}.}
    \label{fig:densenet_arch}
\end{figure*}

The first convolutional neural network, namely \emph{LeNet-5}~\cite{Lecun:1998:LeNet}, consisting of 3 convolutional layers, 2 sub-sampling layers, and 2 fully connected layers, was applied to document recognition. The CNN has advantages compared to other approaches, as they are shift, scale, and distortion invariant with its receptive field, shared weights, and sub-sampling. However, it was with the AlexNet model, the deep learning heat picks up in different fields. Since then, various CNN models have been proposed, including VGG-16, VGG-19, Inception, ResNet, and DenseNet. One of the most important CNN model which has widely been used in medical imaging community is \emph{DenseNet-121}.  A typical DenseNet model consists of multiple densely connected convolutional layers, which improve the flow of information and gradients through the network, making it converge better and overcoming the issue of vanishing gradients. The DenseNet model uses concatenation of the output of previous layers to the output of the current layer. The use of feature maps of all previous layers as inputs to the current layer helps to achieve \emph{feature reuse} capability, helping in reducing the number of training parameters despite having a large and deeper network. Moreover, having access to features of previous layers, helps model to learn from data during training. Given that thoracic diseases are located on small regions within medical images, the overall data shows similar global features with small local differences. Using DenseNet-121 as backbone network could help the identification of small disease regions due to its capability of combining both local and global features. Fig.~\ref{fig:densenet_arch} shows architecture of DenseNet-121 model which consists of 4 consecutive dense blocks connected by 3 transition layers. The DenseNet-121 model has provided significant improvement on various medical imaging tasks, including classification, detection, and segmentation.

\subsection{Multi-head self-attention}
Convolutional networks have been the paradigm of choice for many computer vision applications. The design of the convolutional layer imposes: 1) locality via a limited receptive field, and; 2) translation equivariance via weight sharing. Both these properties prove to be crucial inductive biases when designing models that operate over images. However, the local nature of the convolutional kernel prevents it from capturing global contexts in an image, often necessary for better recognition of objects in images. Self-attention~\cite{vaswani:2017:attention_is_all_you_need} has emerged as a recent advancement to capture long-range interactions, but has mostly been applied to sequence modelling and generative modelling tasks. While reading any text, humans don't pay same attention everywhere. Similarly, while recognising an image, human pay more attention to certain regions. Therefore, there is a need to given different importance to different features. Recent studies~\cite{bello2019,Parmar2019_stand-alone_self-attention} proved that \emph{transformer-based} architectures can beat state-of-the-art convolutional networks. In ~\cite{cordonnier2020}, authors investigated how a self-attention layer can learn convolutional filter. They found that a \emph{multi-head self-attention} layer with sufficient number of heads can be at least as expressive as any convolution layer. The key idea behind self-attention is to produce a weighted average of values computed from hidden units. Unlike the pooling or the convolutional operator, the weights used in the weighted average operation are produced dynamically via a similarity function between hidden units. As a result, the interaction between input signals depends on the signals themselves rather than begin predetermined by their relative location like in convolutions. In particular, this allows self-attention to capture long-range interactions without increasing the number of parameters. In ~\cite{bello2019}, authors develop a novel two-dimensional relative self-attention mechanism that maintains translation equivariance while being infused with relative position information, making it well suited for images. The proposed self-attention mechanism proves competitive for replacing convolutions entirely, however in controlled experiments, authors found that the best results are obtained when they combine self-attention mechanism with convolutions. Hence, authors proposed to augment convolutions with the self-attention mechanism by concatenating convolutional feature maps, which enforce locality, to self-attention maps, capable of modelling longer range dependencies. 

\begin{figure*}
    \centering
    \includegraphics[scale=0.4]{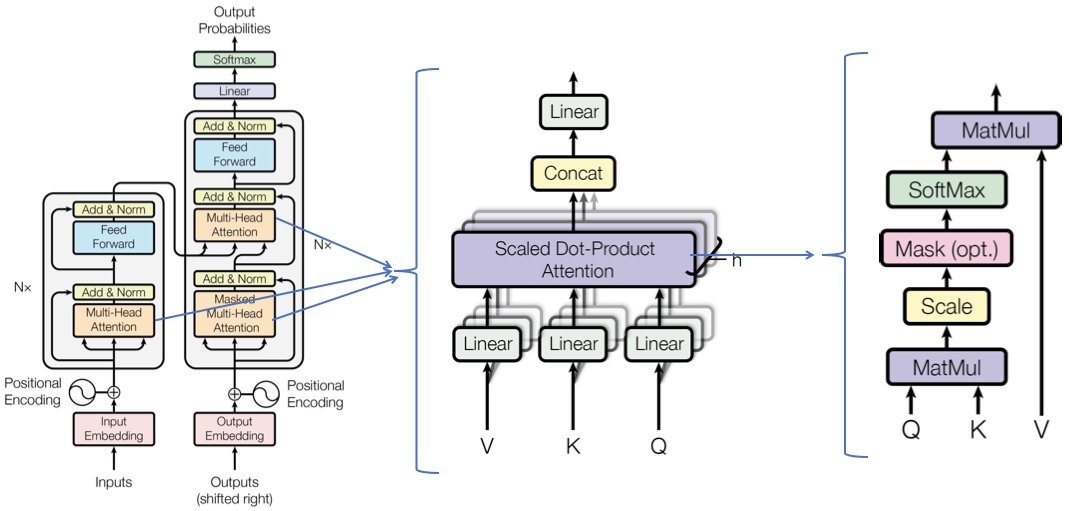}
    \caption{Transformer architecture, multi-head attention block, and scaled dot product attention~\cite{vaswani:2017:attention_is_all_you_need}.}
    \label{fig:transformer_illustration}
\end{figure*}

The main difference between CNN and self-attention layers is that the new value of a pixel depends on every other pixel of the image. As opposed to convolution layer whose receptive field is the $K \times K$ neighbourhood grid, the self-attention's receptive field is always the full image. A self-attention layer is defined by a \emph{key/query} size $D_k$, a head size $D_h$, a number of heads $N_h$, and an output dimension $D_{\textit{out}}$. The layer is parameterised by a key matrix $\mathbf{W}^{(h)}_{\!\textit{key}}$, a query matrix $\mathbf{W}^{(h)}_{\!\textit{qry}}$ and a value matrix $\mathbf{W}^{(h)}_{\!\textit{val}}$ for each head $h$, along with a projection matrix $\mathbf{W}_{\!\textit{out}}$ used to assemble all the heads together. Scaled dot-product is one of the most commonly used attention functions.
\begin{equation}
    \text{Attention}(Q, K, V) = \text{softmax}(QK^T/\sqrt{d_k})V
\end{equation}
where Q, K, and V are matrices of queries, keys, and values, respectively, and $d_k$ refers to the dimension of the key. If Q, K, and V are equal, the function is called self-attention. Similar to filters in CNN, a single attention is not enough for learning multiple features' weights. To allow the model to learn information from different representation subspaces, the MHSA mechanism is presented. Fig.~\ref{fig:transformer_illustration} shows transformer architecture and the multi-head attention block.

\begin{equation}
    \text{MultiHead}(Q, K, V) = \text{Concat}(head_1, head_2, \ldots, head_h)W^o
\end{equation}

\begin{equation}
    \text{Head}_i = \text{Attention}(QW_i^Q, KW_i^K, VW_i^V)
\end{equation}
where $W^o$, $W_i^Q$, $W_i^K$, and $W_i^V$ are parameter matrices. 

The advantage of combining CNN with MHSA is to get merits of both approaches as well reducing the number of parameters drastically~\cite{bello2019}. This study heavily draws from the original work~\cite{bello2019} where authors originally proposed attention-augmented CNN architecture combining the benefits of convolutions and self-attention. 

\subsection{Data augmentation}
Deep convolutional neural networks showed remarkable performance on many computer vision tasks. However, deep neural networks need large-scale annotated datasets to train them. Deep neural networks generally show overfitting on small-scale datasets, where a network learns perfectly on the training data but do not generalise well on the testing set. Various application domains, including medical image analysis do not have large-scale annotated datasets because it is time-consuming, costly, and requires experts to annotate medical datasets. Data augmentation, which is the process of synthetically generating samples during model training, helps in preventing model overfitting. Data augmentation techniques create variations in images, in turn helping model become more robust to generalise well on test images as well as improving the performance of the model. Data augmentation has proven to be useful in problems having class-imbalance in datasets. In this study, we use different augmentation techniques, including \emph{horizontal flip}, \emph{rotation}, \emph{scaling}, \emph{random crop}, \emph{normalising using mean and standard deviation}. We do not used \emph{vertical flip} because the processed image is not what radiologists often see in medical scans. 

\subsection{Transfer learning}
\emph{Transfer learning}, which makes use of pre-trained model is useful in transferring the knowledge learned by the first model to the second one. Transfer learning plays key role in providing better models since medical datasets are smaller in size. In order to use transfer learning in our experiments, we initialised our CNN model with pre-trained weights that are trained on a large-scale image dataset, namely the ImageNet. 

\subsection{CheXpert-An open-source radiology report labeler}
The CheXpert labeler~\cite{Irvin:2019:CheXpert} parses the input sentences into dependency structures and runs a series of surface and syntactic rules to extract the presence status of $14$ clinical observations seen in chest radiology reports. It was evaluated to have over $95\%$ overall $F1$ when compared against oracle annotations from multiple radiologists on a large-scale radiology report dataset. Fig~\ref{fig:chexpert_labeler} shows a sample radiology report along with its output for different observations. 

\begin{figure*}
\small
{%
\begin{minipage}{0.45\textwidth}
    \textbf{Sample radiology report} \\
    \\
    1. Unremarkable cardiomediastinal silhouette. \\  \\
    2. diffuse reticular pattern, which can be seen with a atypical infection or chronic fibrotic change. no focal consolidation. \\ \\
    3. no pleural effusion or pneumothorax. \\ \\
    4. mild degenerative changes in the lumbar spine and old right rib fractures. \\ \\
  \end{minipage}}
\hspace{\fill} 
{%
\begin{minipage}[c]{0.25\textwidth}
\begin{small}
\textbf{Labeler output} \\
        No Finding \\
        Enlarged Cardiom. \\
        Cardiomegaly \\
        Lung Opacity \\
        Lung Lesion \\
        Edema \\
        Consolidation \\
        Pneumonia \\
        Atelectasis \\
        Pneumothorax \\
        Pleural Effusion  \\
        Pleural Other \\
        Fracture  \\
        Support Devices  \\
\end{small}
\end{minipage}}
\hspace{\fill} 
{%
\begin{minipage}[c]{0.25\textwidth}
\begin{small}
\textbf{Observation} \\ 
         \\
        0 \\
         \\
        1 \\
         \\
         \\
        0 \\
        u \\
         \\
        0 \\
        0  \\
         \\
        1  \\
          \\
\end{small}
\end{minipage}}
\caption{CheXpert labeler output on various observations on a sample radiology report.}
 \label{fig:chexpert_labeler}
\end{figure*}

There are possibly four labels for each of the fourteen observations. For each observation, CheXpert labeler parses the report and outputs $0$, $1$, $u$ or $blank$. If no mention of observation is made in the text, it is left $blank$. If the mention of observation indicates presence of abnormality, the output is $1$, indicating positive. If the mention of observation indicates absence of abnormality, the output is $0$, indicating negative. For uncertain cases, the labeler output is $u$, indicating uncertain case. 

\subsection{Handling uncertain labels by CheXpert labeler}
The CheXpert labeler which is a rule-based system provides four labels for each of the $14$ medical observations in the radiology report. These four labels are $blank$ for unmentioned, $0$ for negative, $1$ for positive, and $-1$ for uncertain. Various policies have been proposed in \cite{Irvin:2019:CheXpert} to deal with these uncertain labels. In order to apply multi-label classification algorithms on the dataset, the uncertain labels can all be \emph{ignored} (U-Ignore), mapped to \emph{negative} (U-Zeros), or mapped to \emph{positive}(U-Ones). If we apply U-Ignore policy (deleting all rows having uncertain observations), we could not make use of the whole dataset, giving us a reduced dataset to work on. The U-Zeros mapping will definitely misguide the model training since uncertain cases are treated as negative cases. In order to handle uncertain cases correctly so that radiologist can have their expert look, it is logical to map them to U-Ones. With this motivation, we map all uncertain labels to $1$, treating them as positive labels in our study.

\section{Experimental setup}
In this section, we formulate thoracic disease identification as multi-label classification problem. We provide brief overview of four of the largest chest X-ray datasets, namely IU-CXR, ChestX-ray14, CheXpert, and the MIMIC-CXR, along with their labels and distribution in terms of training, validation, and testing. We then, provide data analysis. Finally, we provide evaluation measures to report performance of the model.

\subsection{Problem definition}
Given each chest X-ray can have one or more pathologies (or labels) present, the task can be formulated as a \emph{multi-label} image classification problem. In the multi-label settings, a training dataset $D=\{(x^{(i)}, y^{(i)}; i=1 \ to \ N\}$ that contains $N$ chest X-rays where each input image $x^{(i)}$ is associated with label $y^{(i)} \in \{0,1\}^{14}$, where $0$ and $1$ correspond to \emph{negative} and \emph{positive} observations respectively. During training, the goal is to train a model, parameterised by weights $\theta$, that maps $x^{(i)}$ to a prediction $\hat{y}^{(i)}$ such that the \emph{binary cross-entropy} loss function is minimised over the training set $D$. In a multi-label classification problem, since we need probability for each of the labels, we use \emph{sigmoid} activation function defined below.
\begin{equation}
    \hat{y}^k = \frac{1}{1+exp(-z_k)}, k=1, \ldots,  14,
\end{equation}
where $z_k$ are the logits at the last layer of the CNN which outputs the probability of each of the $14$ labels.
The loss function is then given by
\begin{equation}
    l(\theta) = \sum_{i=1}^N \sum_{k=1}^{14} log y_k^{(i)} + (1 - y_k^{(i)}) log (1-\hat{y}_k^{(i)})
\end{equation}

The binary cross-entropy (BCE) loss function is minimised to obtain optimal classification performance.
\begin{equation}
    BCE(y, \hat{y}) = - \frac{1}{N} \sum_{i=1}^{N} [y_i log(\hat{y_i}) + (1-y_i) log(1 - \hat{y_i})]
\end{equation}

where $N$ is the number of classes (\textit{i.e.}, N = 14), $y_i$ is the ground-truth and $\hat{y_i}$ is the predicted probability.

\subsection{Datasets}
In order to check the effectiveness of our proposed methodology for the diagnosis of thoracic diseases in Chest X-rays, we did experiments on $4$ large-scale datasets having chest X-rays along with their labels. All datasets use rule-based Natural Language Processing (NLP) labeler to extract $14$ common mentions from raw radiology reports. The main difference between these datasets lies in the inclusion of uncertainty label and the disease categories. The CheXpert labeler extracted uncertain findings of the disease that are denoted as \textit{u}, while the ChestX-ray14 omitted such mentions. These four datasets have $7$ disease labels in common: \emph{Atelectasis}, \emph{Cardiomegaly}, \emph{Effusion}, \emph{Pneumonia}, \emph{Pneumothorax}, \emph{Consolidation}, and \emph{Edema}. All of these datasets are de-identified to satisfy the US Health Insurance Portability and Accountability Act of 1996 (HIPAA) Safe Harbor requirements. The protected health information (PHI) is removed from all of these datasets. In order to provide patient specific information, randomly generated identifiers are use to group distinct reports and patients. 

\begin{figure*}[tb]
  \centering
  \subfigure[Atelectasis]{\includegraphics[width=3.25cm, height=3.5cm]{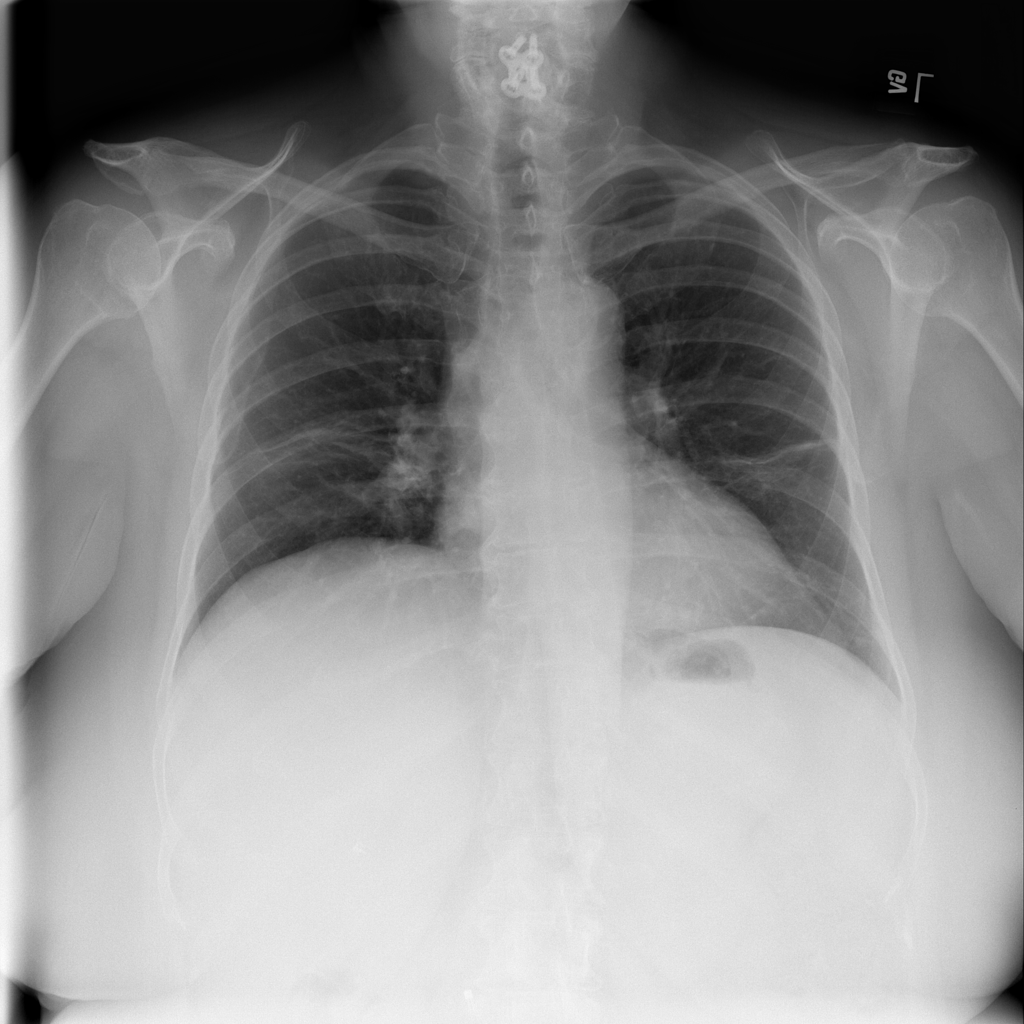}}\quad
  \subfigure[Cardiomegaly]{\includegraphics[width=3.25cm, height=3.5cm]{chestX-ray14-cardiomegaly.png}}\quad
  \subfigure[Effusion]{\includegraphics[width=3.25cm, height=3.5cm]{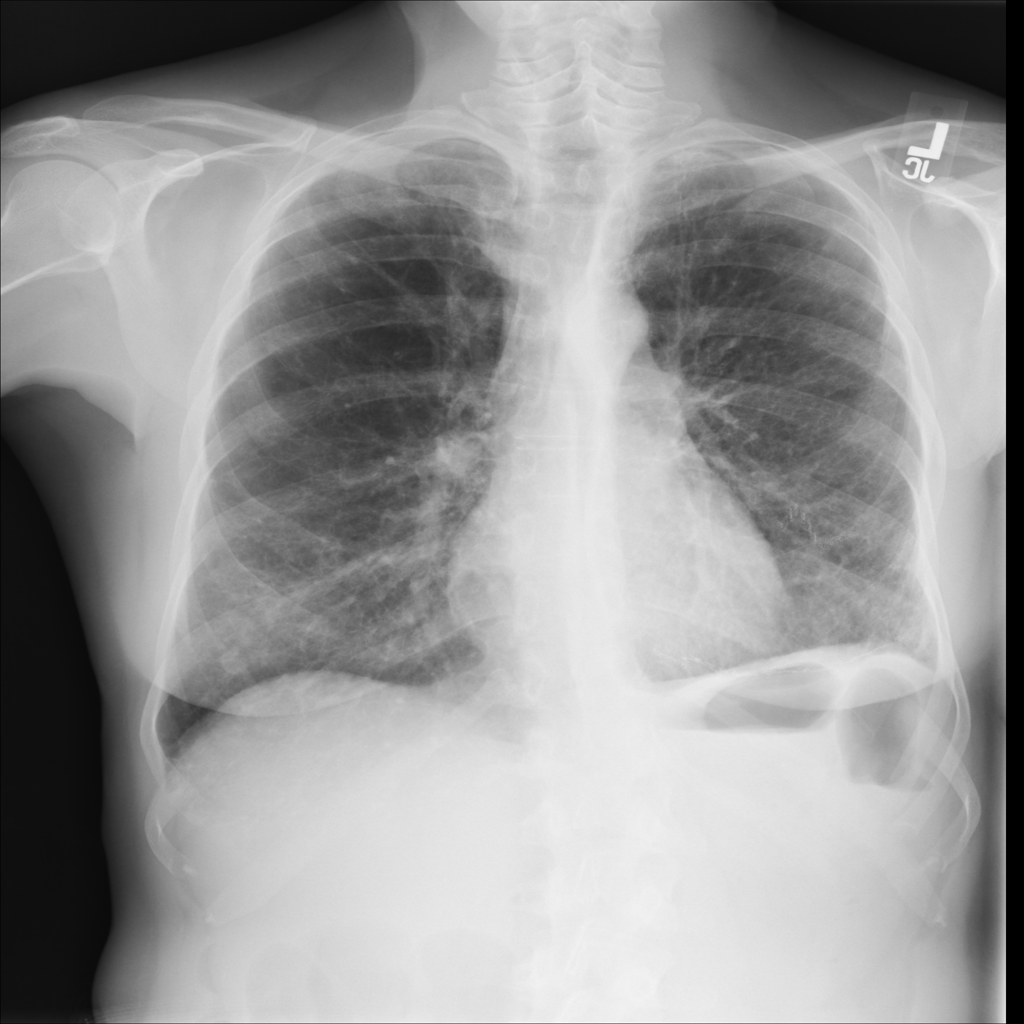}}\quad
  \subfigure[Infiltration]{\includegraphics[width=3.25cm, height=3.5cm]{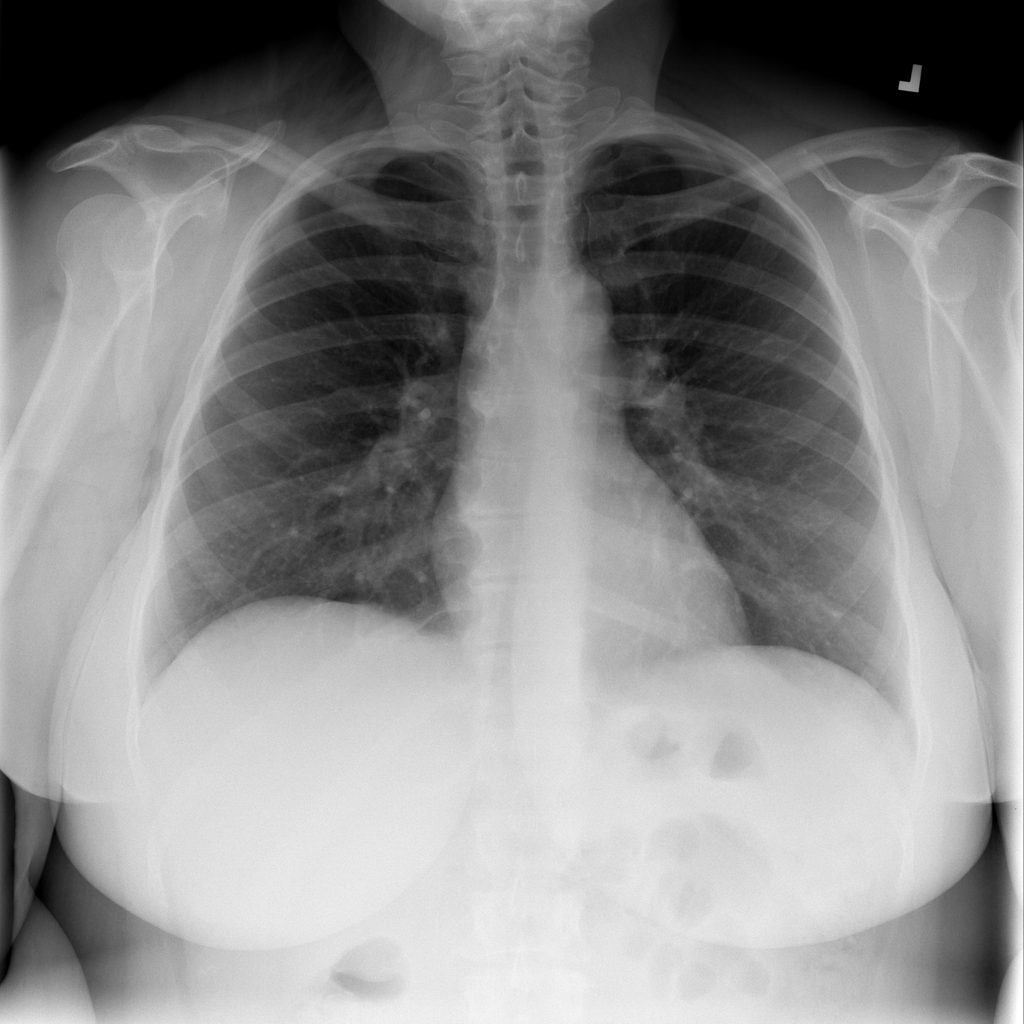}}\quad \\
  \subfigure[Mass]{\includegraphics[width=3.25cm, height=3.5cm]{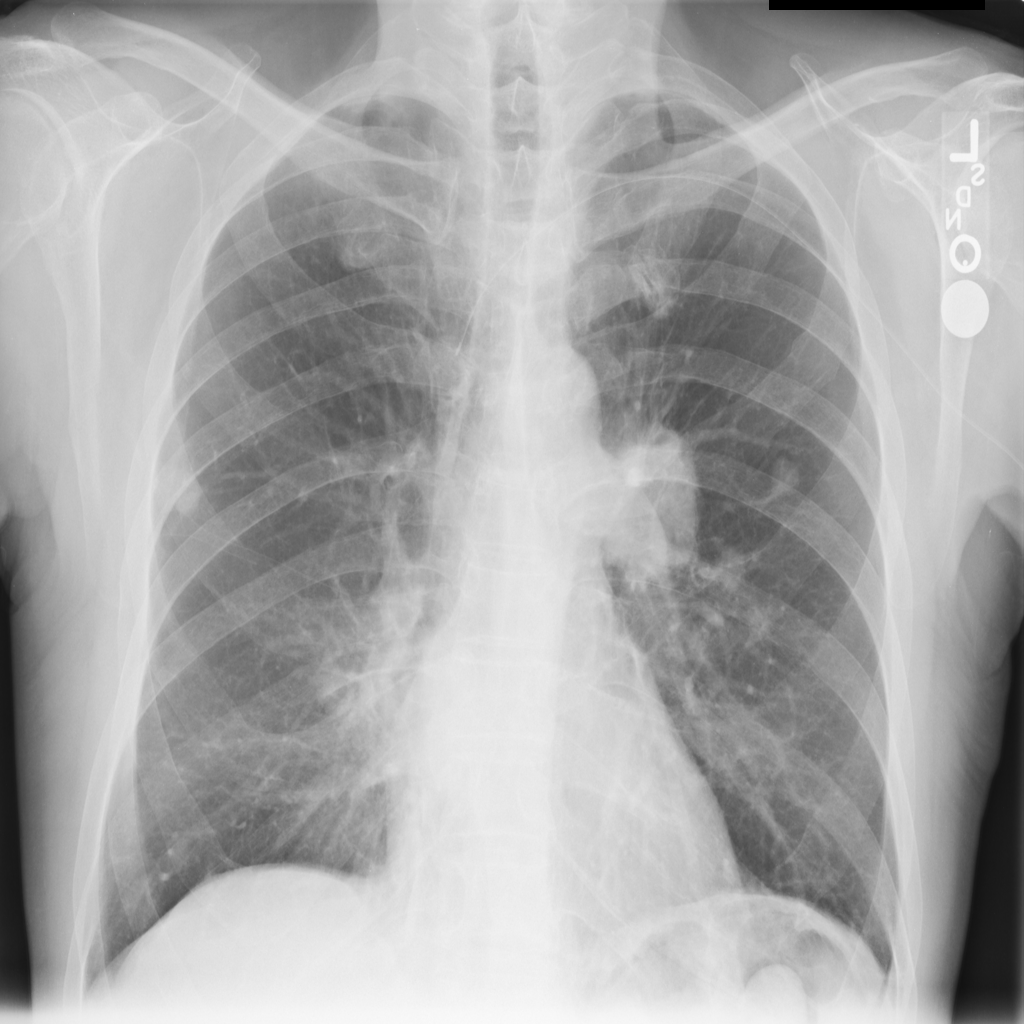}}\quad
  \subfigure[Nodule]{\includegraphics[width=3.25cm, height=3.5cm]{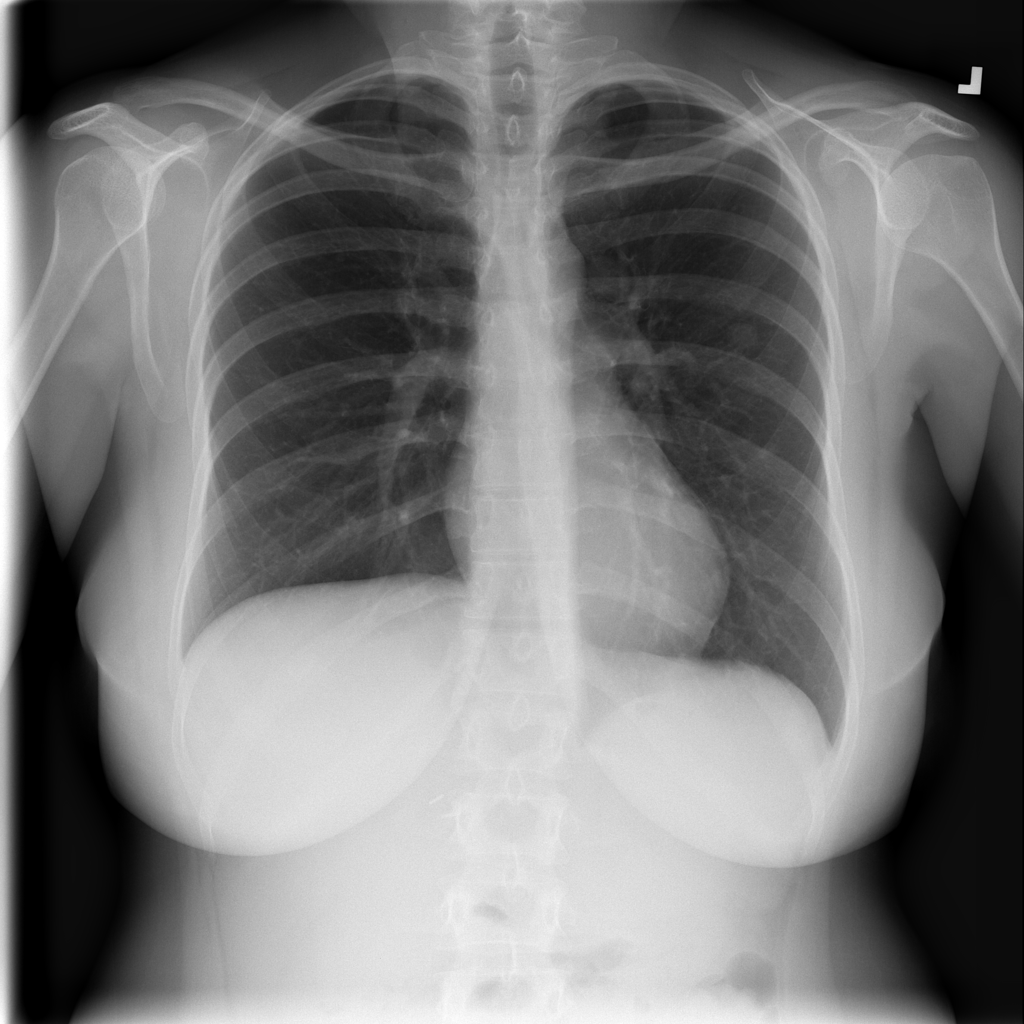}} \quad
  \subfigure[Pneumonia]{\includegraphics[width=3.25cm, height=3.5cm]{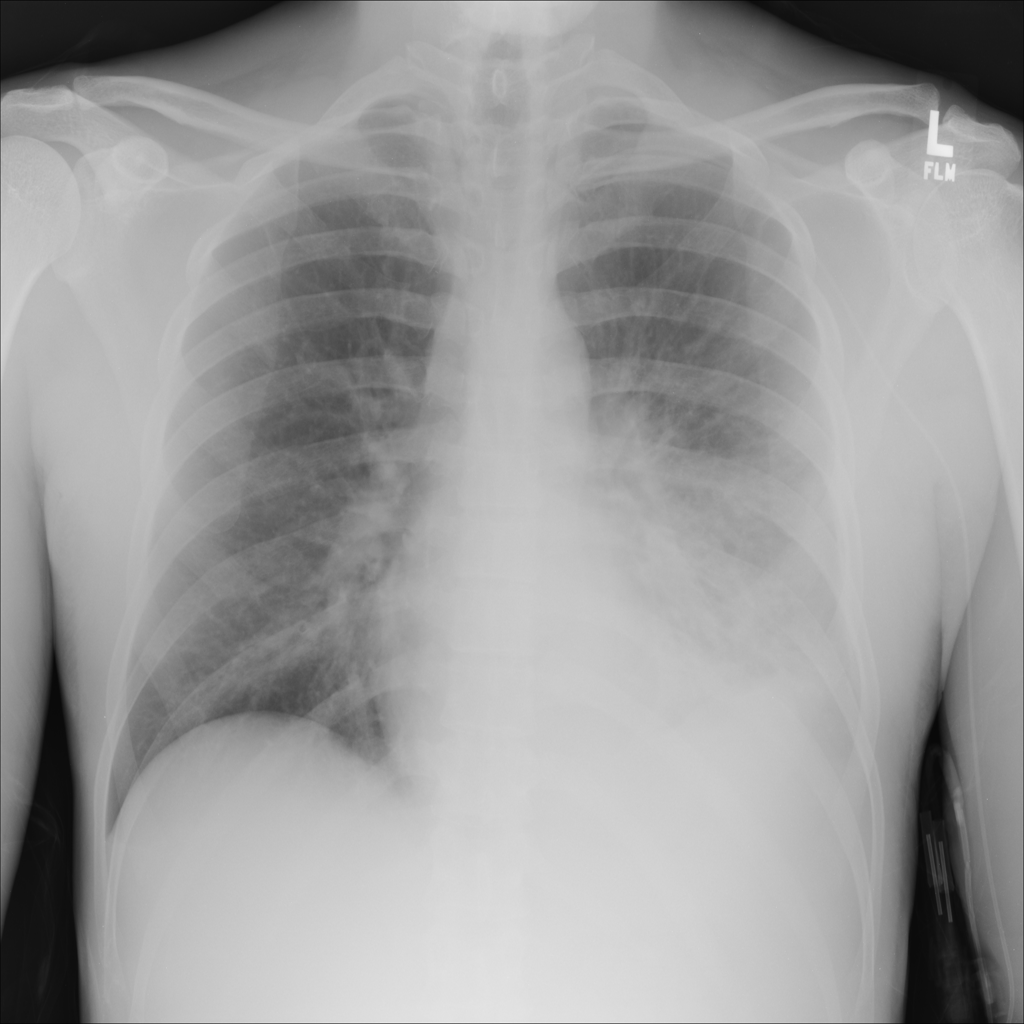}} \quad
  \subfigure[Pneumothorax]{\includegraphics[width=3.25cm, height=3.5cm]{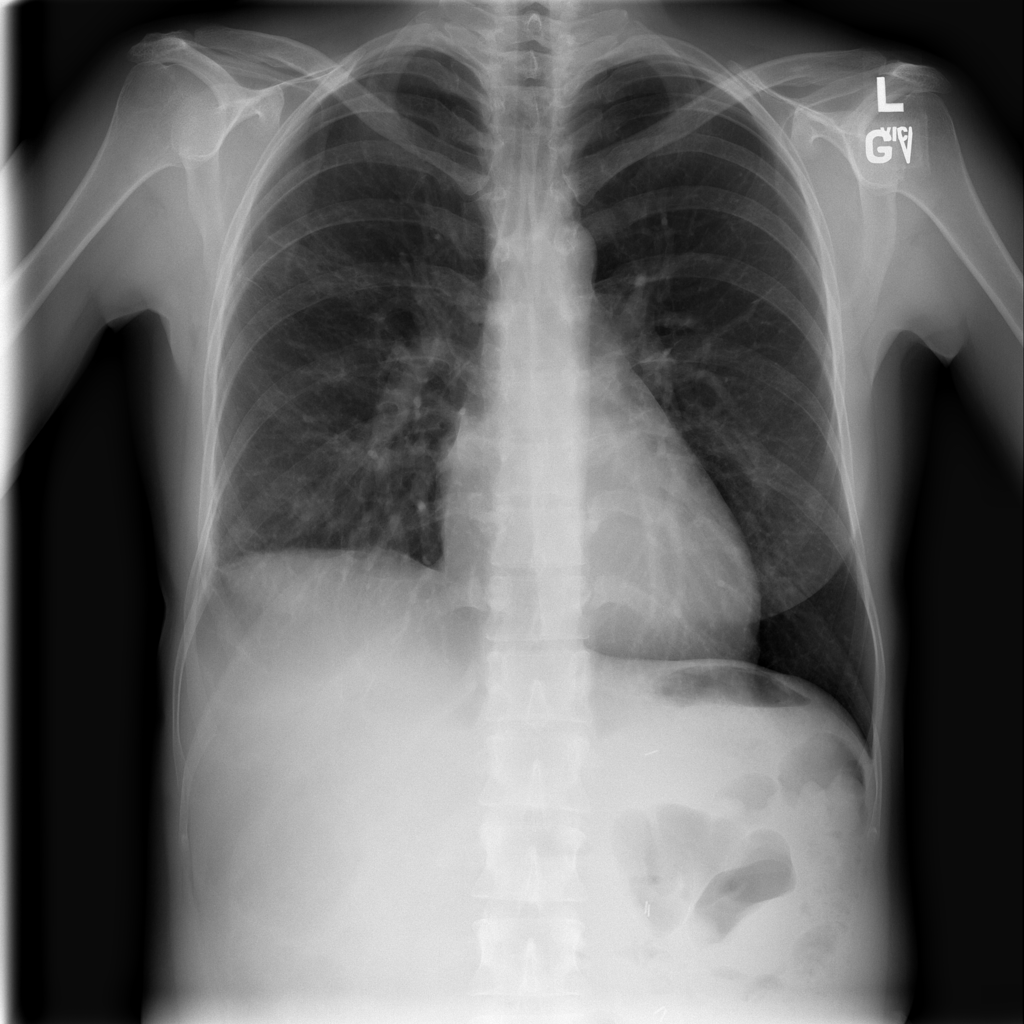}} \quad \\
  \subfigure[Consolidation]{\includegraphics[width=3.25cm, height=3.5cm]{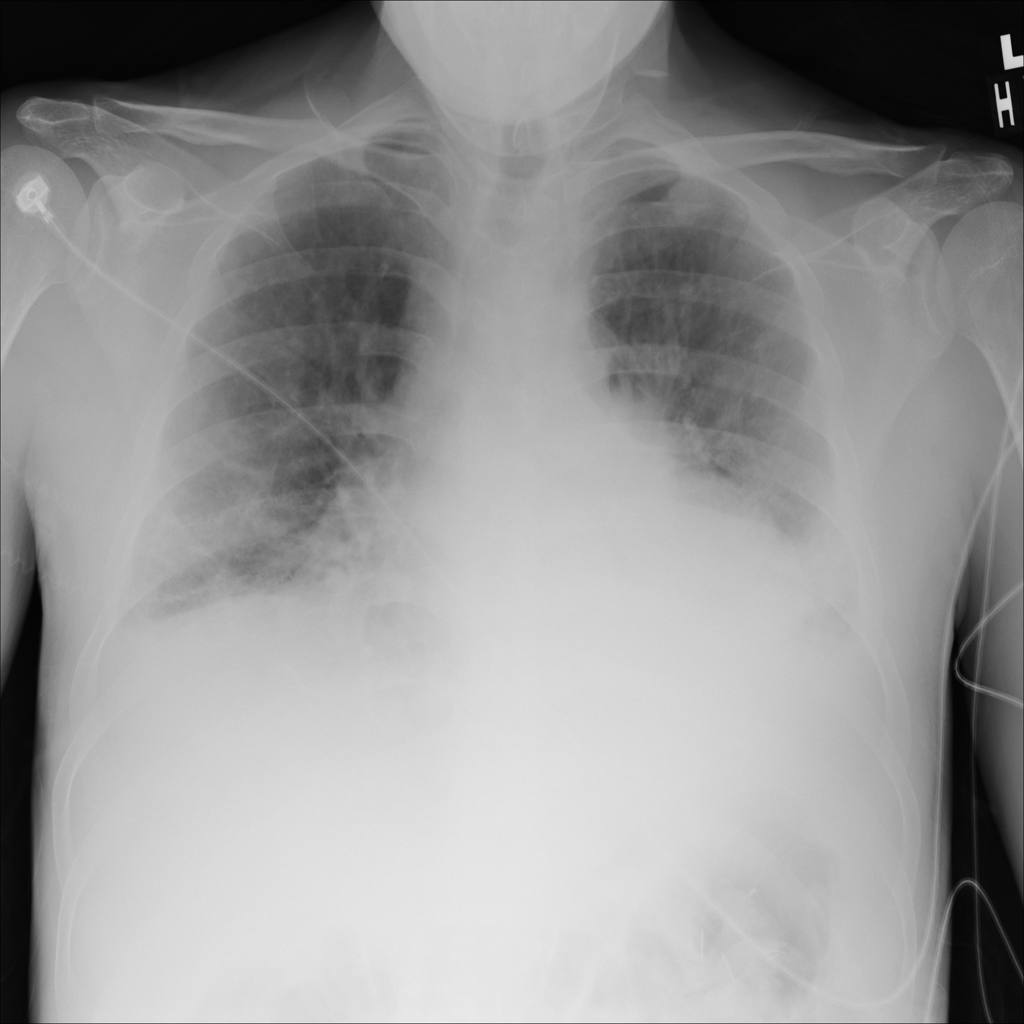}}\quad
  \subfigure[Edema]{\includegraphics[width=3.25cm, height=3.5cm]{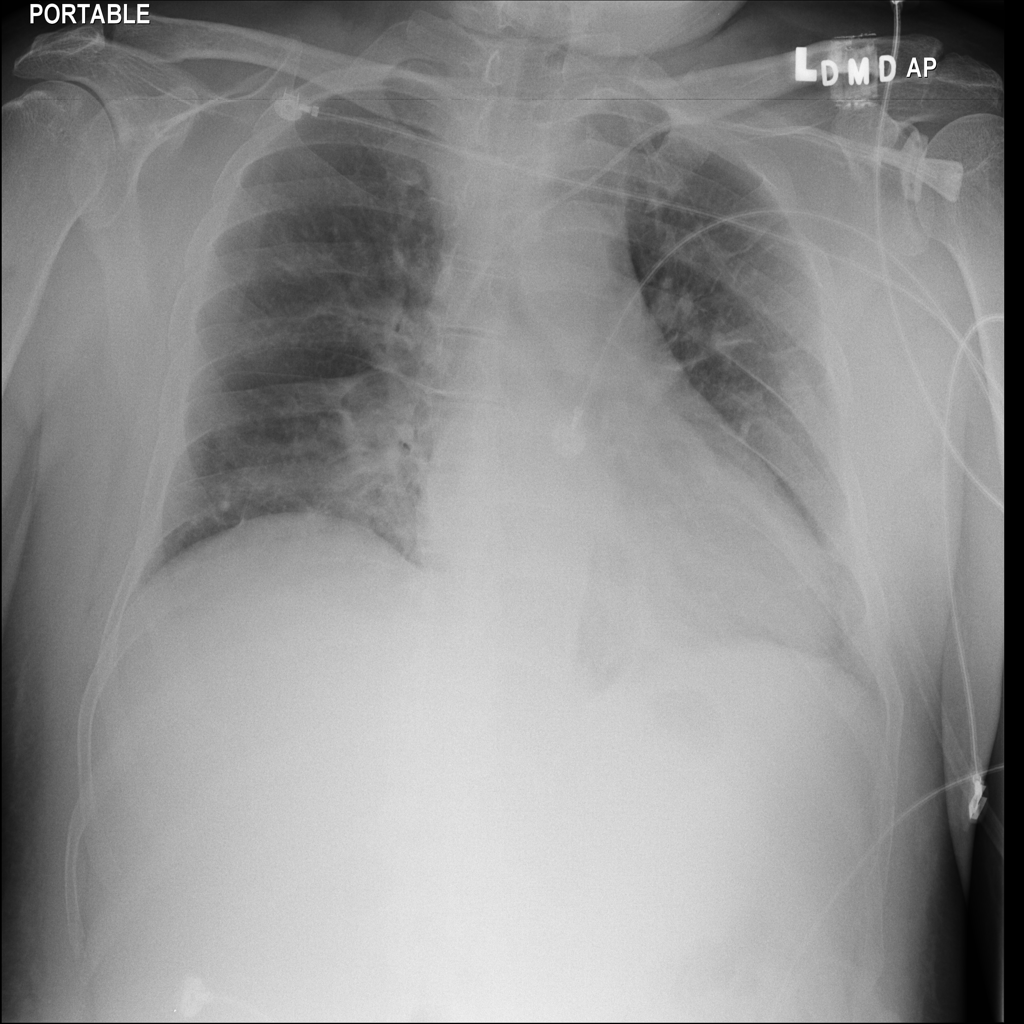}} \quad
  \subfigure[Emphysema]{\includegraphics[width=3.25cm, height=3.5cm]{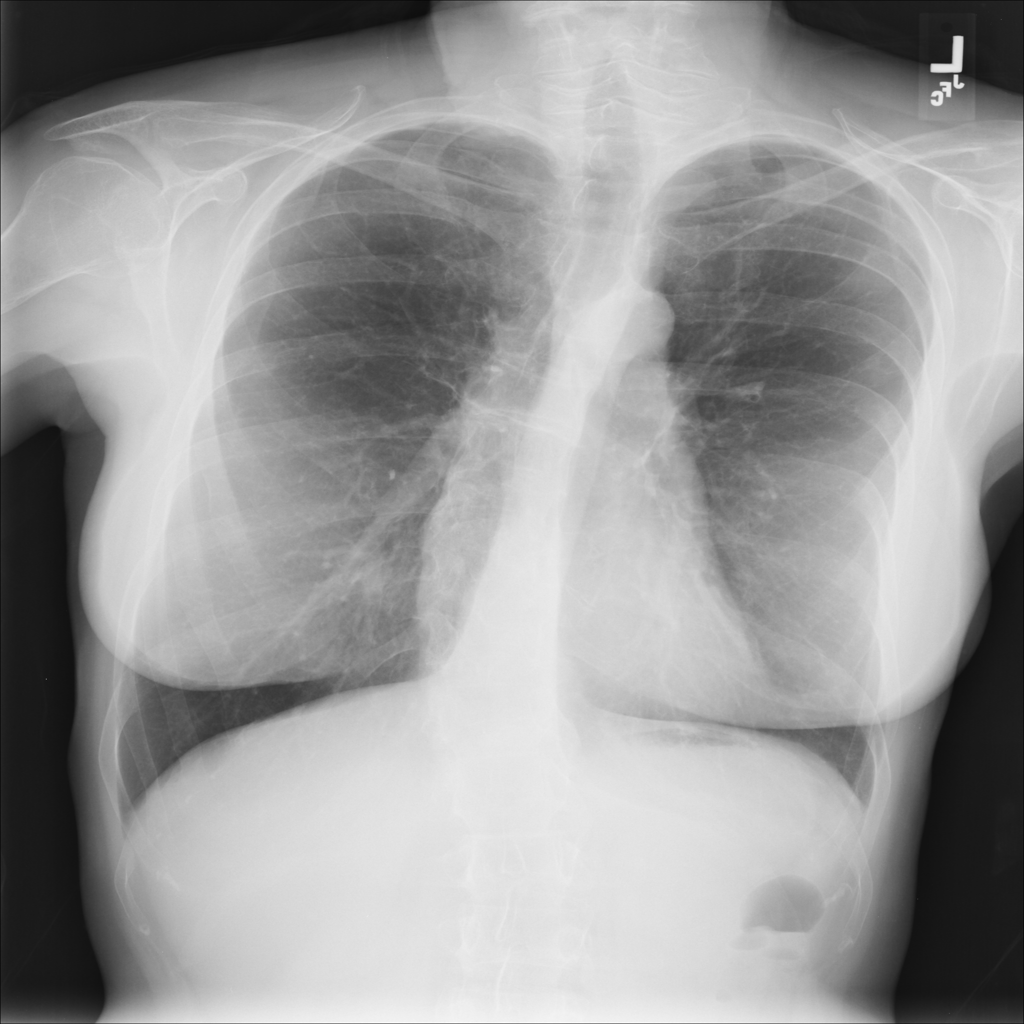}} \quad
  \subfigure[Fibrosis]{\includegraphics[width=3.25cm, height=3.5cm]{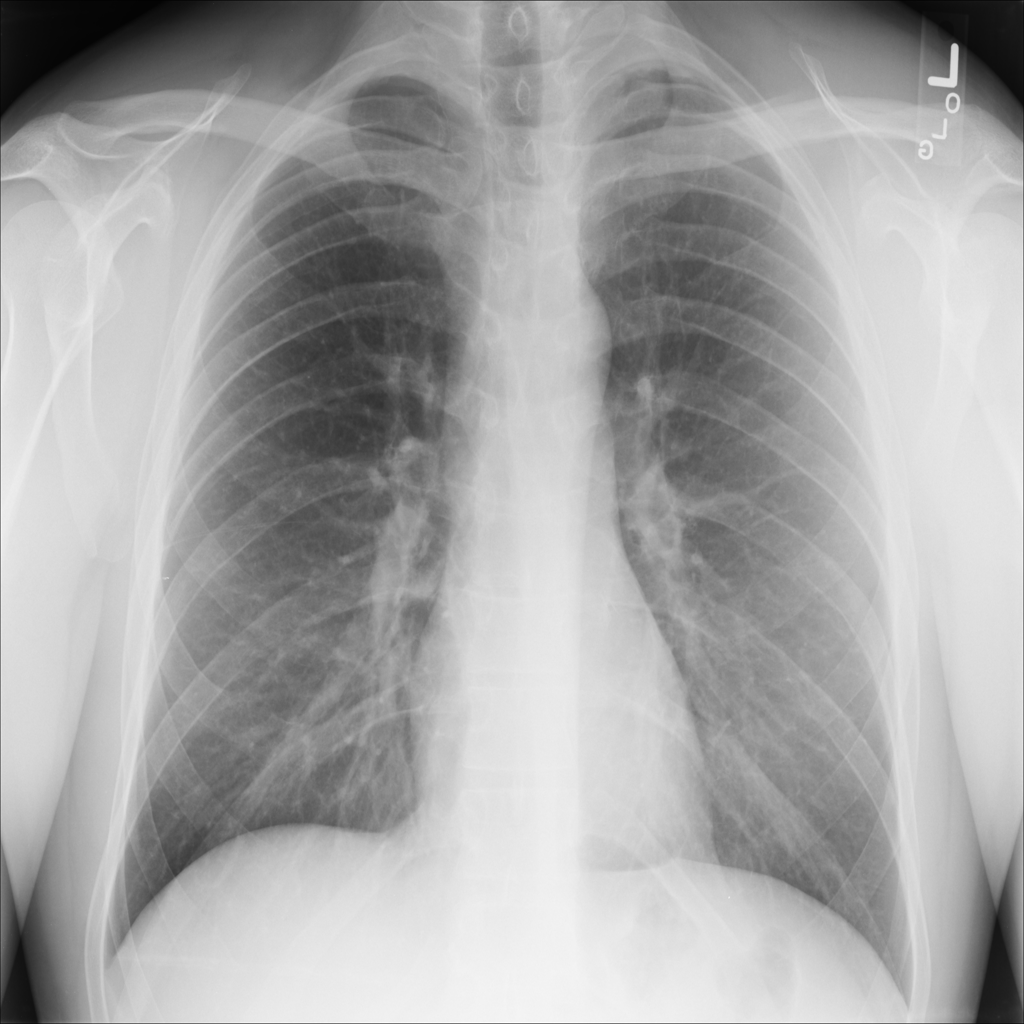}} \quad \\
  \subfigure[Pleural Thickening]{\includegraphics[width=3.25cm, height=3.5cm]{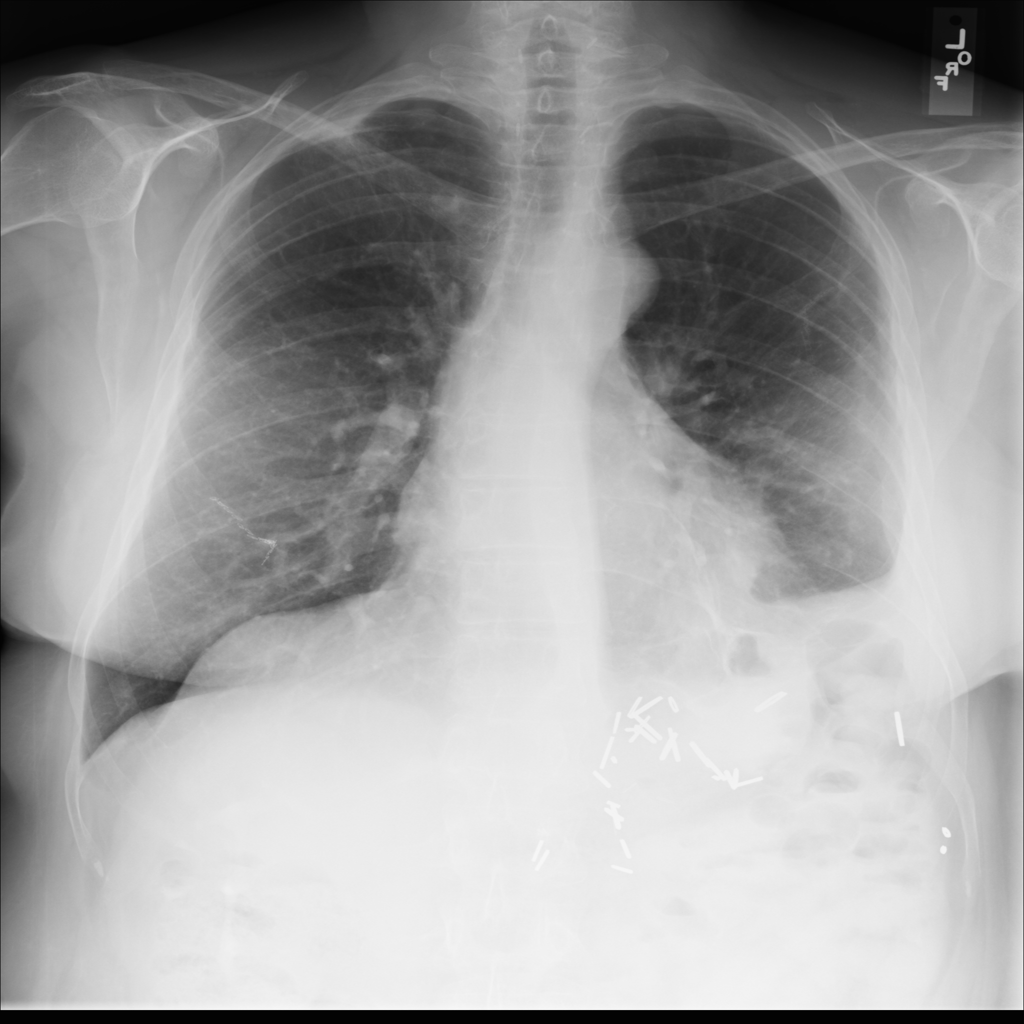}}\quad
  \subfigure[Hernia]{\includegraphics[width=3.25cm, height=3.5cm]{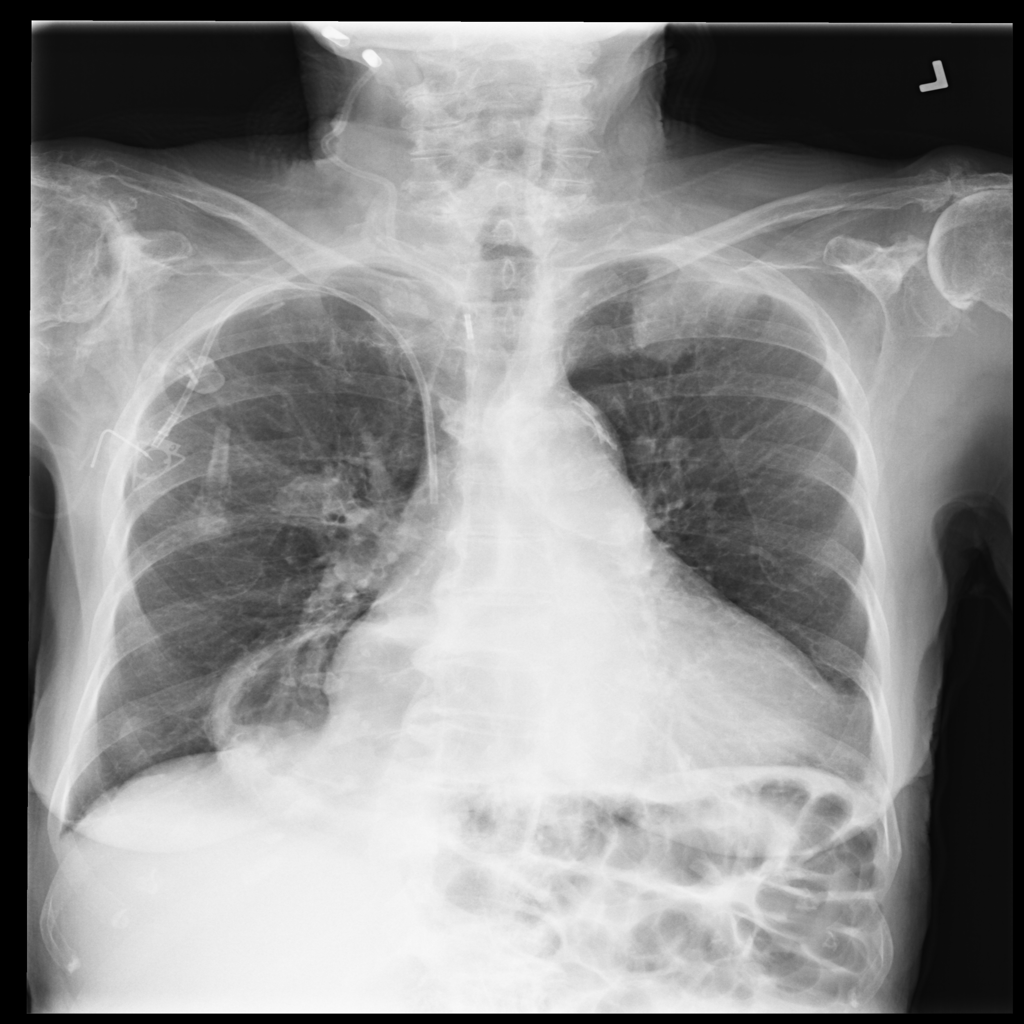}} \quad
  \subfigure[No Finding]{\includegraphics[width=3.25cm, height=3.5cm]{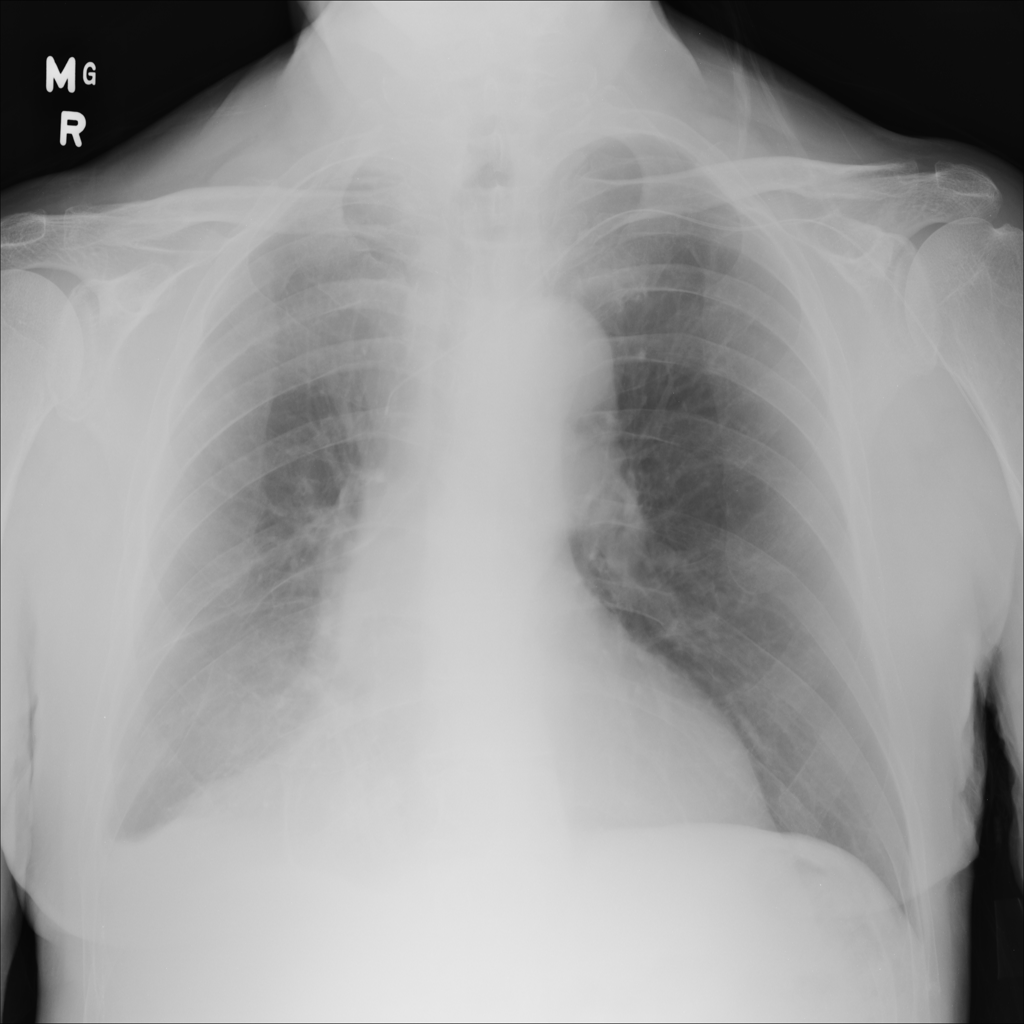}} \quad \\
  \caption{Sample image of each of the $14$ observations and ``No Finding" in the ChestX-ray14 dataset.} \label{fig:chestXray14-sampleimages}
\end{figure*}

\begin{figure*}[tb]
  \centering
  \subfigure[No Finding]{\includegraphics[width=3.25cm, height=3.5cm]{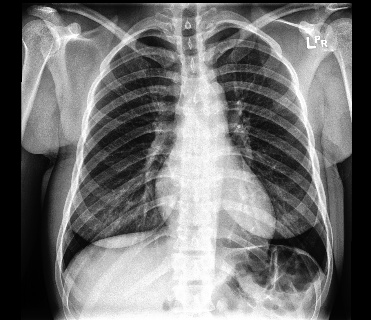}}\quad
  \subfigure[Enlarged Cardiomediastinum]{\includegraphics[width=3.25cm, height=3.5cm]{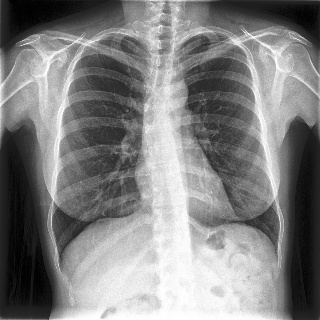}}\quad
  \subfigure[Cardiomegaly]{\includegraphics[width=3.25cm, height=3.5cm]{cheXpert-Cardiomegaly.jpg}}\quad
  \subfigure[Lung Lesion]{\includegraphics[width=3.25cm, height=3.5cm]{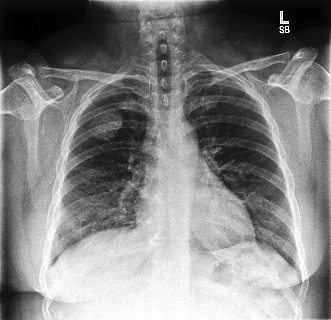}}\quad \\
  \subfigure[Lung Opacity]{\includegraphics[width=3.25cm, height=3.5cm]{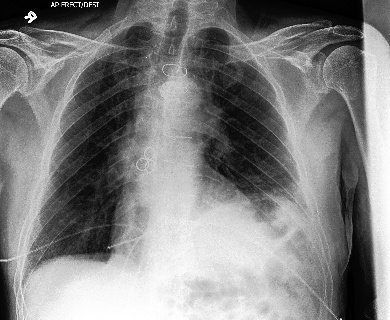}}\quad
  \subfigure[Edema]{\includegraphics[width=3.25cm, height=3.5cm]{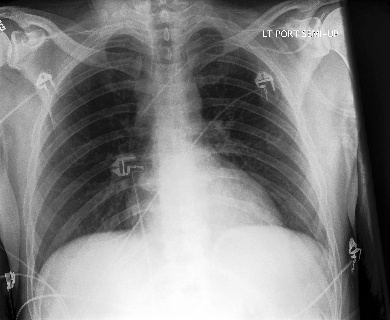}} \quad
  \subfigure[Consolidation]{\includegraphics[width=3.25cm, height=3.5cm]{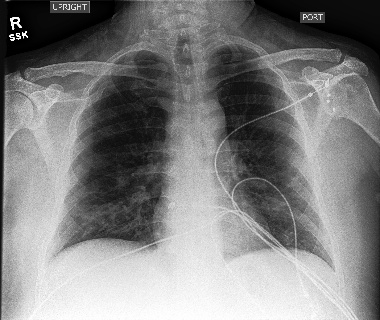}} \quad
  \subfigure[Pneumonia]{\includegraphics[width=3.25cm, height=3.5cm]{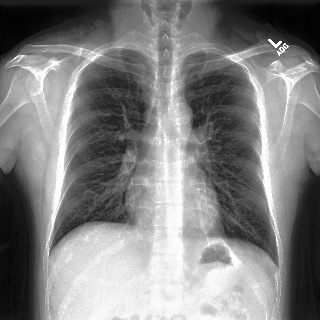}} \quad \\
  \subfigure[Atelectasis]{\includegraphics[width=3.25cm, height=3.5cm]{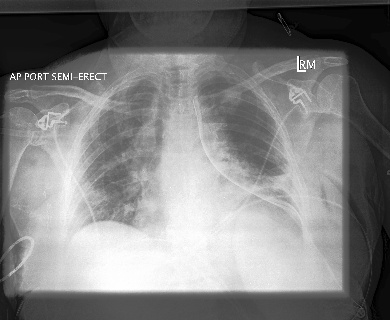}}\quad
  \subfigure[Pneumothorax]{\includegraphics[width=3.25cm, height=3.5cm]{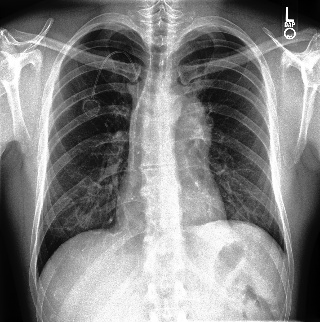}} \quad
  \subfigure[Pleural Effusion]{\includegraphics[width=3.25cm, height=3.5cm]{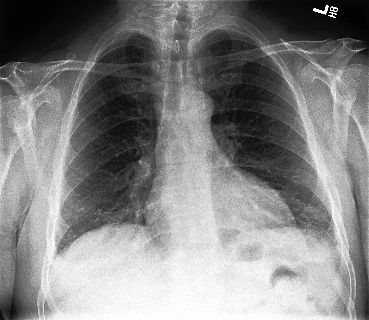}} \quad
  \subfigure[Pleural Other]{\includegraphics[width=3.25cm, height=3.5cm]{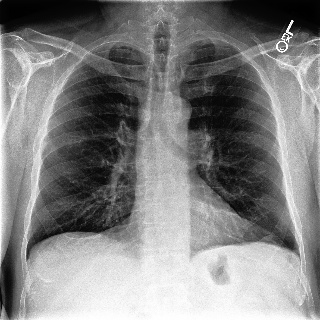}} \quad \\
  \subfigure[Fracture]{\includegraphics[width=3.25cm, height=3.5cm]{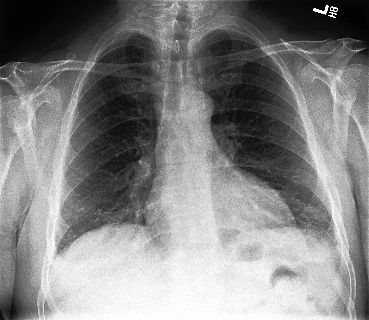}}\quad
  \subfigure[Support Devices]{\includegraphics[width=3.25cm, height=3.5cm]{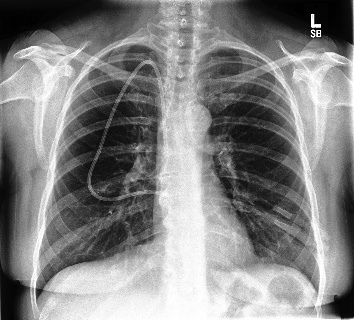}} \quad \\
  \caption{Sample image of each of the $14$ observations in the CheXpert dataset.} \label{fig:CheXpert-sampleimages}
\end{figure*}

\begin{figure*}[tb]
  \centering
 \subfigure[No Finding]{\includegraphics[width=3.25cm, height=3.5cm]{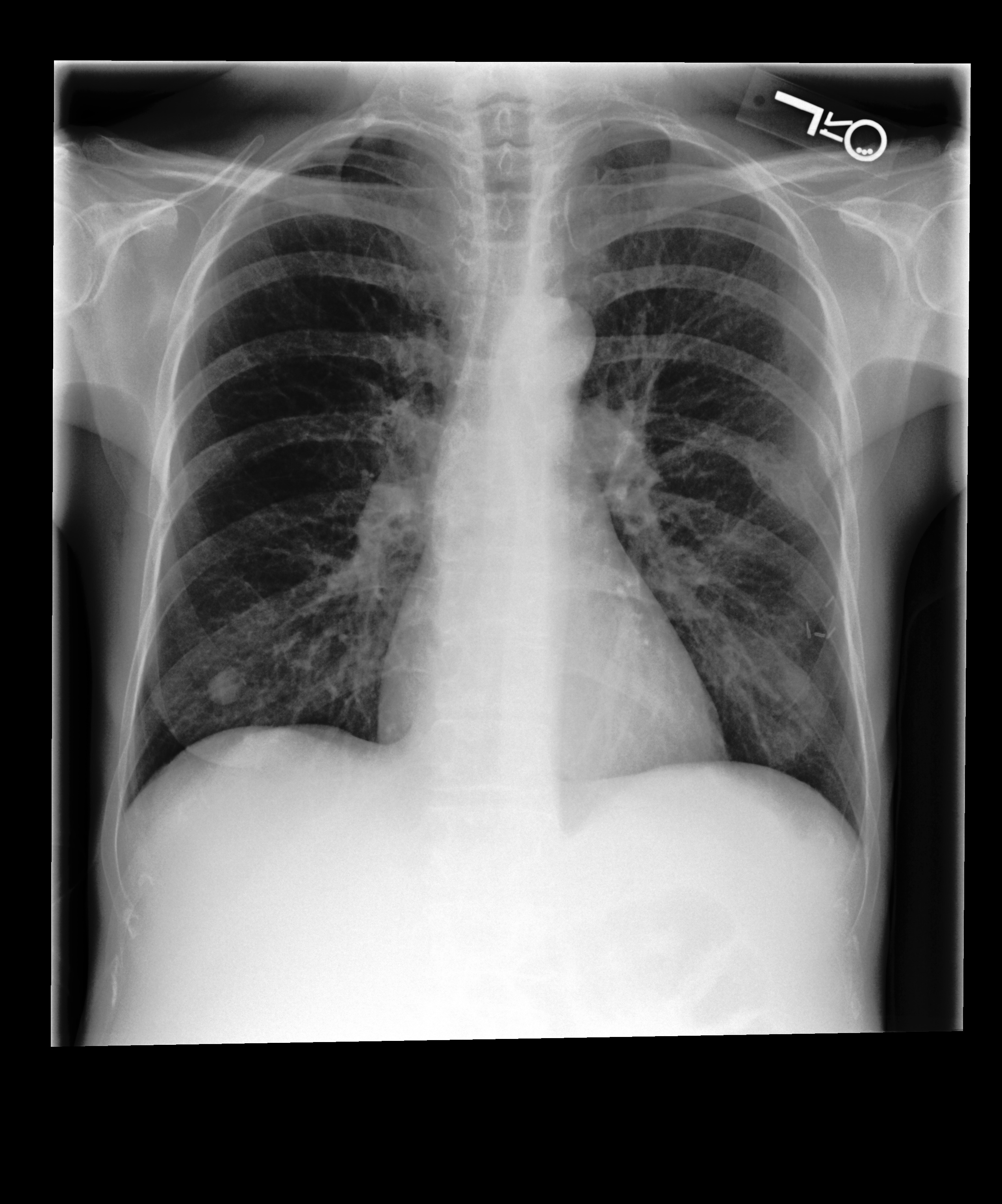}}\quad
  \subfigure[Enlarged Cardiomediastinum]{\includegraphics[width=3.25cm, height=3.5cm]{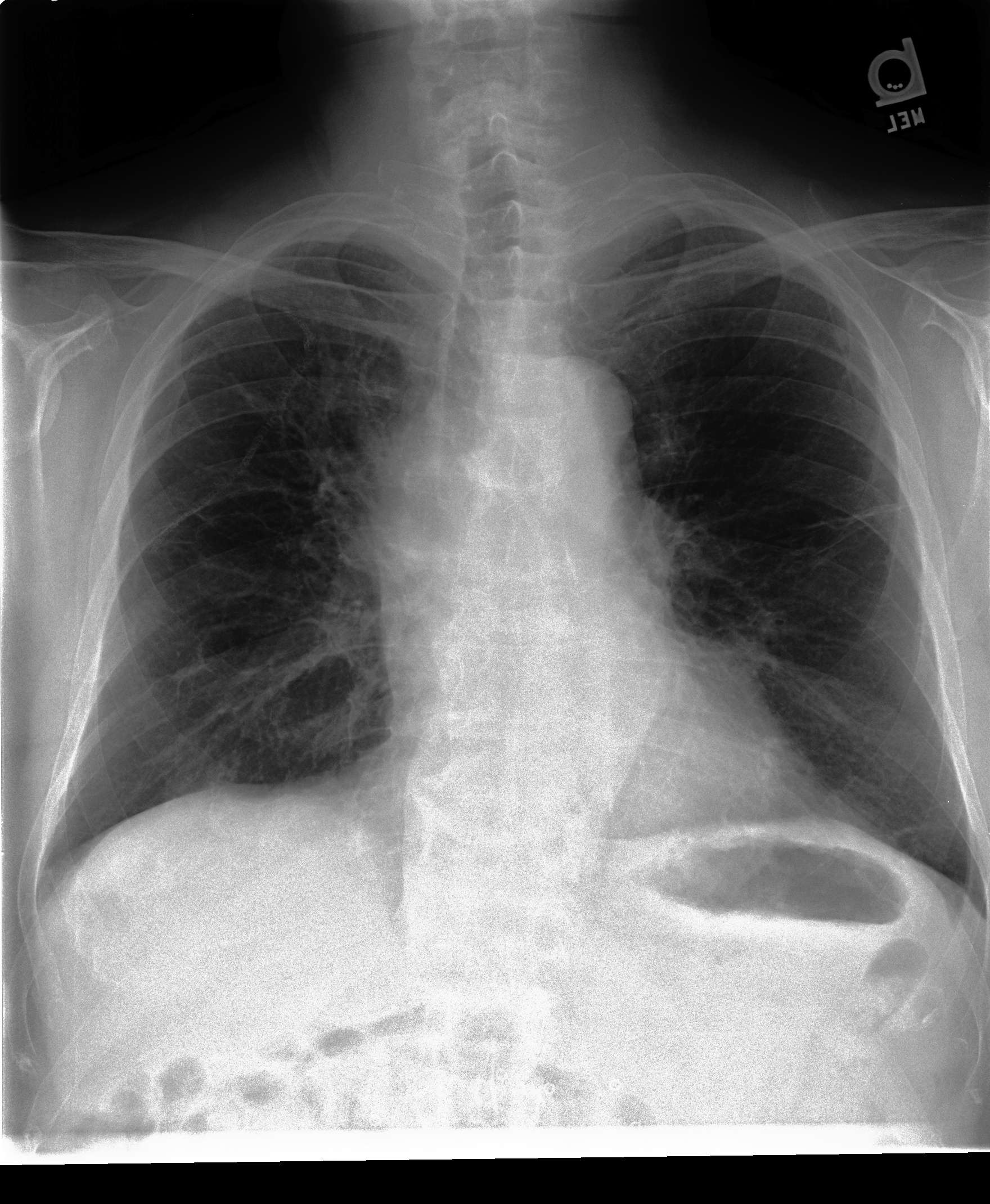}}\quad
  \subfigure[Cardiomegaly]{\includegraphics[width=3.25cm, height=3.5cm]{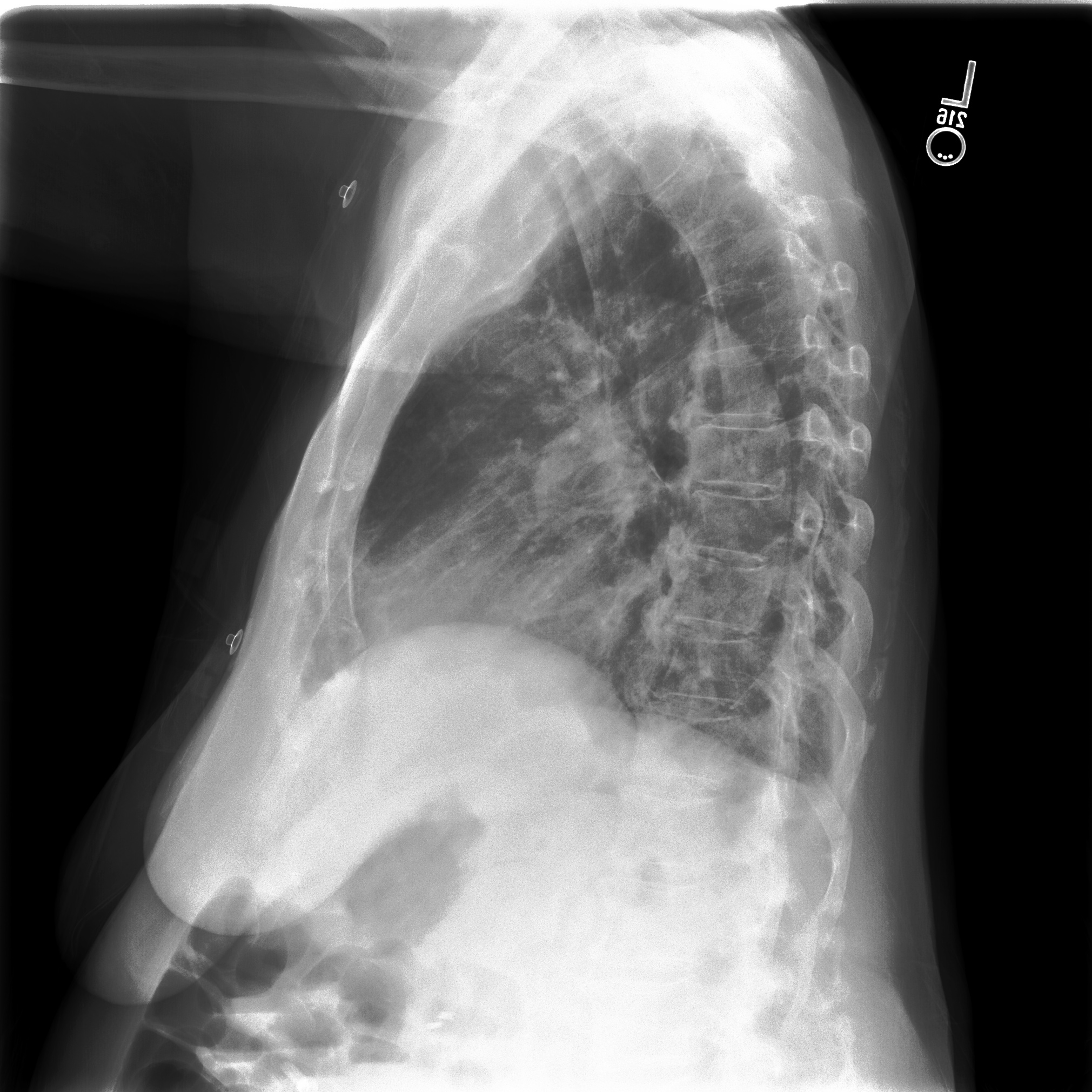}}\quad
  \subfigure[Lung Lesion]{\includegraphics[width=3.25cm, height=3.5cm]{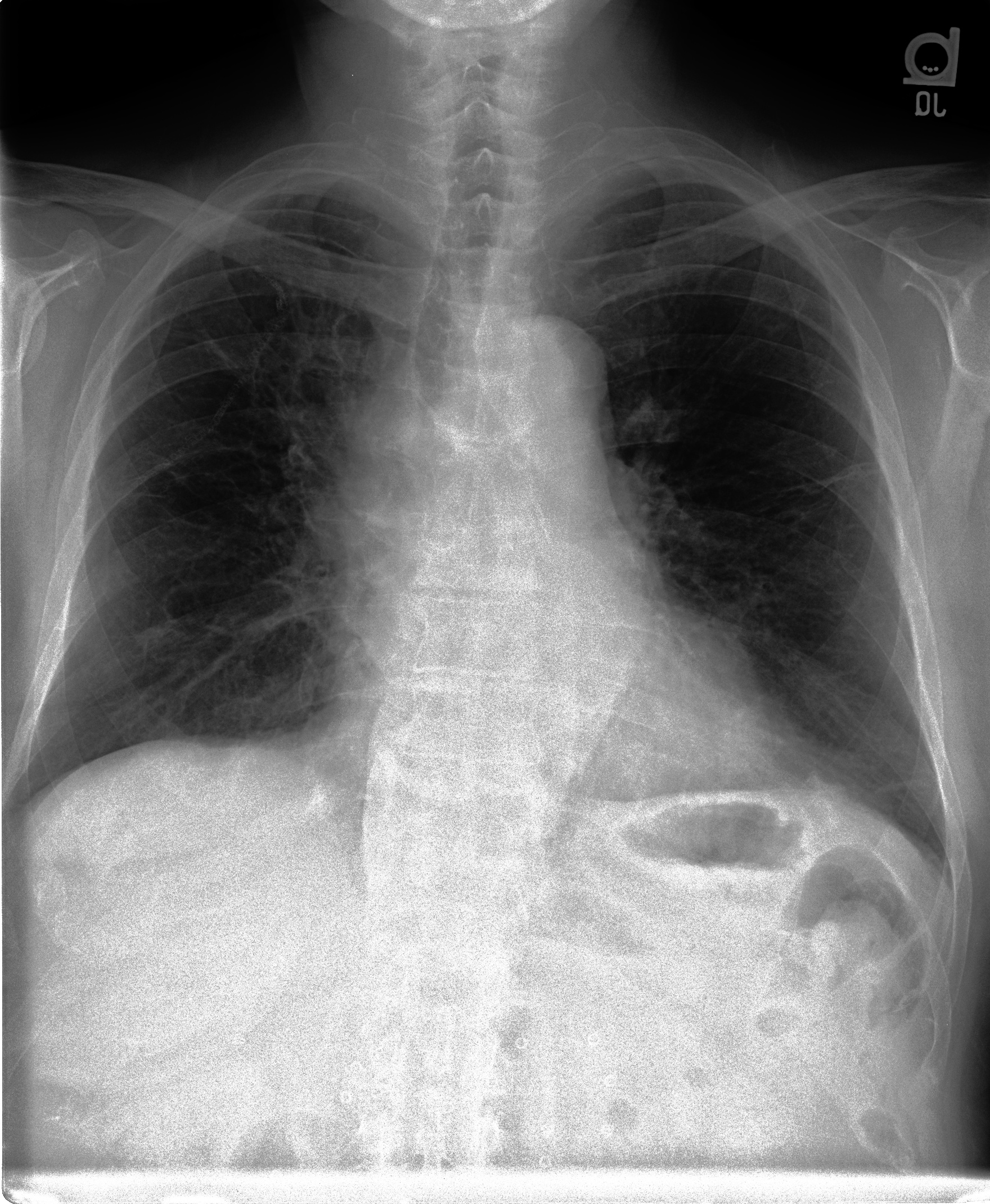}}\quad \\
  \subfigure[Lung Opacity]{\includegraphics[width=3.25cm, height=3.5cm]{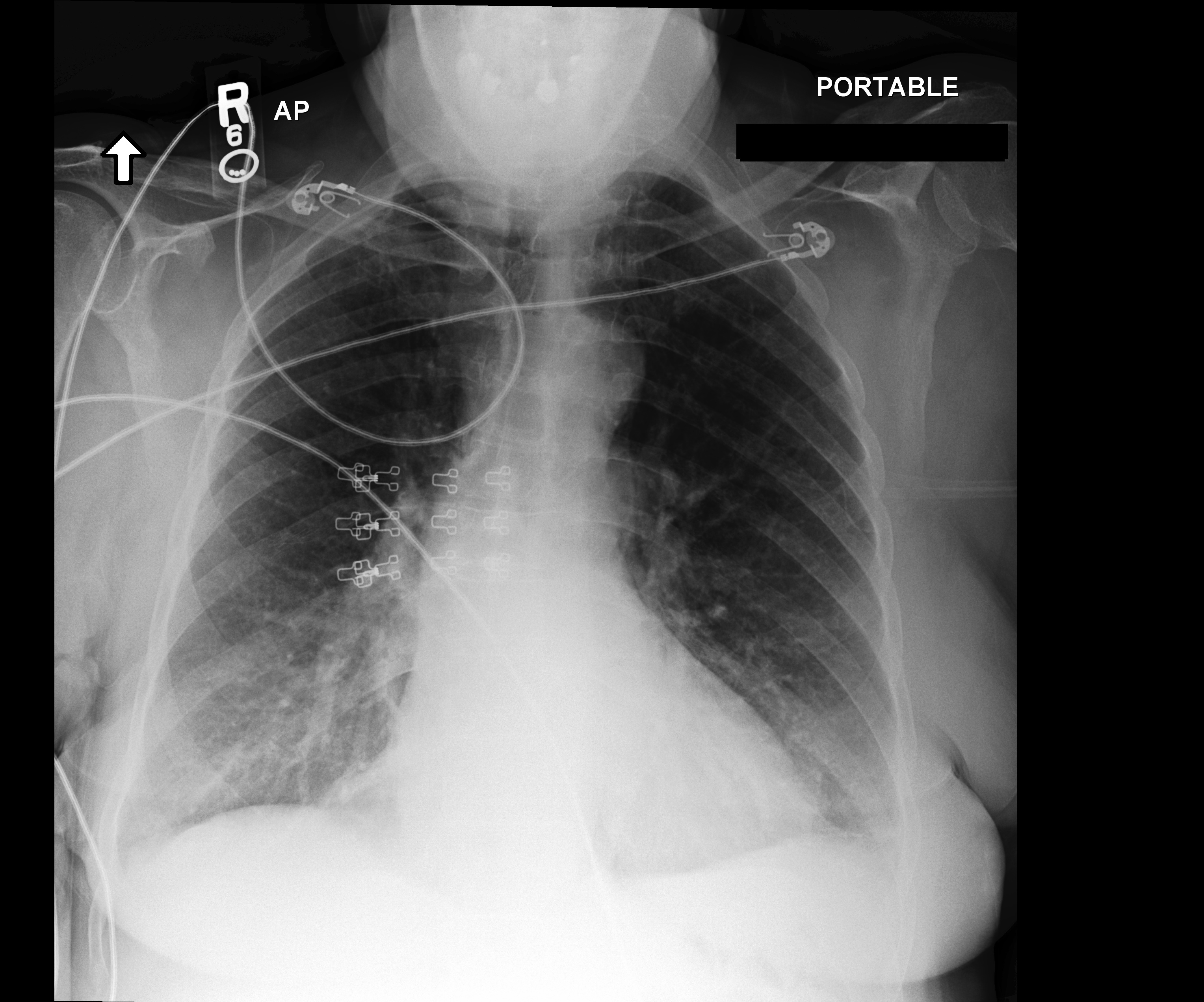}}\quad
  \subfigure[Edema]{\includegraphics[width=3.25cm, height=3.5cm]{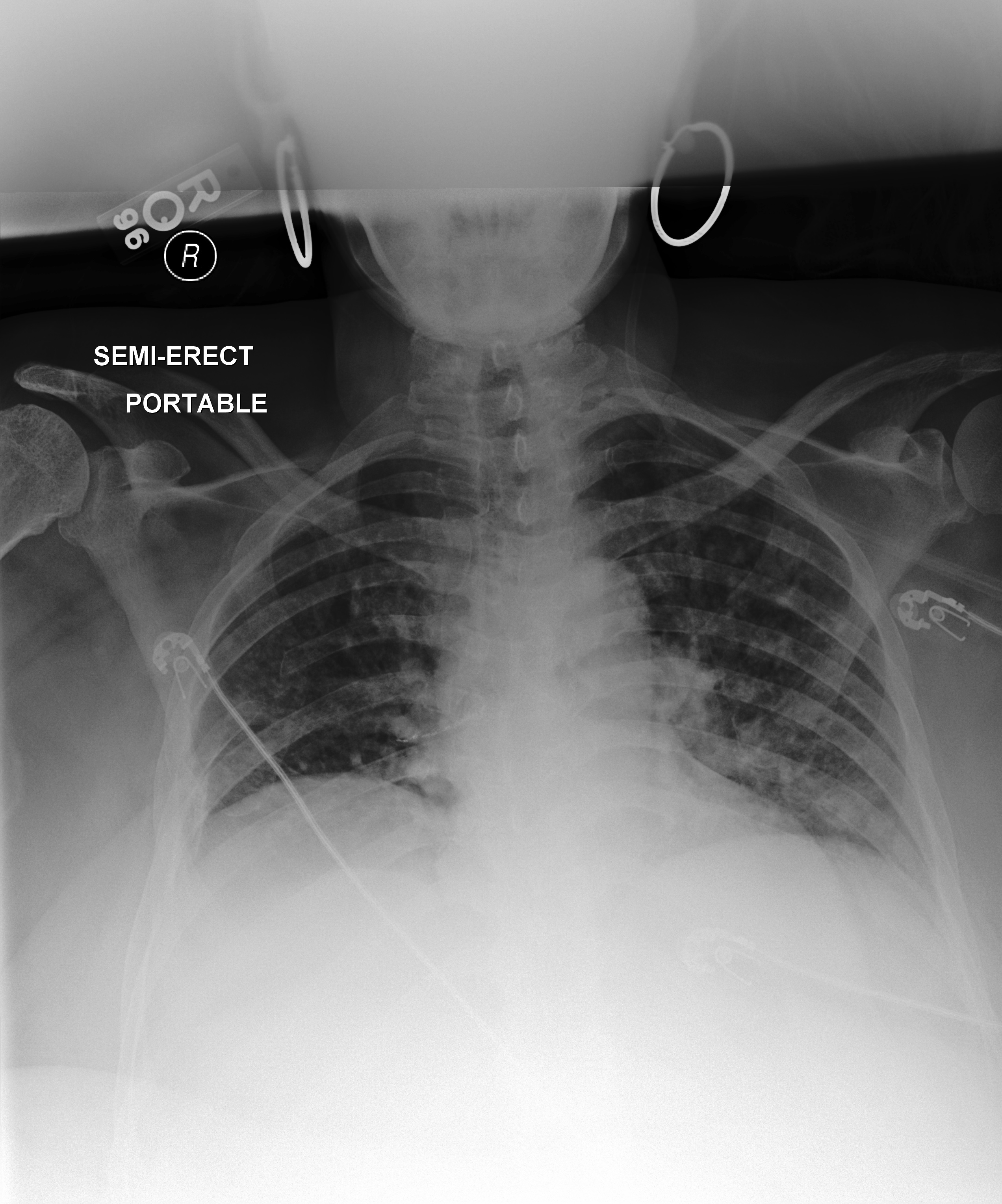}} \quad
  \subfigure[Consolidation]{\includegraphics[width=3.25cm, height=3.5cm]{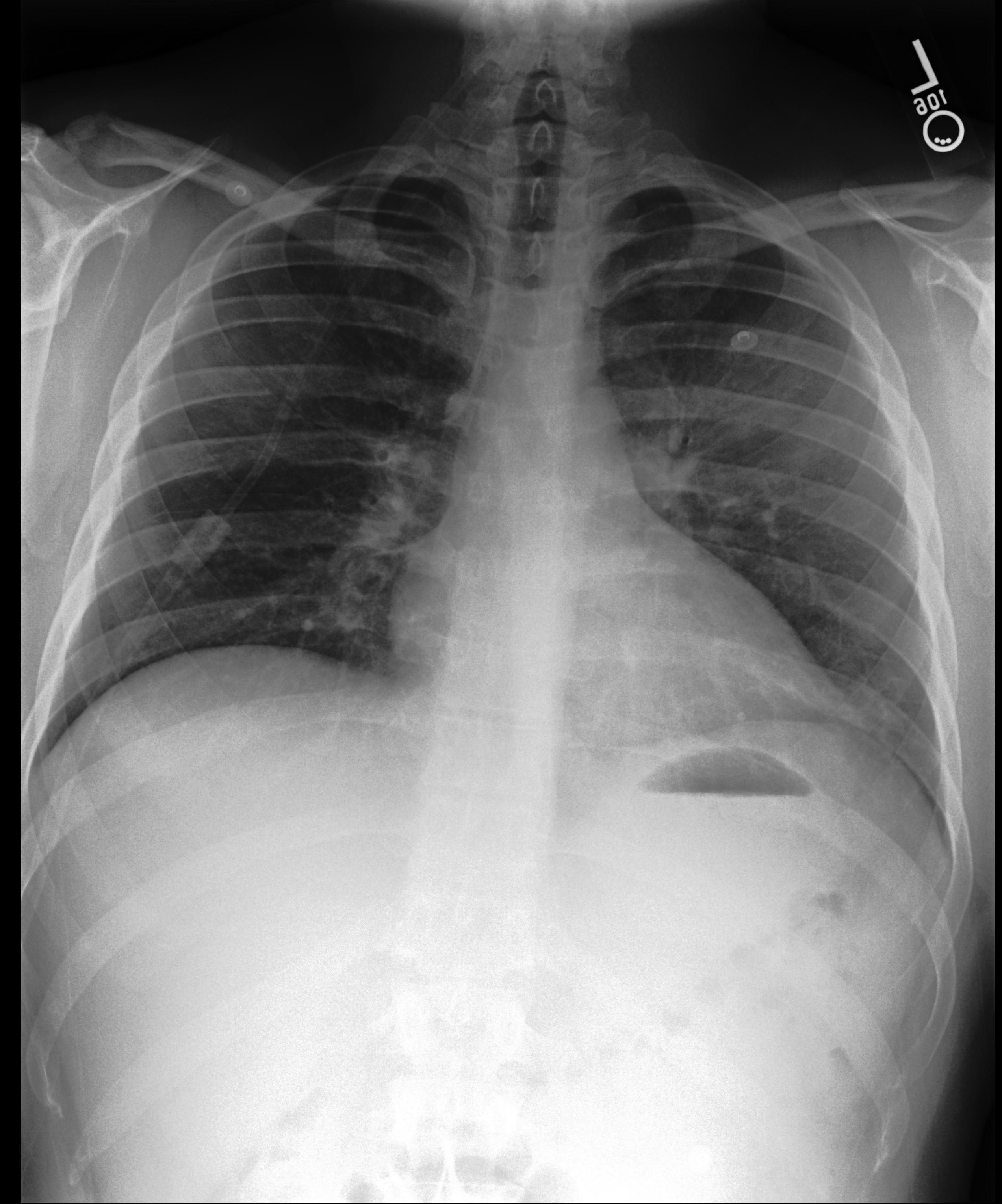}} \quad
  \subfigure[Pneumonia]{\includegraphics[width=3.25cm, height=3.5cm]{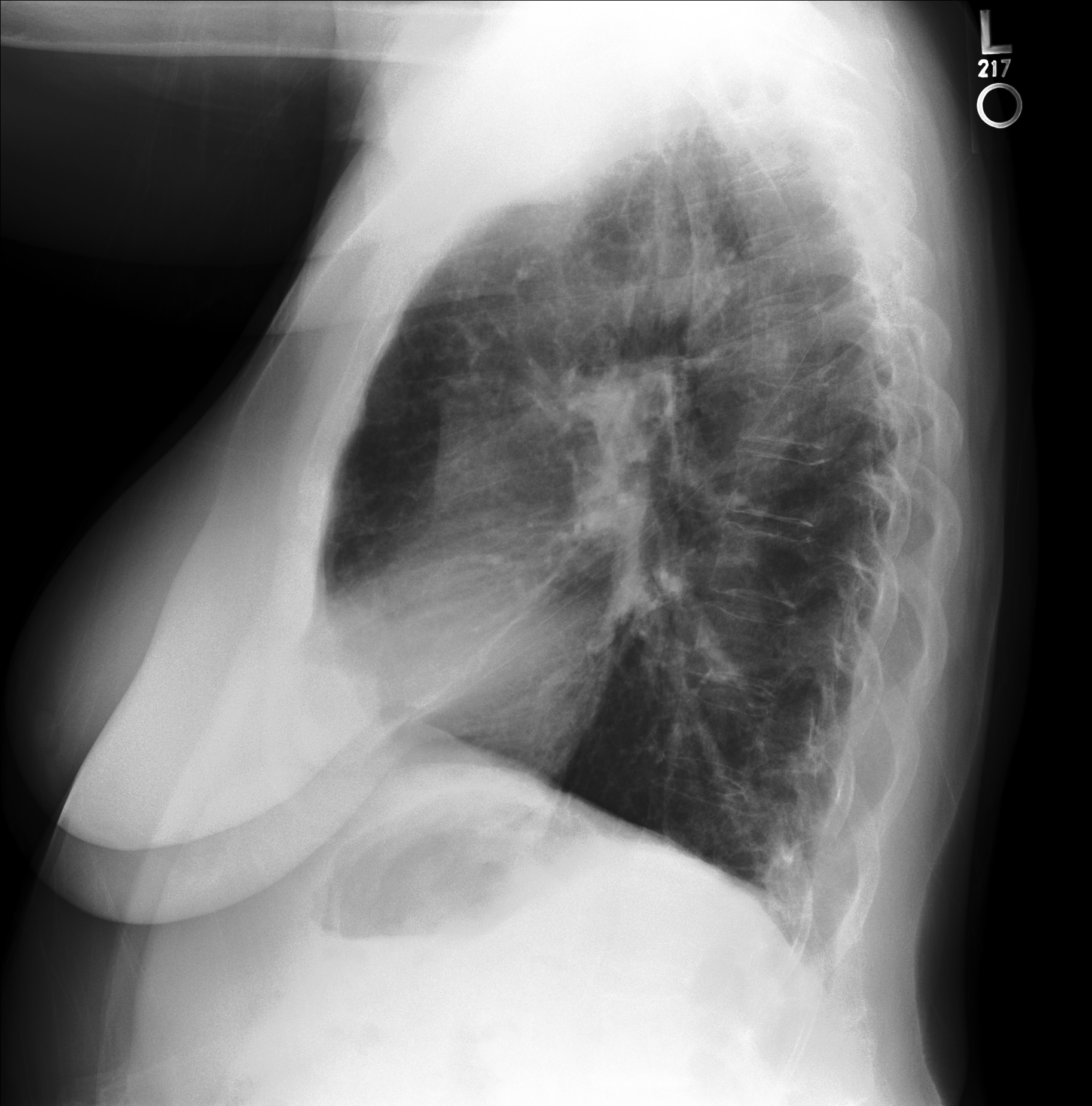}} \quad \\
  \subfigure[Atelectasis]{\includegraphics[width=3.25cm, height=3.5cm]{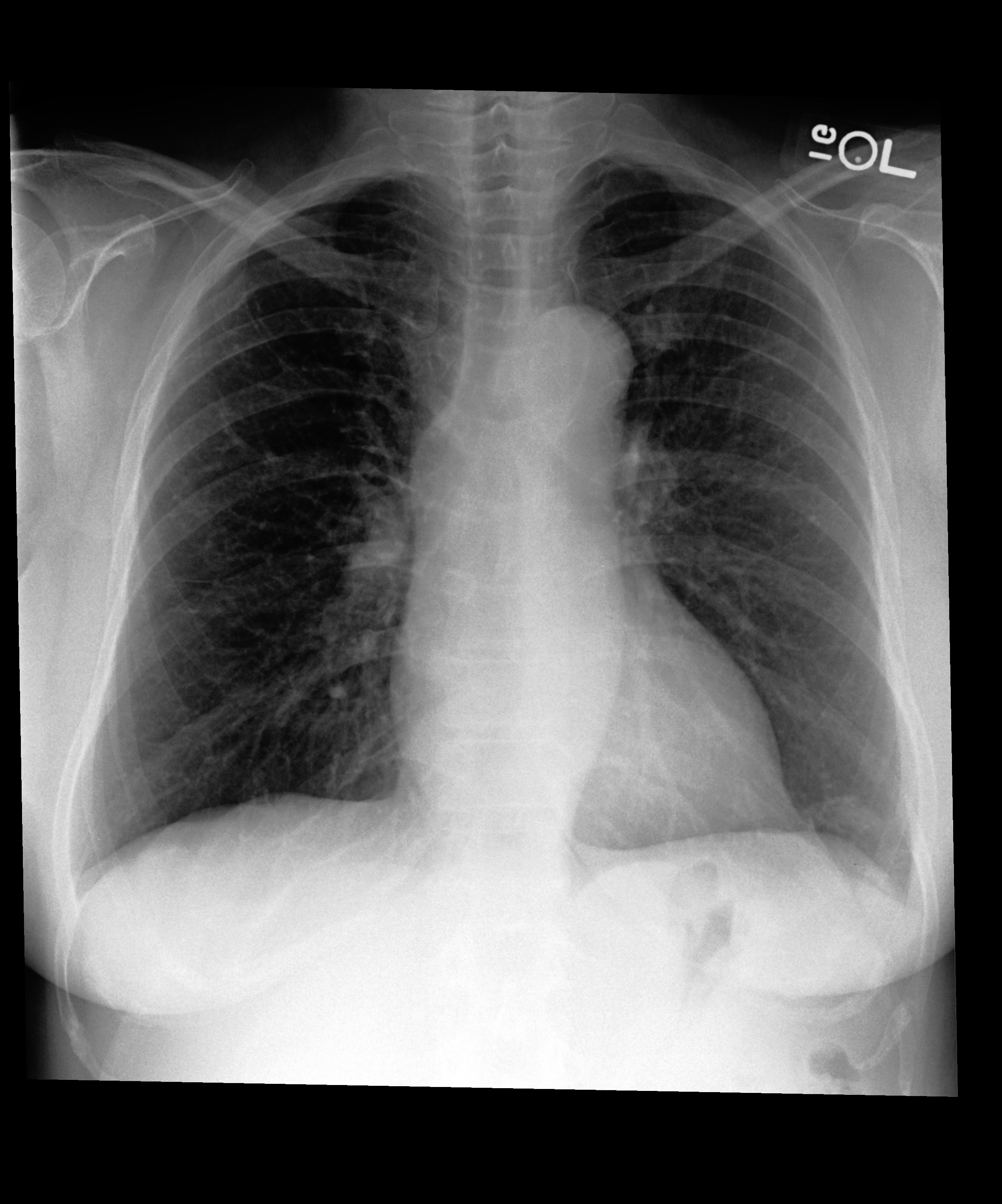}}\quad
  \subfigure[Pneumothorax]{\includegraphics[width=3.25cm, height=3.5cm]{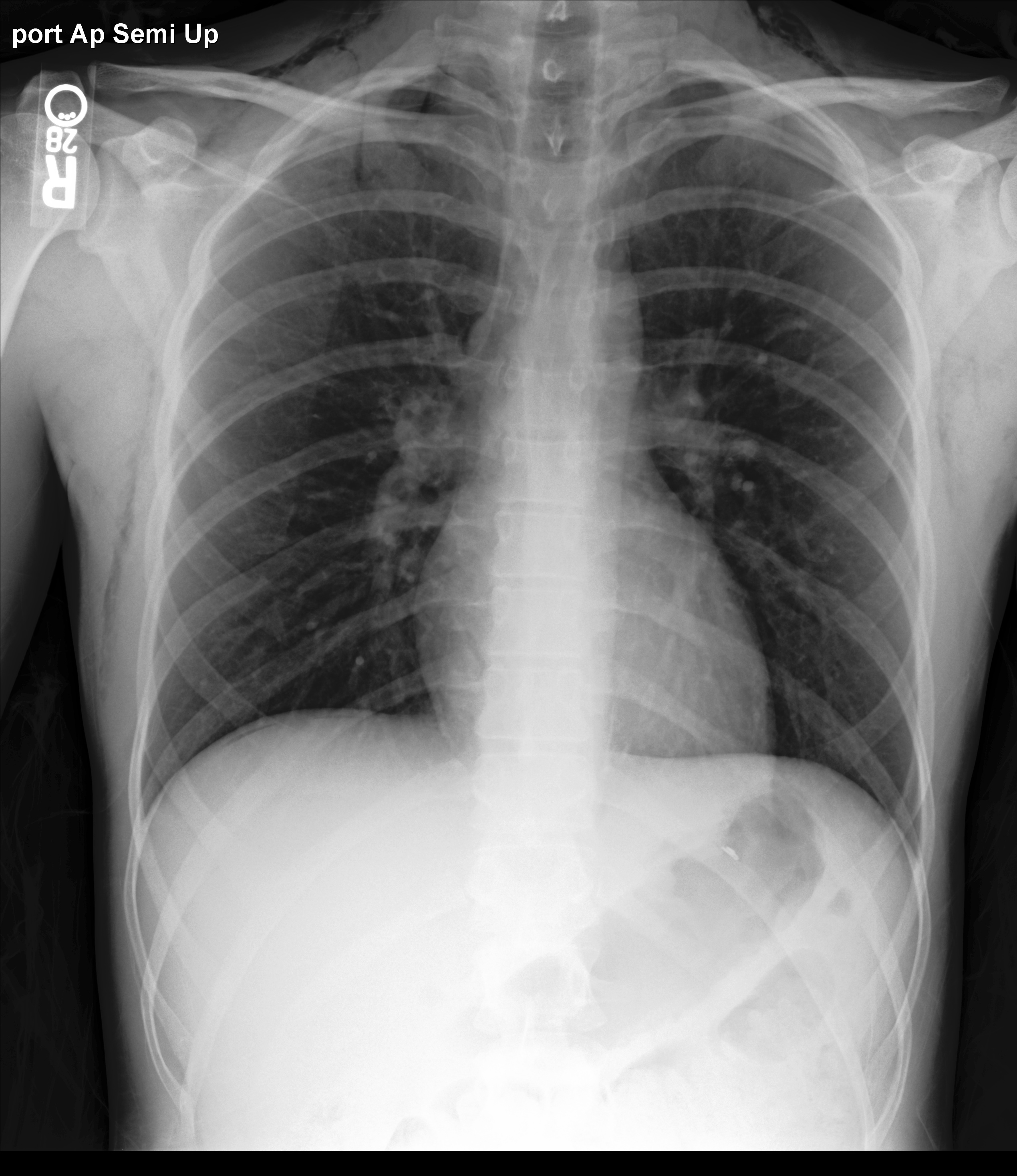}} \quad
  \subfigure[Pleural Effusion]{\includegraphics[width=3.25cm, height=3.5cm]{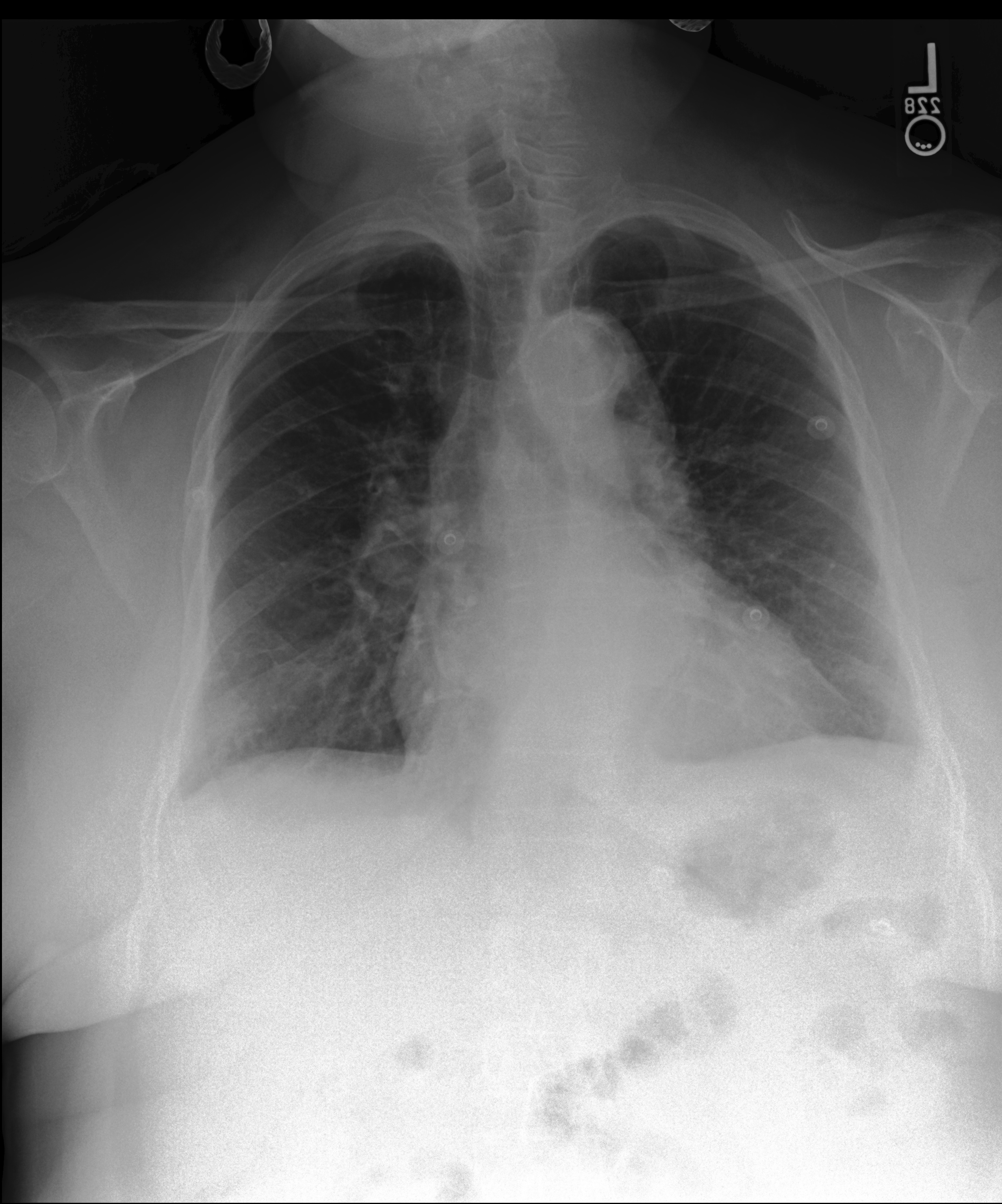}} \quad
  \subfigure[Pleural Other]{\includegraphics[width=3.25cm, height=3.5cm]{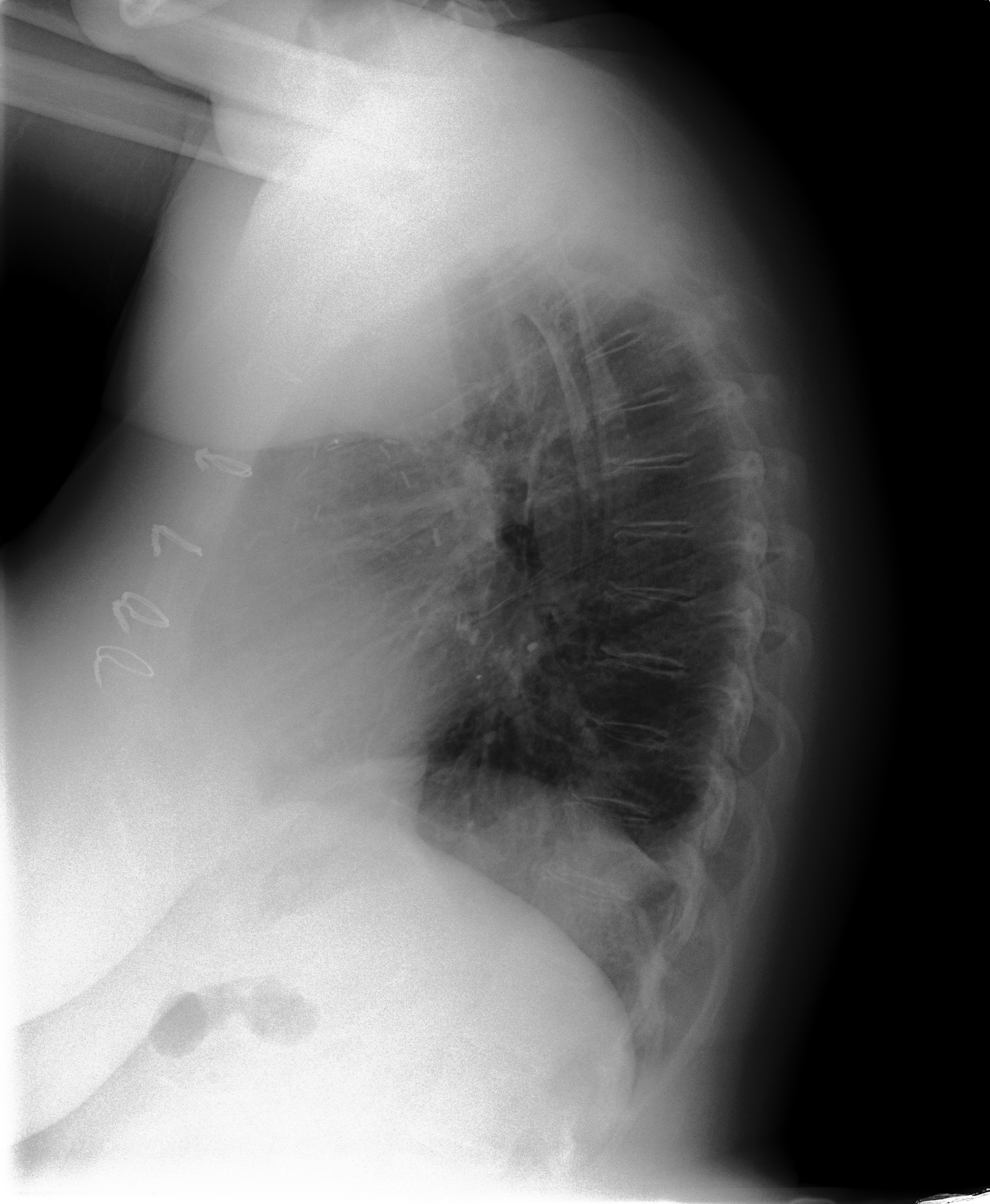}} \quad \\
  \subfigure[Fracture]{\includegraphics[width=3.25cm, height=3.5cm]{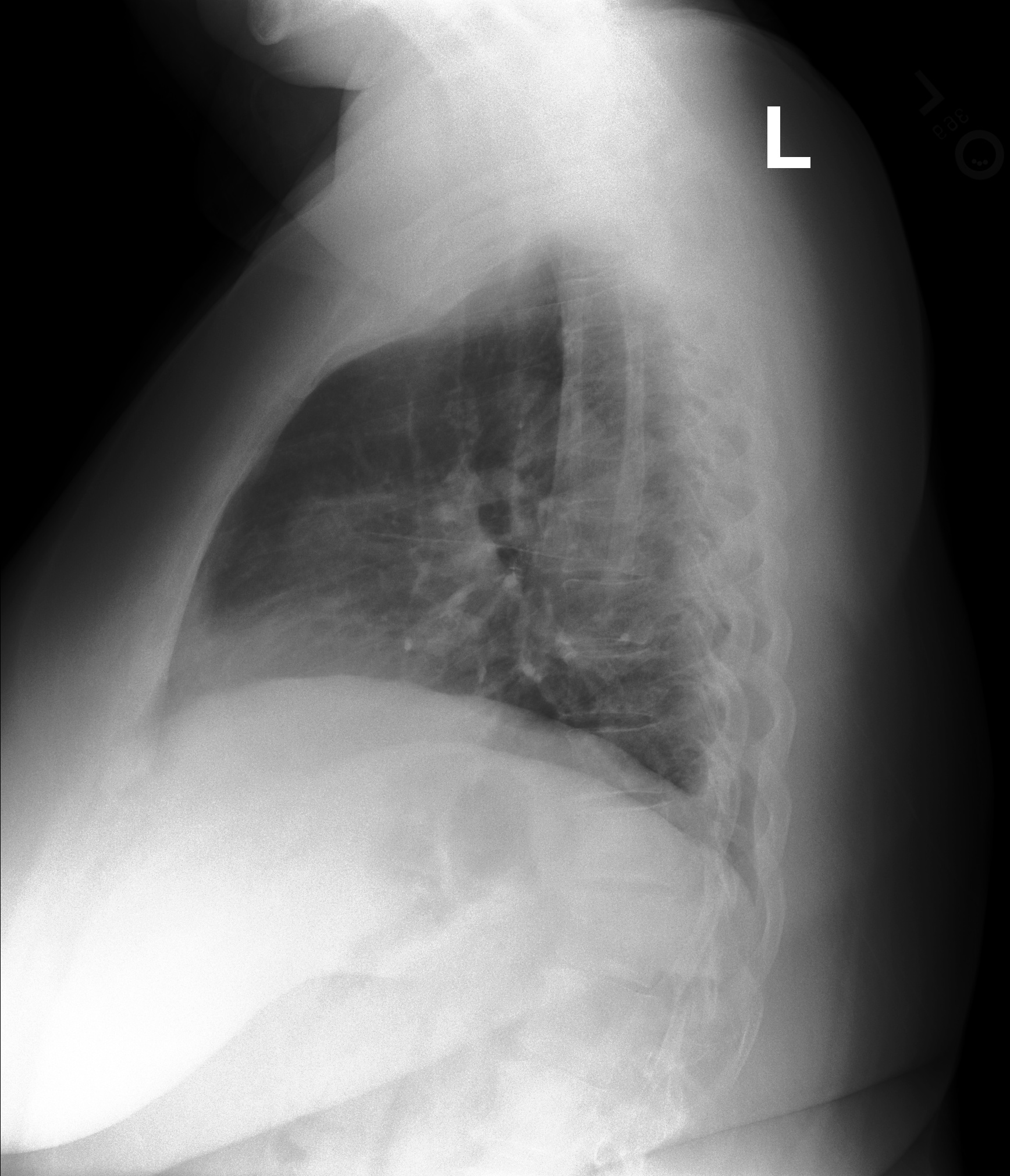}}\quad
  \subfigure[Support Devices]{\includegraphics[width=3.25cm, height=3.5cm]{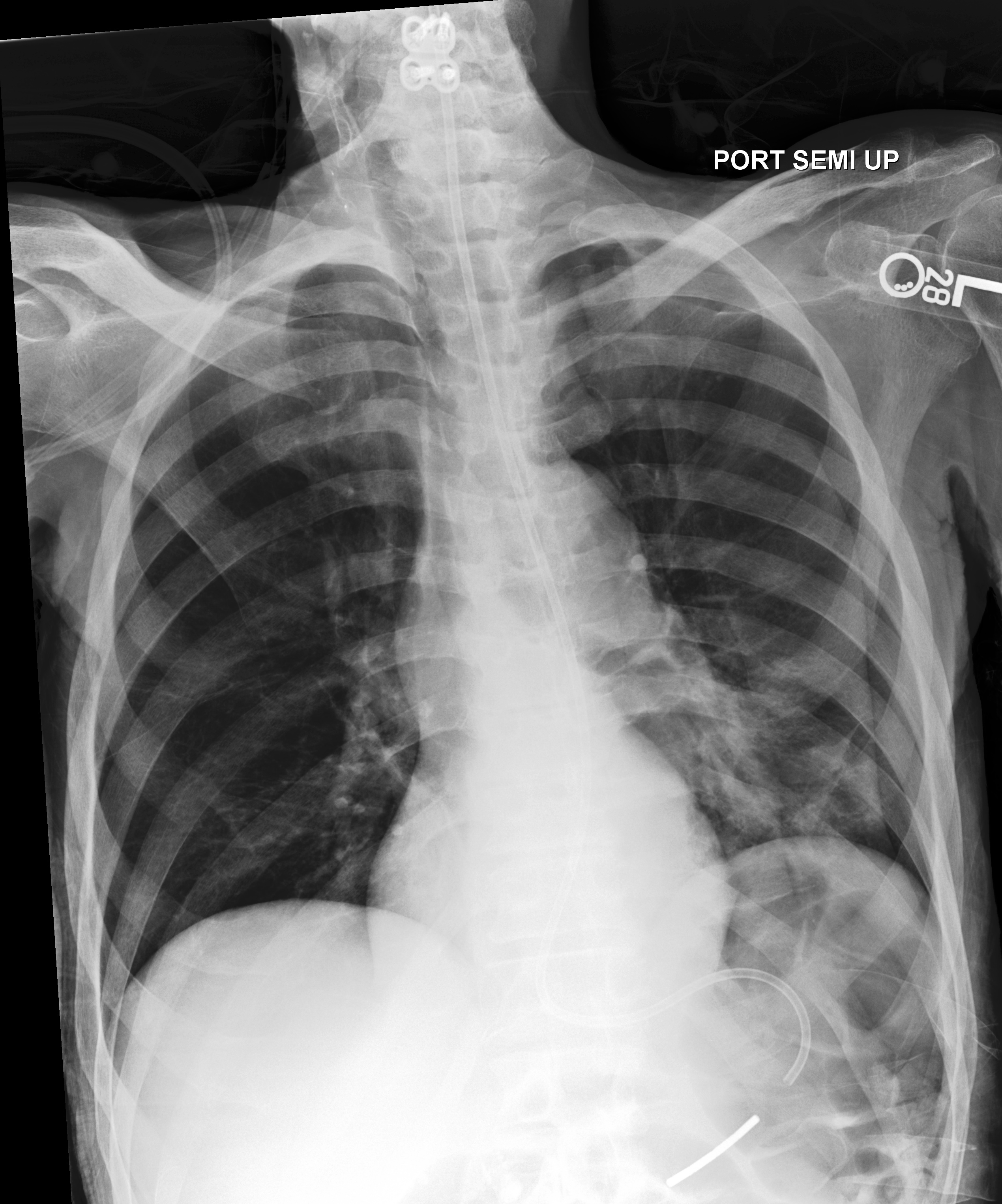}} \quad \\
  \caption{Sample image of each of the $14$ observations in the MIMIC-CXR-JPG dataset.} \label{fig:mimic-cxr-sampleimages}
\end{figure*}

\begin{figure*}[tb]
  \centering
   \subfigure[No Finding]{\includegraphics[width=3.25cm, height=3.5cm]{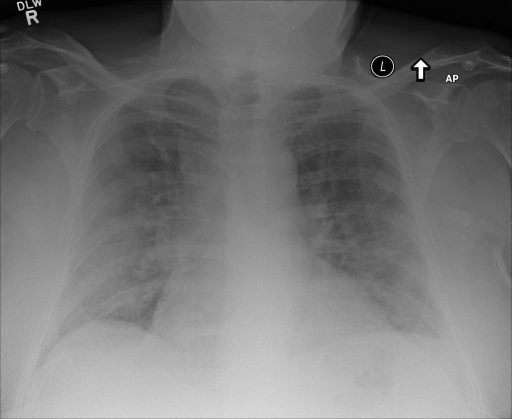}}\quad
  \subfigure[Enlarged Cardiomediastinum]{\includegraphics[width=3.25cm, height=3.5cm]{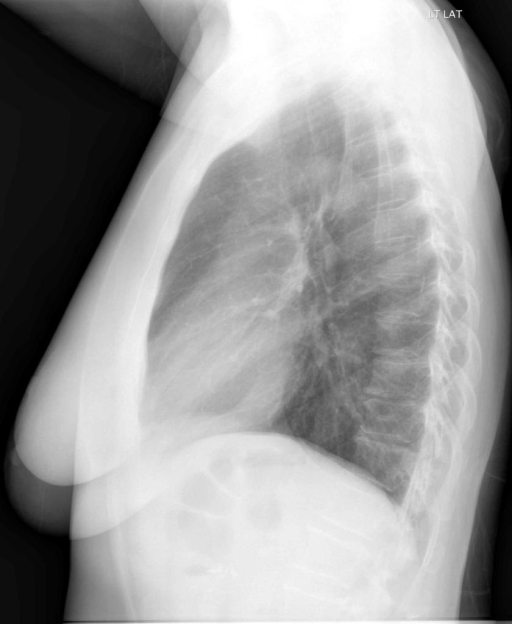}}\quad
  \subfigure[Cardiomegaly]{\includegraphics[width=3.25cm, height=3.5cm]{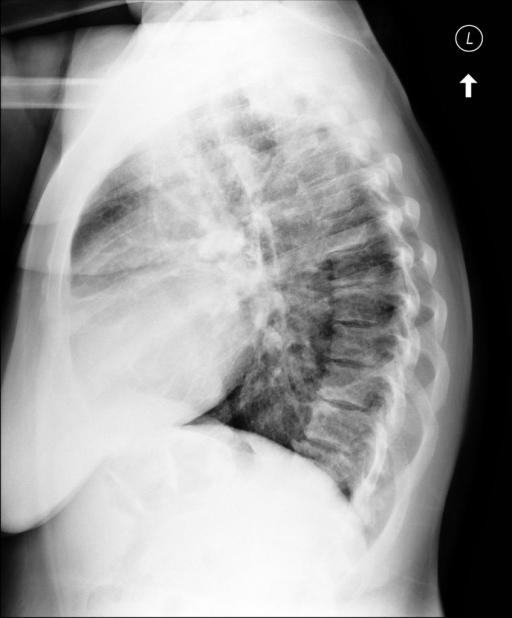}}\quad
  \subfigure[Lung Lesion]{\includegraphics[width=3.25cm, height=3.5cm]{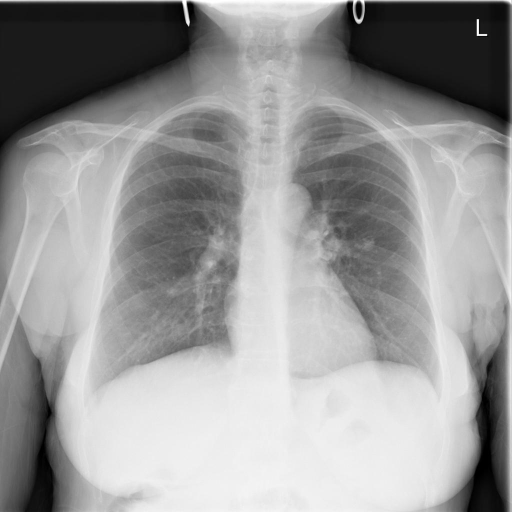}}\quad \\
  \subfigure[Lung Opacity]{\includegraphics[width=3.25cm, height=3.5cm]{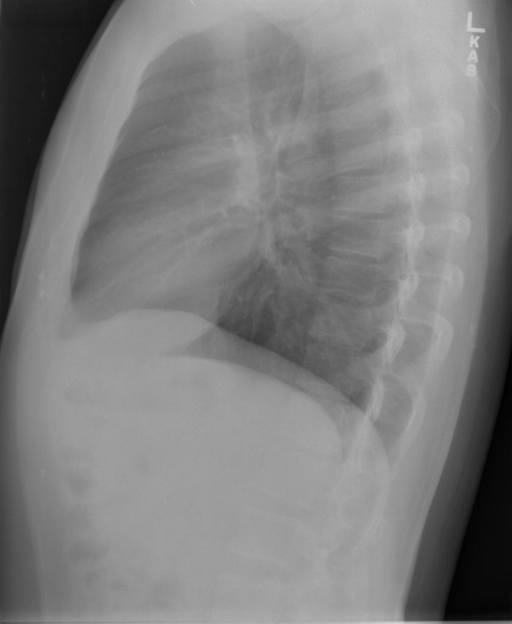}}\quad
  \subfigure[Edema]{\includegraphics[width=3.25cm, height=3.5cm]{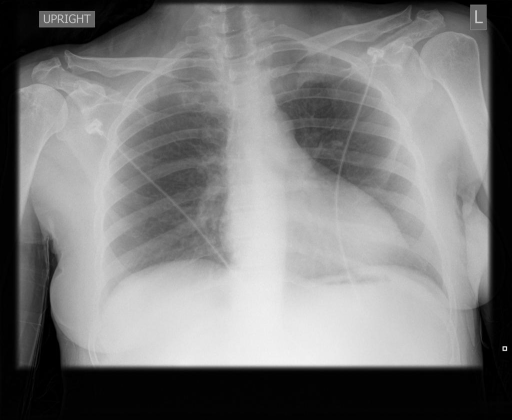}} \quad
  \subfigure[Consolidation]{\includegraphics[width=3.25cm, height=3.5cm]{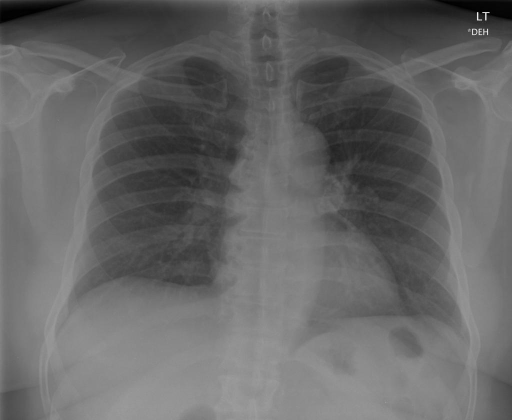}} \quad
  \subfigure[Pneumonia]{\includegraphics[width=3.25cm, height=3.5cm]{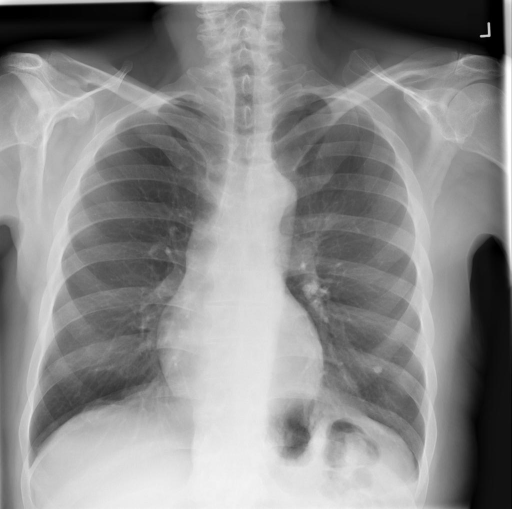}} \quad \\
  \subfigure[Atelectasis]{\includegraphics[width=3.25cm, height=3.5cm]{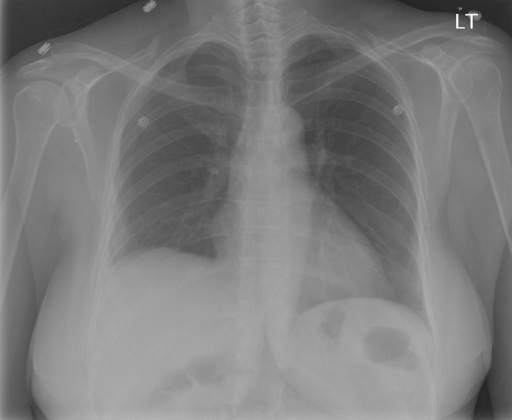}}\quad
  \subfigure[Pneumothorax]{\includegraphics[width=3.25cm, height=3.5cm]{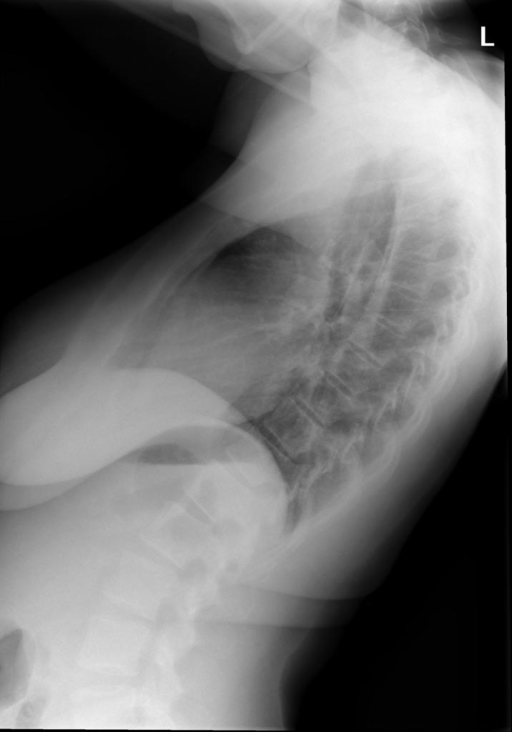}} \quad
  \subfigure[Pleural Effusion]{\includegraphics[width=3.25cm, height=3.5cm]{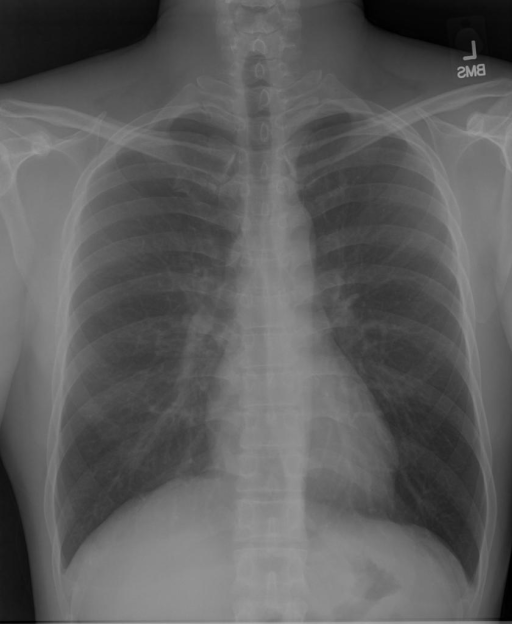}} \quad
  \subfigure[Pleural Other]{\includegraphics[width=3.25cm, height=3.5cm]{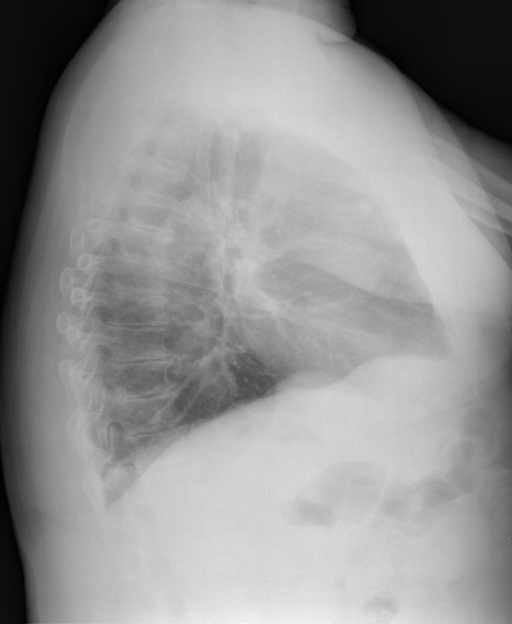}} \quad \\
  \subfigure[Fracture]{\includegraphics[width=3.25cm, height=3.5cm]{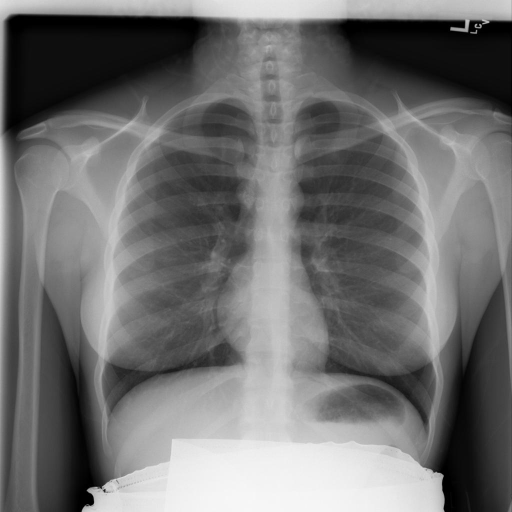}}\quad
  \subfigure[Support Devices]{\includegraphics[width=3.25cm, height=3.5cm]{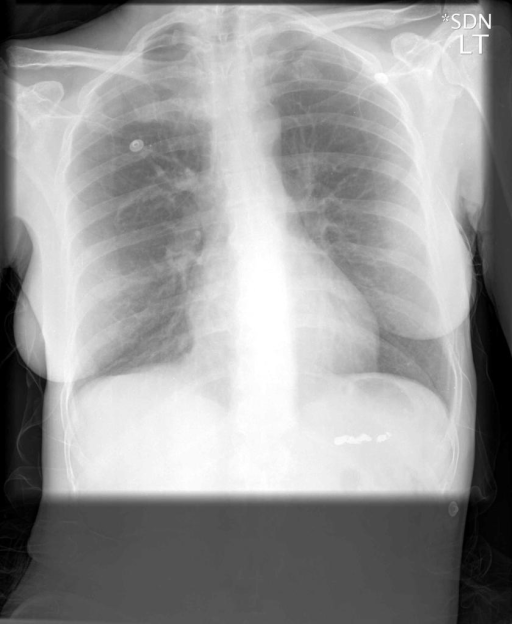}} \quad \\
  \caption{Sample image of each of the $14$ observations in the IU-CXR dataset.} \label{fig:iucxr-sampleimages}
\end{figure*}

\subsubsection{ChestX-ray14}
The \emph{ChestX-ray14} is a large-scale dataset publicly released by the National Institute of Health. It contains $112,120$ Chest X-rays from $30,805$ unique patients. Each image is labeled with one or more pathologies, making it a multi-label classification problem. The pathology labels associated with each Chest X-ray are annotated by mining radiology reports applying natural language processing tools. The images present in the dataset are in .png format and are resized to $1024 \times 1024$ without preserving aspect ratio. If there is no pathology present in a Chest X-ray, it is labeled as ``No Finding". The disease labels are expected to have accuracy of above $90\%$. We take the provided labels as ground-truth for training and evaluation in this work. 

The dataset has been divided into training, validation, and testing, consisting of $70\%$, $20\%$, and $10\%$ of the total dataset respectively. The dataset is split in a way to ensure that there is no patient overlap in hte training and testing data. Each image may have one or more labels for Atelectasis, Cardiomegaly, Effusion, Infiltration, Mass, Nodule, Pneumonia, Pneumothorax, Consolidation, Edema, Emphysema, Fibrosis, Pleural Thickening, and Hernia. 

Following the original work by~\cite{Wang:2017:ChestX-ray14}, where this dataset was introduced, there are several notable studies that utilised the same dataset. Unfortunately, each of these studies performed their own splits, making it hard to establish consistent benchmark. Recently, an official train and test split is provided by~\cite{Wang:2017:ChestX-ray14}. These standard train and test splits are publicly available and the split was done per patient rather than per image so that images of the same patient do not occur in both train and test at the same time as it will leak information and can provide spurious results. Table~\ref{tab:chestX-ray14_num_images} shows the number of images for each pathology in training, validation, and test split as per the official split. This official split of data, having training ($70\%$), validation ($10\%$), and testing ($20\%$), is provided to compare results with the existing work. Out of the 111120 chest X-rays, 51708 contains one or more pathologies. The remaining 60412 images are considered normal. There are in total 15 different labels in the dataset (14 common pathological keywords plus ‘No Finding’).

\begin{table*}[]
    \centering
    \caption{Distribution of Chest X-rays in training, validation, and test split in Chest X-ray14 dataset}
    \begin{tabular}{l|r|r|r|r}
      \toprule
      Pathology & Training & Validation & Testing  & Total  \\ \midrule
      Atelectasis & 6,168 & 2,112 & 3,255 & 11535  \\
      Cardiomegaly & 1,273 & 434 & 1,065 & 2772  \\
      Effusion & 6,537 & 2,122 & 4,648 & 2516  \\
      Infiltration & 10,244 & 3,538  & 6,088 & 19871  \\
      Mass & 3,012 & 1,022 & 1,712 & 5746  \\
      Nodule & 3,501 & 1,207 & 615  & 6323 \\
      Pneumonia & 655 & 221 & 477 & 1353 \\
      Pneumothorax & 1,939 & 698 & 2,661 & 5298  \\
      Consolidation & 2,114 & 738 & 1,815 & 4667  \\
      Edema & 1,027 & 361 & 925 & 2303  \\
      Emphysema & 1,075 & 348 & 1,093  & 2516  \\
      Fibrosis & 963 & 288 & 435 & 1686  \\
      Pleural Thickening & 1,693 & 549 & 1,143 & 3385  \\
      Hernia & 105 & 36 & 86 & 227  \\
      No Finding & & &  & 60412  \\\midrule
      \textbf{Totals} & 78,468 & 11,219 & 22,433 & 112,120  \\ \bottomrule
    \end{tabular}
    \label{tab:chestX-ray14_num_images}
\end{table*}

\subsubsection{CheXpert}
The CheXpert dataset~\cite{Irvin:2019:CheXpert} is a large-scale public dataset for chest radiograph interpretation, consisting of $224,316$ chest X-rays of $65,240$ patients which is curated from chest radiographic examinations from Stanford Hospital between October 2002 and July 2017 in both inpatient and outpatient centres, along with their associated radiology reports. Each chest X-ray has an associated radiology report. An automated rule-based NLP labeler is applied to extract the presence of $14$ thoracic conditions as \emph{positive}, \emph{negative} or \emph{uncertain}. Based on the Fleischner Society's recommended glossary~\cite{Hansell:Bankier:2008:Fleischner_society} of terms for thoracic diseases and their clinical relevance, labels for only $14$ observations are extracted from radiology reports in the form of structured labels. Each image is labeled for the presence of 14 findings: \emph{No Finding}, \emph{Enlarged Cardiomediastinum}, \emph{Cardiomegaly}, \emph{Lung Opacity}, \emph{Lung Lesion}, \emph{Edema}, \emph{Consolidation}, \emph{Pneumonia}, \emph{Atelectasis}, \emph{Pneumothorax}, \emph{Pleural Effusion}, \emph{Pleural Other}, \emph{Fracture}, and \emph{Support Devices}. The whole dataset is divided into a training set of $223,414$ studies, a validation set of $234$ studies, and a test set of $500$ studies. 

Each report was labeled for the presence of 14 observations as positive, negative, or uncertain. We decided on the 14 observations based on the prevalence in the reports and clinical relevance, conforming to the Fleischner Society’s recommended glossary whenever applicable. We then developed an automated rule-based labeler to extract observations from the free text radiology reports to be used as structured labels for the images. The labeler is set up in three distinct stages: mention extraction, mention classification, and mention aggregation. In the mention extraction stage, the labeler extracts mentions from a list of observations from the Impression section of radiology reports, which summarises the key findings in the radiographic study. In the mention classification stage, mentions of observations are classified as negative, uncertain, or positive. In the mention aggregation stage, we use the classification for each mention of observations to arrive at a final label for the 14 observations (blank for unmentioned, 0 for negative, -1 for uncertain, and 1 for positive). Table~\ref{tab:chexpert_dataset} shows number of images for each observation for positive, uncertain and negative attributes. Table~\ref{tab:cheXpert_num_images} shows the distribution of images for each pathology in training, validation, and test split.

\begin{table*}[]
    \centering
    \caption{Distribution of positive, uncertain and negative studies for 14 observations in CheXpert dataset}
    \begin{tabular}{l|r|r|r|r}
      \toprule
      Pathology & Positive & Uncertain & Negative  & Total  \\ \midrule
      No Finding & 16,627 & 0 & 171,014 & 187,641  \\
      Enlarged Cardiomediastinum & 9,020 & 10,148 & 168,473 & 187,641  \\
      Cardiomegaly & 23,002 & 6,597 & 158,042 & 187,641  \\
      Lung Lesion & 6,856 & 1,071 & 179,714 & 187,641 \\
      Lung Opacity & 92,669 & 4,341 & 90,631 & 187,641  \\
      Edema & 48,905 & 11,571 & 127,165 & 187,641  \\
      Consolidation & 12,730 & 23,976 & 150,935 & 187,641  \\
      Pneumonia & 4,576 & 15,658 & 167,407 & 187,641  \\
      Atelectasis & 29,333 & 29,377 & 128,931 & 187,641  \\
      Pneumothorax & 17,313 & 2,663 & 167,665 & 187,641  \\
      Pleural Effusion & 75,696 & 9,419 & 102,526 & 187,641  \\
      Pleural Other & 2,441 & 1,771 & 183,429 & 187,641  \\
      Fracture & 7,270 & 484 & 179,887 & 187,641  \\
      Support Devices & 105,831 & 898 & 80,912 & 187,641  \\ \midrule
      \textbf{Totals} & 452,269 & 117,974 &  2,056,731 & 2,626,974  \\ \bottomrule
    \end{tabular}
    \label{tab:chexpert_dataset}
\end{table*}

\begin{table*}[]
    \centering
    \caption{Distribution of Chest X-rays in training, validation, and test split in CheXpert dataset}
    \begin{tabular}{l|r|r|r|r}
      \toprule
      Pathology & Training & Validation & Testing & Total  \\ \midrule
      No Finding & 22,334 & 38 & 47 & 22,419 \\
      Enlarged Cardiomediastinum & 23,146 & 109 & 55 & 23,310 \\
      Cardiomegaly & 35,005 & 68 & 82 & 35,155 \\
      Lung Lesion & 10,655 & 1 & 19 & 10,675 \\
      Lung Opacity &  110,925 & 126 & 254 & 111,305 \\
      Edema &  65,070 & 45 & 160 & 65,275 \\
      Consolidation & 42,422 & 33 & 103 & 42,558 \\
      Pneumonia & 24,756 & 8 & 53 & 24,817 \\
      Atelectasis & 66,955 & 80 & 160 & 67,195 \\
      Pneumothorax & 22,547 & 8 & 46 & 22,601 \\
      Pleural Effusion &  97,581 & 67 & 234 & 97,882 \\
      Pleural Other & 6,160 & 1 & 16 & 6,177 \\
      Fracture & 9,656 & 0 & 26 & 9,682 \\
      Support Devices & 116,813 & 107 & 267 & 117,187 \\ \midrule
      \textbf{Totals} & 222,914 & 234 & 500 & 223,648 \\ \bottomrule
    \end{tabular}
    \label{tab:cheXpert_num_images}
\end{table*}

\subsubsection{MIMIC-CXR-JPG}
The MIMIC-CXR-JPG dataset v2.0.0~\cite{Johnson:2019:MIMIC-CXR-JPG}, which is a large publicly available dataset, consists of $377,110$ chest X-ray images and $227,827$ reports from $63,478$ patients. The dataset is wholly derived from the MIMIC-CXR dataset~\cite{Johnson:2019:MIMIC-CXR}, having chest X-rays in DICOM format along with accompanied text radiology reports. The structured labels for images in MIMIC-CXR-JPG were labeled from free-text radiology reports using CheXpert. CheXpert is an open source rule based tool that is built on NegBio, and have three stages: extraction, classification, and aggregation. The output of CheXpert was saved to a CSV file with one row per study and one column per finding. Each image is labelled as Positive, Negative, and Uncertain (blank for unmentioned, 0 for negative, -1 for uncertain and 1 for positive). 

\begin{table*}[]
    \centering
    \caption{Statistics about MIMIC-CXR dataset image splits into training, validation and testing sets.}
    \begin{tabular}{l|c|c|c}
    \toprule
    & \textbf{Training} & \textbf{Validation} & \textbf{Testing} \\ \midrule
    Number of images & 368,960 & 2,991 & 5,159 \\
    Number of studies & 222,758 & 1,808 & 3,269 \\
    Number o patients & 64,586 & 500 & 292 \\ 
    \bottomrule
    \end{tabular}
    \label{tab:mimic-cxr-splits}
\end{table*}

\begin{table*}[]
    \centering
    \caption{Distribution of Chest X-rays in training, validation, and test split in MIMIC-CXR-JPG dataset}
    \begin{tabular}{l|c|c|c|c}
      \toprule
      Pathology & Training & Validation & Testing  & Total  \\ \midrule
      No Finding &  141,239 & 1,129 & 984 & 143,352 \\
      Enlarged Cardiomediastinum & 21,926 & 214 & 487 & 22,627 \\
      Cardiomegaly & 70,768 & 625 & 1,422 & 72,815 \\
      Lung Lesion & 12,468 & 123 & 237 & 12,828 \\
      Lung Opacity & 80,059 & 596 & 1,684 & 82,339 \\
      Edema & 54,100 & 478 & 1,395 & 55,973 \\
      Consolidation & 20,521 & 184 & 473 & 21,178 \\
      Pneumonia & 53,858 & 410 & 1,104 & 55,372 \\
      Atelectasis & 78,593 & 671 & 1,319 & 80,583  \\
      Pneumothorax & 15,524 & 124 & 171 & 15,819 \\
      Pleural Effusion & 83,086 & 744 & 1,714 & 85,544 \\
      Pleural Other & 4,679 & 33 & 145 & 4,857  \\
      Fracture & 8,445 & 39 & 174 & 8,658  \\
      Support Devices & 82,240 & 731 & 1,463 & 84,434 \\ \midrule
      \textbf{Totals} & 368,945 & 2,991 & 5,159 & 377,095  \\ \bottomrule
    \end{tabular}
    \label{tab:mimic-cxr-jpg_num_images}
\end{table*}

\subsubsection{IU-CXR}
The Indiana University Chest X-ray Collection (IU-CXR) is a public radiography dataset collected by the Indiana University with $7,470$ chest X-ray images and $3,955$ radiology reports. Each radiology report is associated with a pair of images having frontal and lateral views. The reports are in XML format and include pre-parsed sections. Each radiology report consists of sections namely, \emph{comparison}, \emph{indication}, \emph{findings} and \emph{impression}. The Comparison section indicates whether the current imaging study is compared with any of the patient’s previous imaging studies. The Indication section lists patient information such as age, gender and relevant clinical information, including any existing disease and symptoms. The Findings section indicates whether each area in the image is normal, abnormal or potentially abnormal. The Impression section summarises the findings, patient clinical history and indication for the study and is considered to be the most indicative and important part of a radiology report for decision making. The dataset is fully de-identified and no patient specific information can be extracted.

Given no observation labels are available for this dataset, we applied \emph{CheXpert}, NLP labeler to get structured labels from radiology reports. We concatenated \emph{findings} and \emph{impression} section to form text of radiology reports and applied CheXpert labeler to them. The CheXpert labeler gives output in the form of $blank$ for no mention of observation in the report; $0$ for negative; $1$ for positive; and $u$ for uncertain mentions of each of the $14$ observations. We applied $U-Zeros$ mapping to convert $blank$ label to $0$ and $U-Ones$ mapping to convert $u$ to $1$. 

\begin{table*}[]
    \centering
    \caption{Distribution of Chest X-rays in training, validation, and test split in IU-CXR dataset}
    \begin{tabular}{l|r|r|r|r}
      \toprule
      Pathology & Training & Validation & Testing  & Total  \\ \midrule
      No Finding & 2,209 & 115 & 110 & 2,434 \\
      Enlarged Cardiomediastinum &  693 & 35 & 41 & 769  \\
      Cardiomegaly & 1,153 & 58 & 58 & 1,269  \\
      Lung Lesion & 411 & 27 & 28 & 466  \\
      Lung Opacity & 1,546 & 86 & 74 & 1,706  \\
      Edema &  344 & 10 & 22 & 376  \\
      Consolidation &  368 & 18 & 21 & 407  \\
      Pneumonia & 288 & 9 & 15 & 312  \\
      Atelectasis & 624 & 31 & 36 & 691  \\
      Pneumothorax & 391 & 26 & 15 & 432  \\
      Pleural Effusion & 681 & 35 & 40 & 756  \\
      Pleural Other & 101 & 3 & 3 & 107  \\
      Fracture &  258 & 11 & 18 & 287  \\
      Support Devices & 402 & 20 & 18 & 440  \\ \midrule
      \textbf{Totals} & 6,718 & 350 & 350 & 7418  \\ \bottomrule
    \end{tabular}
    \label{tab:iucxr_num_images}
\end{table*}

\subsection{Data analysis}
\begin{figure*}
    \centering
    \includegraphics[scale=0.5]{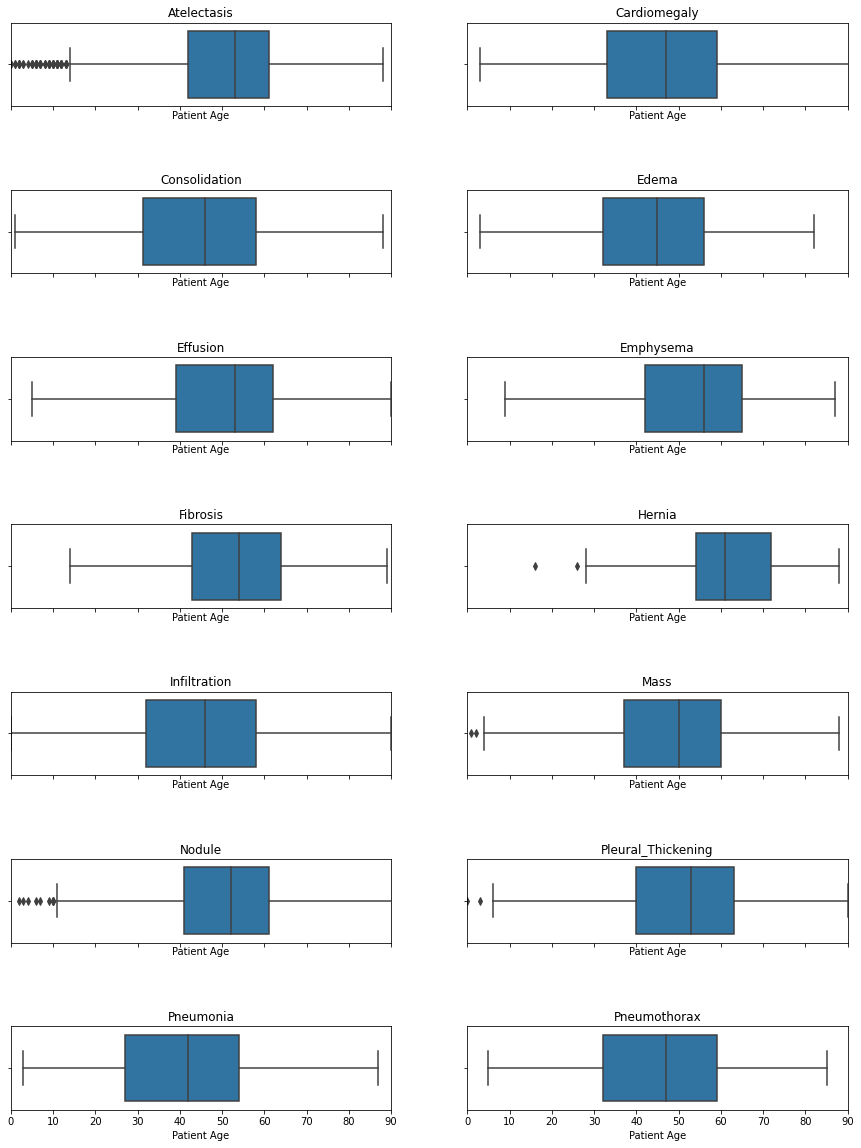}
    \caption{Boxplot for showing distribution of age for different pathologies in the ChestX-ray14 dataset.}
    \label{fig:ChestX-ray14-boxplot-age-distribution}
\end{figure*}

\begin{figure*}
    \centering
    \includegraphics[scale=0.5]{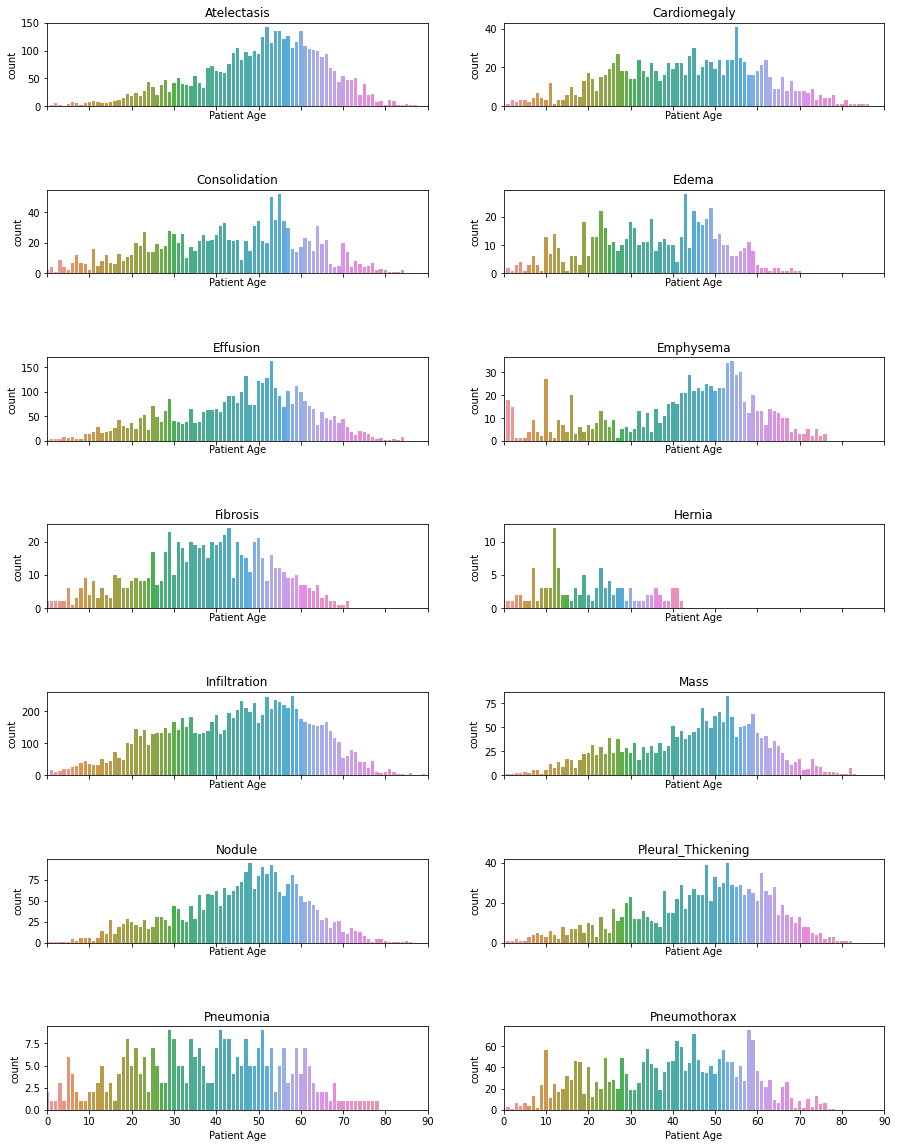}
    \caption{Distribution of age for different pathologies in the ChestX-ray14 dataset.}
    \label{fig:ChestX-ray14-plot-age-distribution}
\end{figure*}

\begin{figure*}
    \centering
    \includegraphics[scale=0.20]{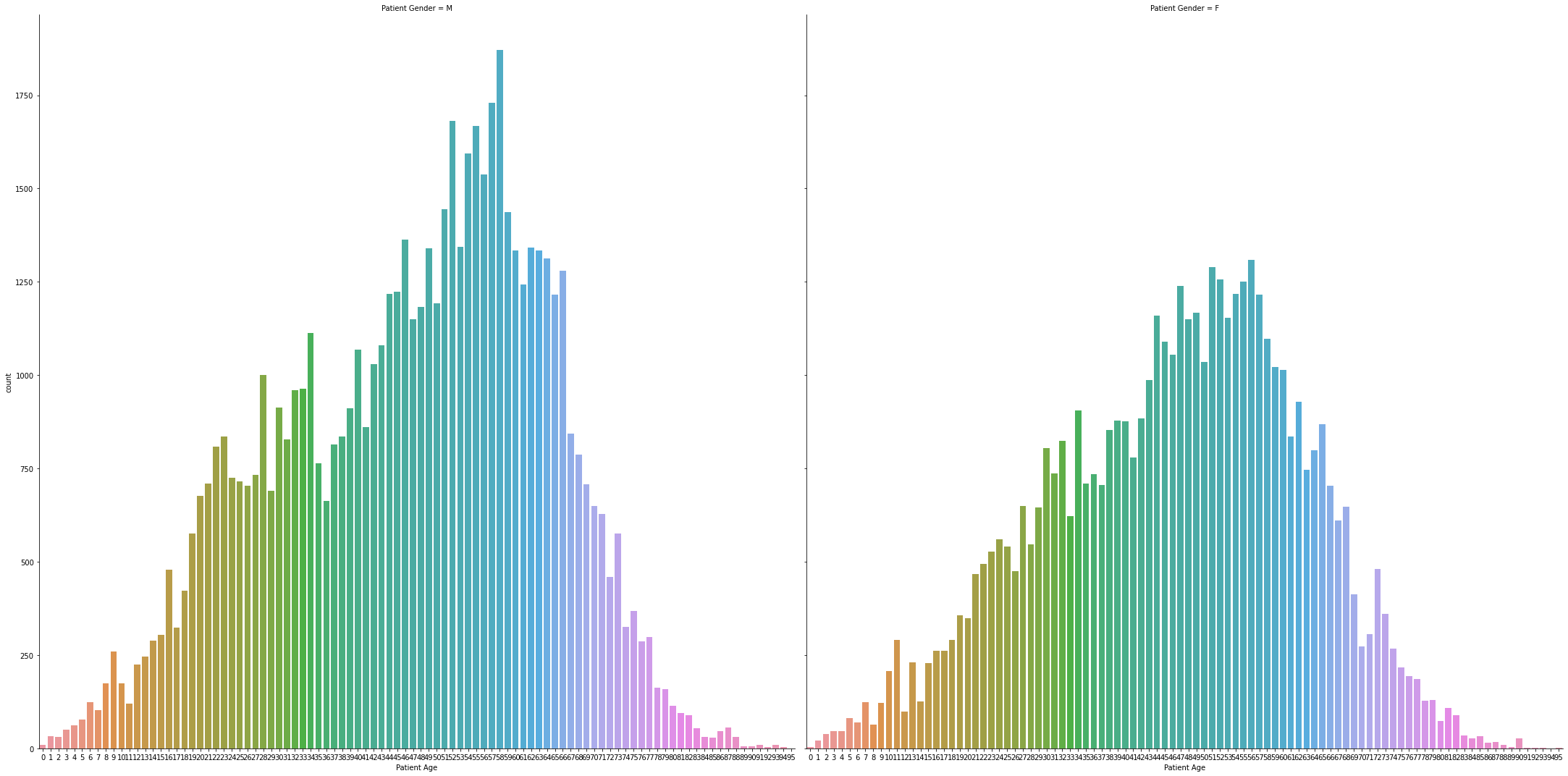}
    \caption{Gender-wise distribution of age in the ChestX-ray14 dataset.}
    \label{fig:ChestX-ray14-factorplot-age}
\end{figure*}

\begin{figure*}
    \centering
    \includegraphics[scale=0.40]{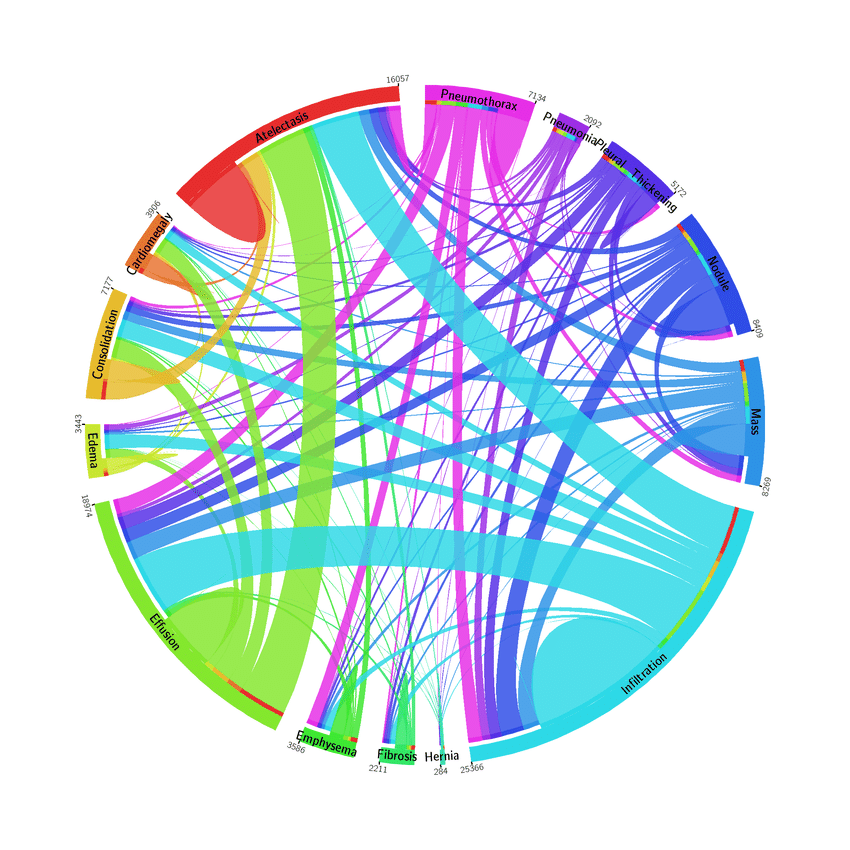}
    \caption{Chord chart showing pathology co-occurrence statistics in the ChestX-ray14 dataset~\cite{Wang:2017:ChestX-ray14}.}
    \label{fig:ChestX-ray14-chord-chart}
\end{figure*}

Before we start the data modelling, we did data analysis to have closer look to the distribution of pathologies in each of the four large scale chest X-ray datasets. The observation \emph{Infiltration} has largest number of samples in the dataset. However, \emph{Hernia} has only 227 samples in the entire ChestX-ray14 dataset. Fig.~\ref{fig:ChestX-ray14-boxplot-age-distribution} shows box plot providing the distribution of patients' age for each of the fourteen pathologies. Fig.~\ref{fig:ChestX-ray14-plot-age-distribution} gives histogram showing the distribution of patient age in the ChestX-ray14 dataset. Fig.~\ref{fig:ChestX-ray14-factorplot-age} shows gender-wise distribution of age of patients. It indicates that male patients outnumbered female patients. \cite{Larrazabal:2020:gender_imbalance_medical_imaging_datasets} studied gender imbalance in the chest X-ray datasets are reported that gender imbalance in medical imaging datasets may produce biased classifiers. They suggested having gender-balanced dataset along with diversity can increase generalisation capability of CAD systems. Fig.~\ref{fig:ChestX-ray14-chord-chart} shows chord chart showing co-occurrence of pathologies in the ChestX-ray14 dataset. This shows that many pathologies can co-occur and hence a single chest X-ray can be associated with multiple pathological labels. For example, Atelectasis often occurs with Infiltration and Effusion. We see similar patterns in the remaining three datasets.

\subsection{Evaluation metrics}
The \emph{multi-label} classification problem differs from standard classification problem in the sense that model outputs the probability of each pathology. Therefore, we use area under the receiver operating characteristics (AUC) as a metric to measure the performance of our proposed model. The AUC measures the area under the receiver operating characteristic (ROC) curve, where the x-axis and y-axis are the false positive rate (FPR) and the true positive rate (TPR), respectively. The ROC curve depicts relative trade-off between true positives and false positives. An AUC value have range of 0 to 1, which can be interpreted as the proability of a randomly chosen positive instance being scored higher by the model than a randomly chosen negative instance.  A higher AUC score implies a better classifier. Moreover, since standard accuracy metric is associated with notable difficulties in the context of imbalanced datasets, AUC score provides robust measure of model performance for each of the fourteen disease classes. 

\section{Experimental results}
This section highlights various parameters during model training and presents experimental results on four chest X-ray datasets. 

\subsection{Model training}
The model is implemented in Python with PyTorch~\cite{Paszke:2019:PyTorch} and Tensorflow libraries~\cite{Abadi:2016:Tensorflow}. We also use Pandas~\cite{Pandas} library for reading csv files, OpenCV~\cite{opencv_library} for reading and processing images, and Scikit-learn~\cite{scikit-learn} for in-build methods for model evaluation, including area under the receiver operating characteristic (AUC) score. 

We use DenseNet-121 as the backbone convolutional neural network. We initialise the CNN with the weights of the pre-trained model, trained on the large-scale ImageNet dataset, which helps the model converge faster compared to initialising model from scratch. During training, we fine-tune the model so that model can adapt to feature distribution of medical images. We implemented different data augmentation techniques including horizontal flip, rotation, contrast adjustment, position translation, and adding noise during model training. Data augmentation on the fly helps model to see increased number of images as well as images with variations during training, helping model to overcome overfitting on classes with fewer chest X-rays as well as improving model robustness to data variations. We converted images in datasets to three channel images, down-scaled images to resolution of $224 \times 224$ required for the pre-trained DenseNet-121 model, and normalised images by mean ([0.485, 0.456, 0.406]) and standard deviation ([0.229, 0.224, 0.225]) according to the images from the ImageNet. 

The final fully-connected layer is a 14-dimensional dense layer, followed by a sigmoid activations that were applied to each of the outputs to obtain the predicted probabilities of the presence of the 14 pathology classes. We used Adam optimiser with default parameters $\beta_1 = 0.9$, $\beta_2 = 0.999$ and a batch size of $8$ to find the optimal weights. The learning rate was initially set to $1e-4$ with learning decay factor of $0.97$ to reduce learning rate if no improvement in model is seen during training. During training, we save optimal model providing highest score on the validation set. We also save best checkpoints on the validation set to make them use for ensemble learning. The final results are based on an ensemble of 10 best checkpoints on the validation set and by computing the mean of the output scores over these models. The network was initialised with a pretrained model on ImageNet. During training, our goal is to minimise the binary cross-entropy loss function between the ground-truth labels and the predicted labels output by the network over the training samples.

\subsection{Results}
To evaluate the effectiveness of the proposed methodology, we compare with the baseline method, the \emph{DenseNet121} model. We also compared the results of the proposed methodology with some state-of-the-art methods for multi-label thoracic disease detection on the ChestX-ray14 dataset. Most of the notable work include~\cite{Wang:2017:ChestX-ray14,yao:2018:weakly_supervised_medical_diagnosis,Gundel:2019:Learning_to_recognise_abnormalities,Kumar:2018:Boosted_cascaded_convnets}. Table~\ref{tab:results_official_split} shows results of our proposed model compared to existing baselines, when using the official patient-wise split of the ChestX-ray14 dataset. The proposed methodology outperforms baselines in providing superior results for 11 out of 14 thoracic diseases. In Table~\ref{tab:results_image-wise_random_split}, we also report results for random image-wise split of the ChestX-ray14 dataset. On comparing mean AUC score of the studies, our proposed methodology provides superior results, except for the study by Rajpurkar \textit{et al.}~\cite{Rajpurkar:Irvin:2017:CheXNet}. However, we do emphasise that a fair comparison is not possible due to random split of the dataset. 

\begin{table*}[]
    \centering
    \caption{Area under the ROC curve (AUC) score for diagnosing thoracic diseases when using patient-wise official split of the Chest X-ray14 dataset.}
    \label{tab:results_official_split}
    \begin{tabular}{lccccc}
    \toprule
    \textbf{Pathology} & \cite{Wang:2017:ChestX-ray14} &  \cite{yao:2018:weakly_supervised_medical_diagnosis} & \cite{Gundel:2019:Learning_to_recognise_abnormalities} & ~\cite{wang:jia:2020:thorax-net} & \textbf{SA-DenseNet121}  \\
    \midrule
     Atelectasis &  0.7003 & 0.733 & 0.767 & 0.7505 & 0.8161 \\
      Cardiomegaly & 0.8100 & 0.856 & 0.883 & 0.8710 & 0.9105 \\
      Effusion &  0.7585 & 0.806 & 0.828 & 0.8181 & 0.8839 \\
      Infiltration & 0.6614 & 0.673 & 0.709 & 0.6815 &  0.7077 \\
      Mass &  0.6933 & 0.718 & 0.821 & 0.7994 & 0.8308 \\
      Nodule & 0.6687 & 0.777 & 0.758 & 0.7147 & 0.7748 \\
      Pneumonia &  0.6580 & 0.684 & 0.731 & 0.6938 &  0.7651 \\
      Pneumothorax &  0.7993 & 0.805 & 0.846 & 0.8254 &  0.8739 \\
      Consolidation & 0.7032 & 0.711 & 0.745 & 0.7415 & 0.8008 \\
      Edema &  0.8052 & 0.806 & 0.835 & 0.8354 & 0.8979 \\
      Emphysema & 0.8330 & 0.842 & 0.895 & 0.8428 & 0.9227 \\
      Fibrosis &  0.7859 & 0.743 & 0.818 & 0.8040 & 0.8293 \\
      Pleural Thickening & 0.6835 & 0.724 & 0.761 & 0.7463 & 0.7860 \\
      Hernia & 0.8717 &  0.775 & 0.896 & 0.9022 & 0.9010 \\ \midrule
      \textbf{Mean AUC} & 0.7451 & 0.761 & 0.807 & 0.7876 & 0.8357  \\
    \bottomrule
    \end{tabular}  
\end{table*}

\begin{table*}[]
    \centering
    \caption{Area under the ROC curve (AUC) score for diagnosing thoracic diseases when using image-wise random split of the Chest X-ray14 dataset.}
    \label{tab:results_image-wise_random_split}
    \begin{tabular}{lccccc}
    \toprule
    Pathology & \cite{Wang:2017:ChestX-ray14} & \cite{Yao:2017:learning_to_diagnose} & \cite{Kumar:2018:Boosted_cascaded_convnets} & \cite{Li:Wang:2018:thoracic_disease_identification} & \cite{Rajpurkar:Irvin:2017:CheXNet} \\
    \midrule
    Atelectasis &  0.716 & 0.722 & 0.762 & 0.70 & 0.821 \\
      Cardiomegaly &  0.807 & 0.904 & 0.913 & 0.81 & 0.905 \\
      Effusion &   0.784 & 0.859 & 0.864 & 0.76 & 0.883 \\
      Infiltration &  0.609 & 0.695 & 0.692 & 0.66 & 0.720 \\
      Mass &  0.706 & 0.792 & 0.750 & 0.69 & 0.862 \\
      Nodule &  0.671 & 0.717 & 0.666 & 0.67 & 0.777 \\
      Pneumonia &  0.633 & 0.713 & 0.715 & 0.66 & 0.763 \\
      Pneumothorax &  0.806 & 0.841 & 0.859 & 0.80 & 0.893 \\
      Consolidation &  0.708 & 0.788 & 0.784 & 0.70 & 0.794 \\
      Edema &   0.835 & 0.882 & 0.888 & 0.81 & 0.893 \\
      Emphysema &  0.815 & 0.829 & 0.898 & 0.83 & 0.926 \\
      Fibrosis &  0.769 & 0.767 & 0.756 & 0.79 & 0.804 \\
      Pleural Thickening & 0.708 & 0.765 & 0.774 & 0.68 & 0.814 \\
      Hernia &  0.767 & 0.914 & 0.802 & 0.87 & 0.939 \\ \hline
      \textbf{Mean AUC} & 0.738 & 0.799 & 0.795 & 0.75 & 0.842 \\
    \bottomrule
    \end{tabular}  
\end{table*}

\begin{table*}
    \centering
    \caption{Area under the ROC curve (AUC) score for diagnosing 5 pathologies on validation set when using patient-wise official split of the CheXpert dataset.}
    \begin{tabular}{l|c|c|c|c|c}
      \toprule
      \textbf{Pathology} & \cite{Irvin:2019:CheXpert}  & \cite{Pham:le:2021:interpreting_chest_x-rays} &  \textbf{DenseNet121} & \textbf{SA-DenseNet21}\\ \midrule
      Atelectasis &  0.858 & 0.825 & 0.847 & 0.862\\
      Cardiomegaly &  0.832 & 0.855 & 0.859 & 0.861\\
      Consolidation &  0.899 & 0.937 & 0.900 & 0.916\\
      Edema &  0.941 & 0.930 & 0.936 & 0.936\\
      Pleural Effusion & 0.934 &  0.923 & 0.940 & 0.944\\ \midrule
      \textbf{Mean AUC} & 0.893 & 0.894 &  0.896 & 0.904\\ \bottomrule
    \end{tabular}
    \label{tab:results_5_pathologies_official_split_CheXpert}
\end{table*}

\begin{table*}
    \centering
    \caption{Area under the ROC curve (AUC) score for diagnosing thoracic diseases on test set of the ChestX-ray14 dataset.}
    \begin{tabular}{l|c|c|c|c|c}
      \toprule
      \textbf{Pathology} &  \textbf{DenseNet121} & \textbf{SA-DenseNet121} \\ \midrule
      Atelectasis & 0.8161 & 0.8322 \\
      Cardiomegaly & 0.9105 & 0.9279\\
      Effusion & 0.8839 & 0.9026 \\
      Infiltration & 0.7077 & 0.7468 \\
      Mass & 0.8308 & 0.8642 \\
      Nodule & 0.7748 & 0.7849 \\
      Pneumonia & 0.7651 & 0.7721 \\
      Pneumothorax & 0.8739 & 0.8948 \\
      Consolidation & 0.8008 & 0.8375 \\
      Edema & 0.8979 & 0.9126 \\
      Emphysema & 0.9227 & 0.9238 \\
      Fibrosis & 0.8293 & 0.8435 \\
      Pleural Thickening & 0.7860 & 0.8219 \\
      Hernia &  0.9010 & 0.9136 \\ \midrule
      \textbf{Mean AUC} & 0.8357 & 0.8556 \\ \bottomrule
    \end{tabular}
    \label{tab:results_ChestX-ray14_SADensenetvsDensenet}
\end{table*}

\begin{table*}
    \centering
    \caption{Area under the ROC curve (AUC) score for diagnosing thoracic diseases on validation set of the CheXpert dataset.}
    \begin{tabular}{l|c|c|c|c|c}
      \toprule
      \textbf{Pathology} &  \textbf{DenseNet121} & \textbf{SA-DenseNet121} \\ \midrule
      No Finding & 0.8885 & 0.9123\\
      Enlarged Cardiomediastinum & 0.6114 & 0.6575 \\
      Cardiomegaly & 0.8157 & 0.8246 \\
      Lung Lesion & 0.8071 & 0.8291 \\
      Lung Opacity & 0.7740 & 0.8045 \\
      Edema & 0.8695 & 0.8832 \\
      Consolidation & 0.7198 & 0.7376 \\
      Pneumonia & 0.7241 & 0.7345 \\
      Atelectasis & 0.7088 & 0.7129 \\
      Pneumothorax & 0.8368 & 0.8379 \\
      Pleural Effusion & 0.8747 & 0.8873 \\
      Pleural Other & 0.9212 & 0.9318 \\
      Fracture & 0.7360 & 0.7366 \\
      Support Devices &  0.8819 & 0.8910 \\ \midrule
      \textbf{Mean AUC} & 0.7977 & 0.8129\\ \bottomrule
    \end{tabular}
    \label{tab:results_Chexpert_SADensenetvsDensenet}
\end{table*}

\begin{table*}
    \centering
    \caption{Area under the ROC curve (AUC) score for diagnosing thoracic diseases on test set of the MIMIC-CXR-JPG dataset.}
    \begin{tabular}{l|c|c|c|c|c}
      \toprule
      \textbf{Pathology} &  \textbf{DenseNet121} & \textbf{SA-DenseNet121} \\ \midrule
      No Finding & 0.8051 & 0.8120 \\
      Enlarged Cardiomediastinum & 0.6783 & 0.6815 \\
      Cardiomegaly & 0.7816 & 0.7941 \\
      Lung Lesion & 0.7190 & 0.7184 \\
      Lung Opacity & 0.6945 & 0.7083 \\
      Edema & 0.8392 & 0.8410 \\
      Consolidation & 0.7155 & 0.7117 \\
      Pneumonia & 0.7150 & 0.7184 \\
      Atelectasis & 0.7623 & 0.7660 \\
      Pneumothorax & 0.7997 & 0.8030 \\
      Pleural Effusion & 0.8867 & 0.8976 \\
      Pleural Other & 0.7925 & 0.7987 \\
      Fracture & 0.6604 & 0.6625 \\
      Support Devices &  0.8924 & 0.9134 \\ \midrule
      \textbf{Mean AUC} & 0.7673 & 0.7733 \\ \bottomrule
    \end{tabular}
    \label{tab:results_official_split_MIMIC-CXR-JPG}
\end{table*}

\begin{table*}
    \centering
    \caption{Area under the ROC curve (AUC) score for diagnosing thoracic diseases on test set of the IU-CXR dataset.}
    \begin{tabular}{l|c|c|c|c|c}
      \toprule
      \textbf{Pathology} &  \textbf{DenseNet121} & \textbf{SA-DenseNet121} \\ \midrule
      No Finding & 0.7152 & 0.7301 \\
      Enlarged Cardiomediastinum & 0.7013 & 0.7264 \\
      Cardiomegaly & 0.7179 & 0.7360 \\
      Lung Lesion & 0.5130 & 0.5439 \\
      Lung Opacity & 0.6924 & 0.6941 \\
      Edema & 0.7331 & 0.7422 \\
      Consolidation & 0.4973 & 0.5218 \\
      Pneumonia & 0.7003 & 0.7160 \\
      Atelectasis & 0.7707 & 0.7873 \\
      Pneumothorax & 0.5974 & 0.6204 \\
      Pleural Effusion & 0.5783 & 0.5829 \\
      Pleural Other & 0.8473 & 0.8485 \\
      Fracture & 0.6021 & 0.6113 \\
      Support Devices &  0.6904 & 0.7170 \\ \midrule
      \textbf{Mean AUC} & 0.6683 & 0.6841 \\ \bottomrule
    \end{tabular}
    \label{tab:results_IU-CXR}
\end{table*}

To extend our proposed methodology on the CheXpert dataset, we first apply it for only five pathologies for which ground-truth annotations are provided in the dataset and also we can compare results with existing studies~\cite{Irvin:2019:CheXpert,Pham:le:2021:interpreting_chest_x-rays}. We consider results from \cite{Irvin:2019:CheXpert} under \emph{U-Ones} policy and from \cite{Pham:le:2021:interpreting_chest_x-rays} under \emph{U-Ones+CT+LSR} policy to have fair comparison of results as we consider \emph{U-Ones} policy in our experimental setup. Table~\ref{tab:results_5_pathologies_official_split_CheXpert} shows results for diagnosing five pathologies for which radiologists' annotations are provided. The experimental results show superior results for the self-attention augmented DenseNet121 model compared to baseline methods.

Having proven the effectiveness of self-attention mechanism comparing baseline methods in diagnosing thoracic diseases on the ChestX-ray14 and the CheXpert dataset, we further extend our experiments on the MIMIC-CXR-JPG and the IU-CXR dataset. Table~\ref{tab:results_ChestX-ray14_SADensenetvsDensenet} results comparing baseline DenseNet121 model with self-attention augmented DenseNet121 model. The experimental results shows the effectiveness of augmenting convolutional network with the self-attention mechanism. Table~\ref{tab:results_Chexpert_SADensenetvsDensenet}, Table~\ref{tab:results_official_split_MIMIC-CXR-JPG}, and Table~\ref{tab:results_IU-CXR} shows results for diagnosing thoracic diseases, comparing the baseline DenseNet121 model with the self-attention augmented DenseNet121 model. The results for self-attention augmented DenseNet121 (\textbf{SA-DenseNet121}) are consistently superior to that of \textbf{DenseNet121} model. For most of the thoracic diseases, the SA-DenseNet121 outperforms DenseNet121 model, except for few observations where scores are close to the DenseNet121 model. On taking a closer look to these results, we find that self-attention provides better increment in results for certain thoracic diseases, such as \emph{lesion} and \emph{consolidation}. Since thoracic diseases such as lesion and consolidation occur in smaller regions, providing higher results using self-attention highlights its ability to capture features at both global and local regions in images. Out of all the four datasets, IU-CXR shows lowest mean AUC score of $0.6841$ compared to mean AUC scores of $0.8556$, $0.7733$, and $0.8129$ on ChestX-ray14, MIMIC-CXR-JPG, and CheXpert dataset respectively. We hypothesise this is due to the small size of the IU-CXR dataset compared to other three datasets which are of large-scale. Given self-attention based models are really complex and have billions of parameters, it is plausible that they overfit on smaller datasets.

\section{Discussion}
Based on the experimental results on all the four large-scale chest X-ray datasets, the self-attention augmented convolutional network (SA-DenseNet121) provides better results compared to the baseline (DenseNet121) model. This indicates that augmenting self-attention to convolutional network helps to capture both global and local features, in turn improving diagnostic ability of the model to recognise thoracic diseases. Compared to three large-scale datasets, namely ChestX-ray14, CheXpert, and MIMIC-CXR-JPG, the overall performance of both the baseline and the SA-DenseNet121 on the IU-CXR dataset is lower. We hypothesise this is due to the small dataset size, having only 7,418 chest X-rays, compared to other three datasets, having more than 100,000 chest X-rays. Taking a closer look of results for each of the pathology labels, we find that model provides higher scores for large objects such as ``Cardiomegaly", ``Emphysema", and ``Pneumothorax" compared to the smaller objects such as ``Mass" and ``Nodule". We hypothesise this is because pathologies which are visible on large regions in the chest X-ray can easily be localised. On the other hand, pathologies such as \emph{mass} and \emph{nodule} occur in smaller regions, making it hard for the model to easily localise in the chest X-ray. The results show that self-attention is complementary to convolutional networks, making the proposed methodology to learn more discriminative features for different pathologies. Taking mean AUC score for all pathologies in dataset as model performance on the dataset, we find variation in results for all the four datasets. This can be attributed to sources of dataset having different acquisition protocols, different scanners, and varying demographics of patients. It is challenging to have a single model which can generalise well on different datasets and providing same level of performance. 

Although the proposed methodology gives promising results in diagnosing thoracic diseases from chest X-rays, there are limitations which we want to highlight. First, all the four chest X-rays datasets considered in this study suffers from the issue of class-imbalance. Although, data augmentation and transfer learning helps to mitigate the effect of class-imbalance problem, specific techniques to handle class-imbalance can be applied. Second, pathology labels such as \emph{pneumonia}, can't be diagnosed solely based on chest radiographs. In actual clinical practice, to diagnose pneumonia, pathology laboratory test as well as chest X-ray studies are evaluated to reach final decision. Therefore, this can affect model performance. Third, the structured labels for chest X-ray datasets used in this study are curated by applying NLP techniques on radiology reports, which are accompanied in hospital PACS. The NLP labelers are build on different methods from parsing, negation detection, mention extraction, mention classification, and aggregation, errors in each stage can provide room for wrong labels. The CheXpert NLP gives more than $95\%$ accuracy, leaving room for $5\%$ of wrong labeling of structured labels. The wrong labeling of chest X-rays can have adverse affect on the model training. Fourth, although different policies have been adopted for mapping uncertain labels to either zeros (U-Zeros policy) or ones (U-Ones policy), still it is not an effective way for model to learn these uncertain cases. One way to handle this is to treat uncertain label as one of the multi-label class, providing exponentially more facets for evaluation. Fifth, incorporating the correlation among different thoracic diseases as a domain prior can possibly further improve the results. 

\section{Conclusions}
The early diagnosis and treatment of thoracic diseases is important for saving human lives. In this paper, we presented a multi-head self-attention augmented convolutional network for the diagnosis of common thoracic diseases from chest X-rays. Experiments on four the largest chest X-ray datasets shows that the proposed model is effective in diagnosing thoracic diseases. Compared with the baseline methods, our proposed approach yielded competitive results. The development of computer-aided systems for the diagnosis of thoracic diseases can augment radiologists by providing second opinion, improve clinical outcomes, reduce diagnostic errors, and fasten clinical workflow. 

\section*{Data availability}
The NIH ChestX-ray14 dataset is publicly available at \url{https://nihcc.app.box.com/v/ChestXray-NIHCC}. The Indiana University Chest X-ray collection is available at \url{https://openi.nlm.nih.gov/}. The CheXpert dataset can be downloaded by filling in the form at \url{https://stanfordmlgroup.github.io/competitions/chexpert/}. The MIMIC-CXR and MIMIC-CXR-JPG datasets can be downloaded after signing of a data use agreement at \url{https://physionet.org/content/mimic-cxr-jpg/2.0.0/}.

\begin{acknowledgements}
This research was undertaken with the assistance of resources from the National Computational Infrastructure (NCI) supported by the Australian Government.
\end{acknowledgements}

\noindent\textbf{Conflict of Interest}: There are no conflicts of interests in this work.

\bibliographystyle{spmpsci}      
\bibliography{references.bib}   
\end{document}